\documentclass{article}

\usepackage{arxiv}

\usepackage[utf8]{inputenc}
\usepackage{amsmath}
\usepackage{mathrsfs}
\usepackage{amsmath,bm}
\usepackage{amssymb}
\usepackage{mathtools}
\usepackage{bbm}
\usepackage{graphicx}
\usepackage{esint}
\usepackage{enumitem}
\usepackage{amsthm}
\usepackage{amsfonts}
\usepackage{booktabs}
\usepackage{caption,subcaption}
\usepackage{epstopdf}
\usepackage[svgnames]{xcolor}
\usepackage{listings}
\usepackage{mathtools}
\usepackage{authblk}
\usepackage{algorithm,algpseudocode}
\usepackage[colorlinks=true,linkcolor=blue,citecolor=blue]{hyperref}
\usepackage[sort, numbers]{natbib}
\usepackage{pgf,tikz}
\usepackage{float}
\usepackage{relsize}

\usetikzlibrary{shapes,arrows,positioning,fit,calc}
\tikzset{block/.style={draw, thick, text width=3cm, minimum height=1.5cm, align=center},   
line/.style={-latex}   
}  
\tikzstyle{block} = [rectangle, draw, 
    text width=5em, text centered, rounded corners, minimum height=4em]
\tikzstyle{line} = [draw, -latex']

\theoremstyle{thmstyleone}%
\theoremstyle{thmstyletwo}%
\newtheorem{remark}{Remark}%

\theoremstyle{thmstylethree}%
\title{Data-driven synchronization-avoiding algorithms in the explicit distributed structural analysis of soft tissue}
\author[]{Guoxiang Grayson Tong}
\author[]{Daniele E. Schiavazzi}
\affil[]{Department of Applied and Computational Mathematics and Statistics\\ University of Notre Dame, Notre Dame, IN, USA}
\date{ }

\begin{document}
\maketitle


\begin{abstract}
\noindent We propose a data-driven framework to increase the computational efficiency of the
explicit finite element method in the structural analysis of soft tissue. An encoder-decoder long short-term memory deep neural network is trained based on the data produced by an explicit, distributed finite element solver. We leverage this network to predict synchronized displacements at shared nodes, minimizing the amount of communication between processors. We perform extensive numerical experiments to quantify the accuracy and stability of the proposed synchronization-avoiding algorithm.
\end{abstract}

\section{Introduction}\label{intro}
A wide range of convergent numerical approaches with rigorously derived error bounds are available from numerical analysis for time integration of ordinary and partial differential equation models.
These methods, combined with the increasing availability of high performance computational resources have significantly contributed to the remarkable realism achievable by modern high-fidelity numerical models in many fields.

This paper focuses on distributed explicit time integrators, where time updates are computed through matrix-vector products and are therefore highly scalable and amenable to efficient GPU implementation. Highly scalable GPU solvers for physics-based modelling are already available in the literature~\cite{bartezzaghi2015explicit, li2021ensemble, strbac2017gpgpu, komatitsch2010high, huthwaite2014accelerated} with GPU-based accelerated explicit finite element structural simulations of soft tissues discussed, for example, in~\cite{li2021ensemble, strbac2017gpgpu,joldes2010real,4388142}.
Unlike implicit time integration, explicit schemes typically do not need element-level quantities to be assembled in a global matrix, leading to memory and runtime savings. 
However, explicit schemes are only conditionally stable~\cite{hughes2012finite,belytschko2014nonlinear,bartezzaghi2015explicit,clough1993dynamics} with time step size a few order of magnitude smaller compared to their implicit counterpart. 
This difference becomes less pronounced for the structural analysis of \emph{biological soft tissue} where explicit approaches have the potential to be competitive with respect to implicit time integration, for example in the context of cardiovascular modeling.

In explicit schemes, the main cost per time step relate to the computation of element-level quantities (mass, stiffness matrix and load vector) and the cost of communication. While the first can be mitigated by reduced numerical integration~\cite{reduced,reduced2} or closed-form representations~\cite{mccaslin2012closed, shiakolas1992closed}, the second remains a fundamental bottleneck, despite several optimized approaches proposed in the context of GPU-based distributed computation~\cite{joldes2010real,aslam2020performance, 4388142, 10.1145/3322813}.
Since synchronization must be performed at every time step, this problem is also exacerbated, in explicit solver, by the small size of the stable time steps.

To alleviate such cost, recently developed data-driven approaches offer a possible solution. 
The expressive power of artificial neural networks has been widely demonstrated in the construction of surrogate models for dynamical systems, producing fast emulators that can be integrated in optimization and UQ design loops.
In this context, extensive recent work include the use of residual networks (ResNet~\cite{he2016deep}) for data-driven generalization of explicit Euler time integrators~\cite{QIN2019620,CHEN2022110782, WU2020109307,churchill2022deep,fu2020learning}, showing promising results for both linear and nonlinear dynamical systems.
Other methods are based, for example, on physics-informed neural networks~\cite{raissi2019physics}, deep operator networks~\cite{lu2021learning} and 
convolutional networks assembled from encoders and decoders~\cite{lauren_mf}.
Others incorporate spectral properties of system dynamics in the design of data-driven models, to realize linearization and handle high-dimensionality~\cite{Kaiser_2021,kutz2016dynamic} or use sparse regression to construct parsimonious surrogates with model complexity from an a-priori selected dictionary (sparse identification of nonlinear dynamics or SINDy~\cite{sindy,sindy2,PhysRevResearch.3.023255}). 
Note that all the approaches above aim to create effective data-driven surrogate models of dynamical systems, rather than levering new advances in data driven architectures to further improve the efficiency of numerical schemes.

In this paper, we combine numerical simulation and data-driven approaches to mitigate the synchronization bottleneck in explicit distributed time integration.
We equip each processor with an independent network which models synchronized displacement solutions for the shared nodes of the respective partition, in order to reduce synchronization frequency and to increase the degree of parallelism.
Using the proposed approach, substantial savings are obtained for the cost of communication without compromising accuracy and long-term stability. In addition, multiple networks are employed to predict the displacements for the same shared nodes, providing a means by which to assess prediction robustness and to bound approximation error.
Our data-driven framework (built based on the \texttt{PyTorch} library~\cite{paszke2019pytorch} and publicly available as a GitHub repository at \url{https://github.com/desResLab/Synchronization-avoiding-algorithms}) uses recurrent neural networks (RNN) due to their ability to handle time series data~\cite{FUNAHASHI1993801,PhysRevE.105.044205,sherstinsky2020fundamentals,RNN-anti}. However, since vanilla RNNs are unable to effectively learn long-term dependence in the data, we employ long short term memory (LSTM) encode-decoder networks~\cite{lstm,lstm1}, that have received significant previous attention in the context of dynamical systems~\cite{Hu2022NeuralPDEAR,lstm1,park2018sequence,goodfellow2016deep,chaotic17}.

This model successfully fits our purpose of approximating dynamical systems only on a subset (shared nodes) of the entire system. 
Learning the partial rather than the full dynamics by a non-recurrent neural network may require, for example, the construction of a memory kernel of the Mori-Zwanzig type~\cite{churchill2022deep, fu2020learning}. This is, however, not required for the proposed LSTM recurrent network which inherently holds a temporal memory due to its sequential input~\cite{sherstinsky2020fundamentals,PhysRevE.105.044205}.
Additional work on learning the dynamics of a subset of degrees of freedom can be found in~\cite{bakarji2022discovering}.

This paper is organized as follows. A brief review of the governing equations for linear elastodynamics and their spatial discretization into finite elements is reported in Section~\ref{chp:elas}, followed by the formulation for our distributed explicit solver in Section~\ref{chp: explicit}.
The proposed data-driven method is discussed by first presenting a single LSTM cell in Section~\ref{chp:lstm}, followed by an encoder-decoder layout in Section~\ref{chp:ED}, with the details of network training summarized in Section~\ref{sample}.
A data-driven synchronization-avoiding algorithm is proposed in Section~\ref{syn-alg}.
Numerical tests are performed in Section~\ref{chp:numerical}, starting with extensive benchmarks on a simple cantilever problem in Section~\ref{canti}, and a realistic coronary model in Section~\ref{cardio}.
In Section~\ref{chp:ec}, we discuss a few criteria to measure the prediction error of our proposed framework.
Performance is assessed in Section~\ref{chp:perf} for a fixed number of cores and fixed mesh size, respectively, while Section~\ref{chp:conclusion} contains the conclusions and addresses possible future research directions.

\section{Governing equations and discretization}
\subsection{Equations of linear elastodynamics}\label{chp:elas}

{\bf Strong form} - Consider the following initial-boundary value problem $(\mathcal{S})$ defined over the domain $\Omega \in \mathbb{R}^3$ with Lipschitz boundary $\partial \Omega = \partial\Omega_{d}\cup\partial\Omega_{n}$ and $\bar{\Omega} = \Omega \cup \partial\Omega$ (see, e.g., \cite{hughes2012finite})
\begin{equation}
{\small
(\mathcal{S}):
\begin{cases}
\textrm{Given} \ f_i, g_i, d_{0i}, \dot{d}_{0i},  \\
\textrm{find} \ d_i,\,\,i,j=\{x,y,z\}, \ s.t.\\[7.5pt]
\begin{aligned}
 \rho \ddot{d}_{i} &= \frac{\partial \sigma_{ij}}{\partial x_j} + f_i  &&\text{in} \ \Omega \times (0, T]\\
d_i &= g_i &&\text{on} \ \partial \Omega_{d} \times (0, T]\\
\sigma_{ij}n_j &= h_i&&\text{on} \ \partial \Omega_{n} \times (0, T]\\
d_i &= d_{0i} &&\text{in} \ \Omega, \ \text{at} \ t=0\\
\dot{d}_{i} &= \dot{d}_{0i} &&\text{in} \ \Omega, \ \text{at} \ t=0,\\
\end{aligned}
\end{cases}}
\label{strong}
\end{equation}
where the quantity of interest $d_i(\boldsymbol{x}, t):\bar{\Omega} \times [0, T]\to \mathbb{R}$ denotes the $i$-th component of the displacement field $\boldsymbol{d}$, $\rho$ is the material density,  $f_i(\boldsymbol{x}, t): \Omega\times (0, T]\to \mathbb{R}$ is the $i$-th component of the body force, $g_i(\boldsymbol{x}, t): \partial \Omega_d \times (0, T]\to \mathbb{R}$,   $h_i(\boldsymbol{x}, t): \partial \Omega_n \times (0, T]\to \mathbb{R}$ are the $i$-th components of the prescribed Dirichlet and Neumann boundary conditions on $\partial\Omega_{d}$ and $\partial\Omega_{n}$, respectively.
In addition, initial values $d_{0i}, \dot{d}_{0i}$ are set for the displacement and velocity component.

We also assume a linear, elastic and isotropic constitutive model in the small strain regime of the form
\begin{equation}
{\small
\sigma_{ij} = 2\mu\epsilon_{ij} + \lambda \delta_{ij}\epsilon_{kk}},\,\,i,j,k = \{x,y,z\},
\label{iso}
\end{equation}
where $\boldsymbol{\sigma} = [\sigma_{ij}]$ is the Cauchy stress tensor, $\boldsymbol{\epsilon} = [\epsilon_{ij}]$ is the infinitesimal strain tensor, $\delta_{ij}$ is the Kronecker delta, and the Lam\'e coefficients $\mu$ and $\lambda$ are defined as
\begin{equation}
\mu =  \displaystyle{\frac{E}{2(1+\nu)}};  \quad \lambda = \dfrac{E\nu}{(1+\nu)(1-2\nu)},
\end{equation}
where $E$ is the material Young's modulus, $\nu$ is the Poisson ratio.

The infinitesimal strain $\boldsymbol{\epsilon}$ is obtained from the symmetric part of the displacement gradient tensor
\begin{equation}
\epsilon_{ij} = d_{(i,j)} = \dfrac{1}{2}\left(\dfrac{\partial d_{i}}{\partial x_j} + \dfrac{\partial d_{j}}{\partial x_i}\right),
\end{equation}
leading to an expression of the Cauchy stress in~\eqref{iso} in terms of displacements as
\begin{equation}
\sigma_{ij} = c_{ijkl} \epsilon_{kl} = c_{ijkl} d_{(k,l)},
\end{equation}
where $\pmb{\mathsf{C}} = [c_{ijkl}]$ is a fourth-order elasticity tensor, defined as (see, e.g., \cite{Scovazzi2016})
\begin{equation}
c_{ijkl} =  \mu (\delta_{ik}\delta_{jl} + \delta_{il}\delta_{jk}) +\lambda \delta_{ij}\delta_{kl}.
\end{equation}

\noindent{\bf Weak form} - A weak or variational formulation for problem~\eqref{strong} can be written as
\begin{equation}
{\small
(\mathcal{W}):
\begin{cases}
\textrm{Given} \ \boldsymbol{f}, \boldsymbol{g}, \boldsymbol{d}_{0}, \dot{\boldsymbol{d}}_0, \ \textrm{find} \ \boldsymbol{d}(\boldsymbol{x},t) \in \mathscr{D}^t, \\
\textrm{that for any} \ \boldsymbol{w} \in \mathscr{W}, \ \ s.t.\\[7.5pt]
\begin{aligned}
   \big(\rho\ddot{\boldsymbol{d}}, \boldsymbol{w}\big)_{\Omega} + a\big(\boldsymbol{d}, \boldsymbol{w} \big)_{\Omega} &= l\big(\boldsymbol{w}\big)_{\Omega} + l_n\big(\boldsymbol{w}\big)_{\partial\Omega_{n}}\\
    \big(\rho\boldsymbol{d}(\boldsymbol{x},0), \boldsymbol{w}\big)_{\Omega} &=  \big(\rho\boldsymbol{d}_0,\boldsymbol{w}\big)_{\Omega}\\
    \big(\rho\dot{\boldsymbol{d}}(\boldsymbol{x},0), \boldsymbol{w}\big)_{\Omega} &=  \big(\rho\dot{\boldsymbol{d}_0},\boldsymbol{w}\big)_{\Omega}\\
\end{aligned}
\end{cases}}
\label{weak}
\end{equation}
where $\big(\cdot, \cdot \big)_{\Omega}$ denotes the standard product in $L^2(\Omega)$, and the linear and bilinear forms $l\big(\cdot \big)_{\Omega}$, $l_n\big(\cdot\big)_{\partial\Omega_{n}}$ and $a\big(\cdot, \cdot \big)_{\Omega}$ are defined, respectively, as
\begin{equation}
a\big( \boldsymbol{d}, \boldsymbol{w}\big)_{\Omega} = \int_{\Omega} w_{(i,j)} c_{ijkl} d_{(k,l)} d\Omega,\,\,l\big(\boldsymbol{w} \big)_{\Omega} = \int_{\Omega} w_i f_i d\Omega,\,\,l_n\big(\boldsymbol{w} \big)_{\partial\Omega_{n}} = \int_{\partial\Omega_{n}} w_i\, h_i\,dA.
\end{equation} 
The weak form~\eqref{weak} relaxes the regularity requirement of displacement solutions for the strong form~\eqref{strong} such that the trial and test spaces only need to satisfy the conditions
\begin{equation}
\begin{split}
\mathscr{D}^t &\coloneqq \Big\{\boldsymbol{d}(\boldsymbol{x},t)\rvert d_i(\boldsymbol{x},t) \in H^1(\Omega); d_i(\boldsymbol{x},t) = g_i(\boldsymbol{x},t), \forall \boldsymbol{x} \in \partial \Omega_d; t\in (0,T]\Big\},\\
\mathscr{W} &\coloneqq \Big\{ \boldsymbol{w}(\boldsymbol{x})\rvert w_i(\boldsymbol{x}) \in H^1(\Omega); w_i(\boldsymbol{x}) = 0, \ \forall \boldsymbol{x} \in \partial \Omega_d\Big\},
\end{split}
\end{equation}
where $H^1(\Omega)$ is the standard Sobolev space of order 1.

\vspace{3pt}

\noindent{\bf Discrete matrix form} - The solution of \eqref{weak} by a Bubnov-Galerkin finite element approach requires the selection of appropriate discrete subspaces $\mathscr{D}^{(n)}_h \subset \mathscr{D}^t$ and $\mathscr{W}_h \subset \mathscr{W}$, such that the \emph{projected} solution $\boldsymbol{d}_h$ converges to the true solution $\boldsymbol{d}$ in $L^2(\Omega)$ with respect to any $\boldsymbol{w}_h \in \mathscr{W}_h$.
We consider the discrete subspace $\mathscr{D}^{(n)}_h$ spanned by the set of linear Lagrange polynomials $\mathcal{P}^1$, leading to the following semi-discrete matrix formulation
\begin{equation}
{\small
(\mathcal{M}):
\begin{cases}
\textrm{Find} \ \boldsymbol{d}_h^{\hspace{0.02cm}(n)} \in \mathscr{D}_h^{(n)},  \ n \in \{1,2,\cdots, n_T\}, s.t.\\[7.5pt]
\begin{aligned}
    \mathbf{M}\ddot{\boldsymbol{d}}_h^{\hspace{0.02cm}(n)} +\mathbf{C}\dot{\boldsymbol{d}}_h^{\hspace{0.02cm}(n)} +\mathbf{K}\boldsymbol{d}_h^{\hspace{0.02cm}(n)} = \boldsymbol{f}^{\hspace{0.02cm}(n)}\\
\end{aligned}
\end{cases}}
\label{d-g}
\end{equation}

\begin{equation}
\begin{split}
\mathscr{D}^{(n)}_h &\coloneqq \Big\{\boldsymbol{d}^{\hspace{0.02cm}(n)}_h(\boldsymbol{x})\rvert d^{\hspace{0.02cm}(n)}_{h,i}(\boldsymbol{x}) \in C^0(\Omega_h); \hspace{0.2cm}d_{h,i}^{{\hspace{0.02cm}(n)}}(\boldsymbol{x}) =  g^{{(n)}}_{h,i}(\boldsymbol{x}), \forall \boldsymbol{x} \in \partial \Omega_h; \hspace{0.2cm} d^{\hspace{0.02cm}(n)} _{h,i}(\boldsymbol{x})\rvert_{e} \in \mathcal{P}^1(\Omega_{h,e}); n \in \{1,2,\cdots, n_T\}\Big\} \\
\mathscr{W}_h &\coloneqq \Big\{\boldsymbol{w}_{h}(\boldsymbol{x})\rvert w_i(\boldsymbol{x}) \in C^0(\Omega_h); w_{hi}(\boldsymbol{x}) = 0, \forall \boldsymbol{x} \in \partial \Omega_h; w_{hi}(\boldsymbol{x})\rvert_{e} \in \mathcal{P}^1(\Omega_{h,e})\Big\},\\
\end{split}
\end{equation}
where the subscripts $(\cdot)\rvert_e$, $(\cdot)_e$ indicate restriction to a single finite element and $n_T$ is the total number of time steps. The quantities $\mathbf{M},\mathbf{C},\mathbf{K}, \boldsymbol{f}$ denote the mass, damping, stiffness matrices and loading vector, respectively with element-level expressions that are standard in the isoparametric finite element literature (see, e.g.,~\cite{hughes2012finite}) and are therefore omitted. In this paper, we consider mass-proportional damping, i.e., $\mathbf{C} = \alpha\, \mathbf{M}$ with damping factor $\alpha \in \mathbb{R}$ (see, e.g.,~\cite{hughes2012finite}).

\subsection{A distributed explicit structural finite element solver}\label{chp: explicit}

\noindent{\bf Explicit time integration} - The algebraic system \eqref{d-g} is discretized in time using an explicit second-order central difference stencil, where structural accelerations and velocities are approximated at every time point $n$ as
\begin{equation}
\boldsymbol{a}_{h}^{\hspace{0.005cm}(n)} = \ddot{\boldsymbol{d}}_h^{\hspace{0.02cm}(n)} = \displaystyle{\frac{\boldsymbol{d}^{\hspace{0.02cm}(n+1)}_h - 2\boldsymbol{d}^{\hspace{0.02cm}(n)}_h + \boldsymbol{d}^{\hspace{0.02cm}(n-1)}_h}{\Delta t^2}} + \mathcal{O}(\Delta t^3),\,\,
\boldsymbol{v}_{h}^{\hspace{0.005cm}(n)} = \dot{\boldsymbol{d}}_h^{\hspace{0.02cm}(n)}  = \displaystyle{\frac{\boldsymbol{d}^{\hspace{0.02cm}(n+1)}_h - \boldsymbol{d}^{\hspace{0.02cm}(n-1)}_h}{2\Delta t}}+ \mathcal{O}(\Delta t^3).
\end{equation}

Consistent mass and damping matrices $\bf{M}$ and $\bf{C}$ are replaced by their lumped counterparts $\mathring{\mathbf{M}}$ and $\mathring{\mathbf{C}}$, leading to trivial inversion for fully explicit displacement-based time integrators.
To initiate the time iterations, the quantity $\boldsymbol{d}^{\hspace{0.02cm}(-1)}_h$ is computed using a second order Taylor approximation of $\boldsymbol{d}(-\Delta t)$, for consistency.
Once the initial conditions $\boldsymbol{d}_{h,0}$, $\boldsymbol{v}_{h,0}$ are provided for the displacement and velocity at time $t=0$, an initial acceleration $\boldsymbol{a}_{h,0}$ is computed by solving the following discrete system (note the use of consistent matrices)
\begin{equation}
{\small
    \mathbf{M}\,\boldsymbol{a}_{h,0} +\mathbf{C}\,\boldsymbol{v}_{h,0} +\mathbf{K}\,\boldsymbol{d}_{h,0} = \boldsymbol{f}^{(0)}}.
\end{equation}

The explicit scheme is known to be conditionally stable with respect to the choice of $\Delta t$, consistent with the well known Courant-Friedrichs-Lewy (CFL) condition (see, e.g.,~\cite{belytschko2014nonlinear})
\begin{equation}
{\small
    \Delta t = \alpha_s \frac{ \displaystyle{\min_{e=1,2,\cdots n_e} h_e} }{\sqrt{\frac{E}{\rho(1-\nu^2)}}},
    \label{cfl}}
\end{equation}
where $h_e$ is the diameter of the circumsphere associated with each tetrahedral element, $n_e$ is the total number of elements in the mesh, and the safety factor $\alpha_s \in (0,1)$ is set as $\alpha_s = 0.9$.
The local displacement solution is updated as shown in~\eqref{lumped}, introducing the internal force at step $n$ as $\boldsymbol{f}^{(n),\text{int}} =\mathbf{K}\boldsymbol{d}_h^{(n-1)}$ and renaming $\boldsymbol{f}^{(n)}$ as $\boldsymbol{f}^{(n), \text{ext}}$, i.e., the external force at step $n$. 
This is consistent with our implementation in Algorithm~\ref{alg-para}.

\begin{remark}
Application of initial conditions or loads at $t=0$ in explicit structural dynamics may lead to the excitation of a broad range of frequencies. To prevent this to occur, a \emph{ramp} is applied to the external force, through a time-dependent function $\mathscr{R}(t,t_{\text{end}})$, such that
\begin{equation}
\tilde{\boldsymbol{f}}^{(n)}(t_{\text{end}}) = \boldsymbol{f}^{(n)}\mathscr{R}{(t_{n},t_{\text{end}})}.
\end{equation}
This allows for a smooth and quasi-static application of the external loading until time $t_{\text{end}}$. 
Although many formulations are available in the literature, we select a simple linear ramp function
\begin{equation}
\mathscr{R}(t_n, t_{\text{end}}) = 
\begin{cases}
t_n/t_{\text{end}} \quad t_n\leq t_{\text{end}} \\
1 \quad \text{otherwise}.
\end{cases}
\end{equation}
\label{ramp}
\end{remark}

\begin{figure*}[!ht]
\begin{equation}
\small{
(\Tilde{\mathcal{M}}): 
\begin{cases}
\textrm{Given initial conditions: $\boldsymbol{d}_{h,0},\boldsymbol{v}_{h,0},\boldsymbol{a}_{h,0}$ and time step size $\Delta t$}\\
\textrm{Find} \ \boldsymbol{d}_h^{\hspace{0.02cm}(n)} \in \mathscr{D}_h^{(n)}, \ \textrm{for every discrete step} \ n \in \{1,2,\cdots, n_T\}\ \ s.t.\\[7.5pt]
\begin{alignedat}{2}
    &\boldsymbol{d}_h^{\hspace{0.02cm}(n)} &&= (\mathring{\mathbf{M}} + \frac{\Delta t}{2}\mathring{\mathbf{C}})^{-1}\Big[ \Delta t^2 \boldsymbol{f}^{\hspace{0.02cm}(n)} - (\Delta t^2\mathbf{K}-2\mathring{\mathbf{M}})\boldsymbol{d}^{\hspace{0.02cm}(n-1)}_{h} -(\mathring{\mathbf{M}}-\frac{\Delta t}{2}\mathring{\mathbf{C}})\boldsymbol{d}_h^{\hspace{0.02cm}(n-2)}\Big] \\
    & \ &&= (\mathring{\mathbf{M}} + \frac{\Delta t}{2}\mathring{\mathbf{C}})^{-1} \Big[ \Delta t^2(\boldsymbol{f}^{\hspace{0.02cm}(n),\text{ext}}-\boldsymbol{f}^{\hspace{0.02cm}(n),\text{int}}) + 2\mathring{\mathbf{M}}\boldsymbol{d}^{\hspace{0.02cm}(n-1)}_{h} - (\mathring{\mathbf{M}}-\frac{\Delta t}{2}\mathring{\mathbf{C}})\boldsymbol{d}_h^{\hspace{0.02cm}(n-2)}\Big]\\
    &\boldsymbol{d}_h^{\hspace{0.02cm} (0)} &&= \boldsymbol{d}_{h,0} \\ 
    &\boldsymbol{d}_h^{\hspace{0.02cm}(-1)} &&= \boldsymbol{d}_{h,0} - \Delta t \boldsymbol{v}_{h,0} + \frac{\Delta t^2}{2}\boldsymbol{a}_{h,0} \\ 
\end{alignedat}
\end{cases}}
\label{lumped}
\end{equation}
\end{figure*}

\vspace{3pt}

\noindent{\bf Distributed solver} - The pseudo-code in Algorithm~\ref{alg-para} illustrates how our displacement-based parallel finite element elastodynamics solver is implemented based on element-level computation and communication operations~\cite{bartezzaghi2015explicit, belytschko2014nonlinear,hughes2012finite}. 
Consider a computational mesh partitioned and distributed over $n_c$ processors, labeled as $i = 1,\cdots, n_c$, each containing $n_{e,[i]}$ finite elements. 

The steps in Algorithm~\ref{alg-para} emphasized using boxes denote CPU-to-CPU (or GPU-to-CPU and vice versa) synchronization tasks. These ensure equilibrium to be satisfied within each local partition at every time step, based on communicating internal and external force information at the \emph{shared} nodes (i.e., nodes belonging to multiple mesh partitions).
However, synchronization constitutes one of the main factors responsible for performance degradation in distributed structural analysis codes. This is particularly true for fully explicit time integration, where shared node information needs to be communicated to the root processor at every time step, and therefore millions or tens of millions of times during one simulation.
Thus, development of effective syncronization-avoiding strategies would boost the performance of explicit distributed finite element codes, particularly in the context of ensemble multi-GPU finite element solvers, discussed in our previous work~\cite{li2021ensemble}.

\begin{remark}\label{rmk:pre}
The procedure in Algorithm~\ref{alg-para} generalizes different types of structural problems by forming element stiffness matrix ${\bf{K}}_e$ and external loading $\boldsymbol{f}_e^{(n),\text{ext}}$ at every time step. 
For isotropic linear elastodynamics and constant external loading, it is instead sufficient to generate the local stiffness matrix and loading vector only once, before the beginning of the time loop, and re-use them at every time step.
\end{remark}

\begin{algorithm*}[h!]
\caption{Displacement-based distributed linear elastodynamics solver.}
\begin{algorithmic}
\State Communicate the Dirichlet boundary conditions and initial conditions to each processor 
\State Form and communicate the global lumped mass and damping matrices $\mathring{\mathbf{M}}$, $\mathring{\mathbf{C}}$ to each processor
\For{$n=1,2,\cdots, n_T$} \Comment{Time loop}
\State Initialize local internal and external forces $\boldsymbol{f}^{\hspace{0.02cm}(n), \text{int}}, \boldsymbol{f}^{\hspace{0.02cm}(n), \text{ext}}$ as zero vectors
\For{$e = 1, 2, \cdots, n_{e,[i]}$} \Comment{Element loop}
\State Form element
stiffness matrix $\mathbf{K}_{e}$ and external force $\boldsymbol{f}^{\hspace{0.02cm}(n), \text{ext}}_{e}$
\vspace{.15cm}
\State Calculate element internal force: $\boldsymbol{f}^{\hspace{0.02cm}(n), \text{int}}_{e} = \mathbf{K}_{e}\boldsymbol{d}_{h,e}^{\hspace{0.02cm} (n-1)}$
\vspace{.15cm}
\State Update local forces $\boldsymbol{f}^{\hspace{0.02cm}(n), \text{int}}, \boldsymbol{f}^{\hspace{0.02cm}(n), \text{ext}}$ by $\boldsymbol{f}^{\hspace{0.02cm}(n), \text{int}}_{e}$ and $\boldsymbol{f}^{\hspace{0.02cm}(n), \text{ext}}_{e}$ based on global element label $e$
\vspace{.15cm}
\EndFor
\vspace{.15cm}
\State \fbox{ Send local forces $\boldsymbol{f}^{\hspace{0.02cm}(n), \text{int}}, \boldsymbol{f}^{\hspace{0.02cm}(n), \text{ext}}$ to the root processor}
\vspace{.15cm}
\State  \fbox{  Update local forces $\boldsymbol{f}^{\hspace{0.02cm}(n), \text{int}}, \boldsymbol{f}^{\hspace{0.02cm}(n), \text{ext}}$ based on contributions from the shared mesh nodes}
\vspace{.15cm}
\State  \fbox{ Send updated local forces $\boldsymbol{f}^{\hspace{0.02cm}(n), \text{int}}, \boldsymbol{f}^{\hspace{0.02cm}(n), \text{ext}}$ back to each processor}
\vspace{.15cm}
\State Update local solution $\boldsymbol{d}_{h}^{\hspace{0.02cm} (n)}$ using \eqref{lumped}
\vspace{.15cm}
\State Strongly enforce Dirichlet boundary conditions
\EndFor
\end{algorithmic}
\label{alg-para}
\end{algorithm*}

\noindent{\bf Artificial mass scaling} - Explicit time integration schemes are stable under condition~\eqref{cfl} on $\Delta t$, where the small time steps increase the computational cost for long-term or steady state simulations, and the frequency of synchronization tasks.
In practice, artificial mass scaling is a widely adopted pre-processing technique to increase $\Delta t$, for situations where the choice of the time step size is dictated by a few small elements in the mesh. The pseudocode for a typical implementation is shown in Algorithm~\ref{alg-cms} with more recent approaches discussed, for example, in~\cite{olovsson2005selective,SchillingerMassScaling2022}.

The price to pay for a larger time step is a non-physical increase in the mass of the system that may potentially affect the system dynamics. Therefore, the scaling factor $\beta$ has to be carefully selected not to alter the structural response. In this paper, we consider Algorithm~\ref{alg-cms} applied to the most expensive numerical experiment in Section~\ref{cardio}.

\begin{algorithm*}[h!]
\caption{Artificial mass scaling.}
\begin{algorithmic}
\State Loop through all elements to determine $\Delta t$ by equation \eqref{cfl} 
\State Set a target time step $\widehat{\Delta t} = \beta \Delta t$, $\beta > 1$
\State Initialize the artificial density vector $\widehat{\pmb{\rho}}$
\For{$e= 1, 2, \cdots, n_{e}$} \Comment{Element loop}
\State Calculate the element time step size $\Delta t_{e} = \alpha_s h_e / \sqrt{\frac{E}{\rho(1-\nu^2)}} $
\If{$\Delta t_e < \widehat{\Delta t}$}
   \State $\widehat{\pmb{\rho}}[i] = E \widehat{\Delta t}^2 /\alpha_s^2h_e^2(1-\nu^2)$
\Else  
\State $\widehat{\pmb{\rho}}[i] = \rho$
\EndIf
\State Recompute the mass matrix $\widehat{{\bf{M}}}$ based on $\widehat{\pmb{\rho}}$
\EndFor
\State Compute the total percent mass increase $R_{m}$
\end{algorithmic}
\label{alg-cms}
\end{algorithm*}

\section{Data driven model}
In this section, we introduce data driven models based on artificial neural networks designed to learn the dynamics of discrete systems generated through the finite element method, specifically focusing on LSTM networks. 
Hochreiter and Schmidhuber introduced the LSTM deep neural network in their 1997 seminal paper~\cite{lstm} to overcome the problems with vanishing and exploding gradients in vanilla RNN.

In what follows, we will drop the subscript $(\cdot)_h$ since only discrete solutions will be considered. In addition, we also introduce the notation
\begin{equation*}
\|\boldsymbol{v}\|_2 = \sqrt{\sum_{i=1}^{n} \rvert v_i \rvert^2 }, \quad \boldsymbol{v} \in \mathbb{R}^n,\,\,\|\boldsymbol{V}\|_F = \sqrt{\sum_{i=1}^{n} \sum_{j=1}^{m} \rvert V_{ij} \rvert^2 }, \quad \boldsymbol{V} \in \mathbb{R}^{n\times m}.
\end{equation*}

\subsection{LSTM cell model}\label{chp:lstm}
\begin{figure}[!ht]
    \centering
    \includegraphics[scale=0.405]{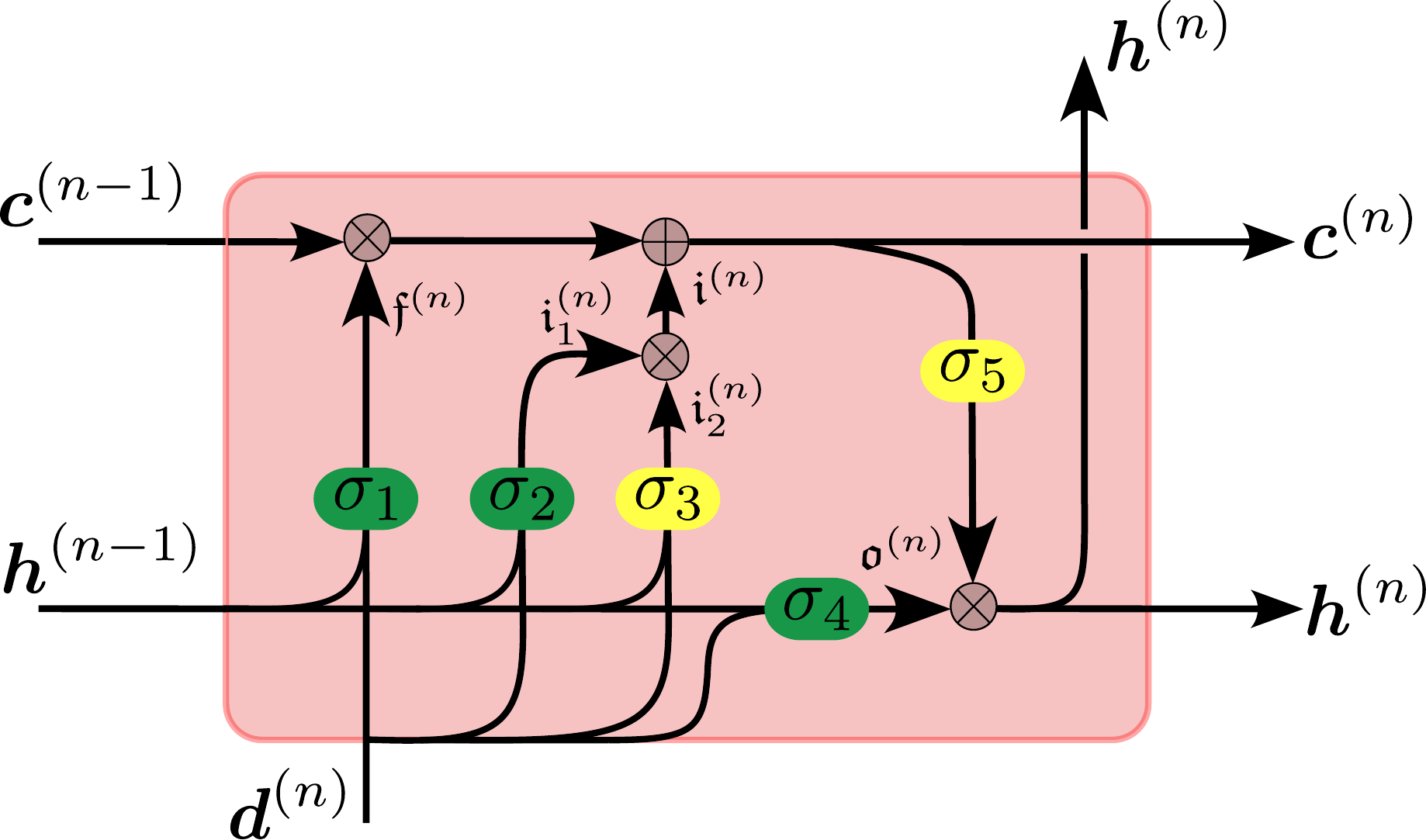}
    \caption{Schematic representation of LSTM cell.}
    \label{lstm cell}
\end{figure}
As illustrated in Fig. \ref{lstm cell}, a classical LSTM \emph{cell} consists of a hidden state $\boldsymbol{h}$, a cell state $\boldsymbol{c}$, an input $\boldsymbol{d}$, the activation functions $\sigma_1, \cdots ,\sigma_5$ and the component-wise operations $\otimes$ and $\oplus$. 
In each cell, the feedback control feature of the deep recurrent neural network is further reinforced by bringing a number of gated flow controls.

The \textbf{Forget gate} learns how information from the previous cell state $\boldsymbol{c}^{\hspace{0.02cm}(n-1)}$ will be kept or discarded. It relates to the weighting matrices $\boldsymbol{W}_{\mathfrak{f},\boldsymbol{d}}$, $\boldsymbol{W}_{\mathfrak{f},\boldsymbol{h}}$, associated with the current input $\boldsymbol{d}^{\hspace{0.02cm}(n)}$ and the previous hidden state $\boldsymbol{h}^{(n-1)}$, respectively, plus the bias $\boldsymbol{b}_{\mathfrak{f}}$. The activation $\sigma_1$ is usually a sigmoid function, which provides an output range from $0$ to $1$. The forget gate implements the expression
\begin{equation}
\small{
\mathfrak{f}^{(n)} = \sigma_1(\boldsymbol{W}_{\mathfrak{f},\boldsymbol{d}} \boldsymbol{d}^{\hspace{0.02cm}(n)} + \boldsymbol{W}_{\mathfrak{f},\boldsymbol{h}} \boldsymbol{h}^{(n-1)}+ \boldsymbol{b}_{\mathfrak{f}})}.
\end{equation}

The \textbf{Input gate} adds new information to the cell state from the current input $\boldsymbol{d}^{\hspace{0.02cm}(n)}$, previous hidden state $\boldsymbol{h}^{(n-1)}$ and their corresponding input gates weighting matrices $\boldsymbol{W}_{\mathfrak{i}_{1,2},\boldsymbol{d}}$, and biases $\boldsymbol{b}_{\mathfrak{i}_{1,2}}$, through the operations

\begin{equation}
\small{
 \begin{aligned}
\mathfrak{i}_{1}^{(n)} &= \sigma_2(\boldsymbol{W}_{\mathfrak{i}_{1},\boldsymbol{d}}\boldsymbol{d}^{\hspace{0.02cm}(n)}+ \boldsymbol{W}_{\mathfrak{i}_{1},\boldsymbol{h}} \boldsymbol{h}^{(n-1)}+ \boldsymbol{b}_{\mathfrak{i}_{1}})\\
\mathfrak{i}_{2}^{(n)} &= \sigma_3(\boldsymbol{W}_{\mathfrak{i}_{2},\boldsymbol{d}} \boldsymbol{d}^{\hspace{0.02cm}(n)}+ \boldsymbol{W}_{\mathfrak{i}_{2},\boldsymbol{h}} \boldsymbol{h}^{(n-1)}+ \boldsymbol{b}_{\mathfrak{i}_{2}} )\\
\mathfrak{i}^{(n)} &= \mathfrak{i}_{1}^{(n)} \otimes  \mathfrak{i}_{2}^{(n)},
\end{aligned}}
\end{equation}
where $\sigma_2, \sigma_3$ are a sigmoid and a hyperbolic tangent activations, respectively, and $\otimes$ is the Hadamard product. Then, the cell state $\boldsymbol{c}^{\hspace{0.02cm}(n)}$ is updated as
\begin{equation}
\small{
    \boldsymbol{c}^{\hspace{0.02cm}(n)} = \boldsymbol{c}^{\hspace{0.02cm}(n-1)} \otimes \mathfrak{f}^{(n)} \oplus \mathfrak{i}^{(n)}},
\end{equation}
where $\oplus$ denotes component-wise sum. Finally, the \textbf{output gate}  updates the hidden state $\boldsymbol{h}^{(n)}$ as
\begin{equation}
\small{
 \begin{aligned}
     \mathfrak{o}^{(n)} &= \sigma_4(\boldsymbol{W}_{\mathfrak{o},\boldsymbol{d}}\boldsymbol{d}^{\hspace{0.02cm}(n)}+ \boldsymbol{W}_{\mathfrak{o},\boldsymbol{h}} \boldsymbol{h}^{(n-1)} + \boldsymbol{b}_{\mathfrak{o}} )\\
\boldsymbol{h}^{(n)} &= \sigma_5(\boldsymbol{c}^{\hspace{0.02cm}(n)}) \otimes \mathfrak{o}^{(n)},
\end{aligned}}   
\end{equation}
where $\sigma_4, \sigma_5$ are again the sigmoid and hyperbolic tangent functions and $\boldsymbol{W}_{\mathfrak{o},\boldsymbol{d}}$, $\boldsymbol{b}_{\mathfrak{o}}$ are weights and bias associated with the output gate. 

In our application, $\boldsymbol{d}^{\hspace{0.02cm}(n)}$ will be the discrete displacement solution at step $n$, at the shared nodes. Since we are interested in sequence to sequence learning, we would have multiple LSTM cells like the one introduced above, sharing the same set of weights and biases.

\subsection{LSTM encoder-decoder network}\label{chp:ED}
\begin{figure*}[!ht]
    \centering
    \scalebox{.9}{\input{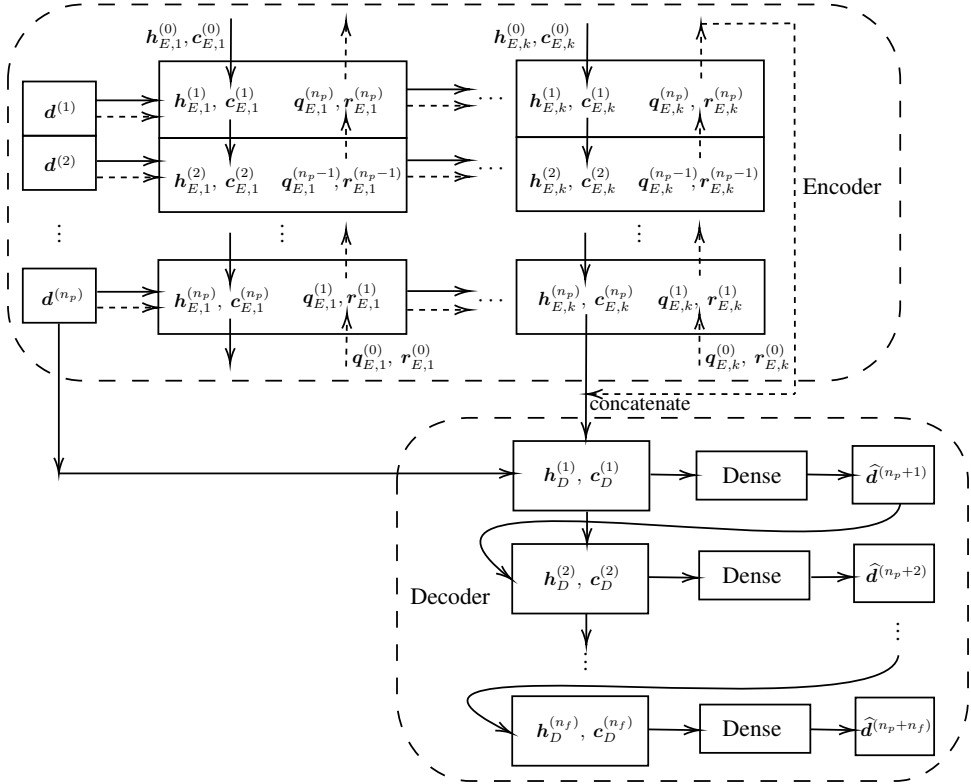}}
    \caption{\centering Schematic representation of the LSTM encoder-decoder model. The Encoder is a stacked bi-directional LSTM network. The Decoder is a single layer uni-directional LSTM network with dense output layers.}
    \label{NN-ED}
\end{figure*}

The design of our deep neural network model is demonstrated in Figure~\ref{NN-ED}. It is the combination of a $k$-layer bi-directional LSTM encoder and a single-layer unidirectional LSTM decoder, inspired and implemented based on~\cite{lstm-ed}. 
Stacking encoder layers ensures that more information is extracted from the input sequence while limiting the total number of parameters. The encoder-decoder structure is also designed to handle variable sequence lengths, which is a distinctive feature in language translation and other sequence-to-sequence models~\cite{goodfellow2016deep}.
In addition, our encoder is enriched by a bidirectional structure, which helps to capture dependency across the whole input sequence~\cite{goodfellow2016deep}. 
Note that a bidirectional LSTM network was also used in~\cite{Hu2022NeuralPDEAR} to approximate time-dependent differential equations over discrete lattices through a \emph{Many-to-Many} recurrent architecture. This is different from the \emph{Many-to-One} architecture used here, thus, as opposed to~\cite{Hu2022NeuralPDEAR}, our decoder remains uni-directional. Further, each item of our input sequence varies spatially, while the input of~\cite{Hu2022NeuralPDEAR} considers time series at each collocation point.

Our proposed deep neural network model can be simply expressed as the following operation with the input sequence $\boldsymbol{X}$ of length $n_p$ and predicted output sequence $\widehat{\boldsymbol{Y}}$ of length $n_f$

\begin{equation}
\small{
    \widehat{\boldsymbol{Y}} = \Big( \mathcal{N}_{D,\boldsymbol{\Theta}_D} \circ \mathcal{N}_{E,\boldsymbol{\Theta}_E} \Big)\boldsymbol{X} = {\mathcal{N}_{\boldsymbol{\Theta}}} \boldsymbol{X}}
\end{equation}
\begin{equation}
\small{
\begin{aligned}
        \boldsymbol{X} &= 
\begin{bmatrix}
\boldsymbol{d}^{\hspace{0.02cm}(1)}, & \boldsymbol{d}^{\hspace{0.02cm}(2)}, & \cdots, & \hspace{-0.1cm}\boldsymbol{d}^{\hspace{0.02cm}(n_p)}
\end{bmatrix} \in \mathbb{R}^{n_{\text{dof}} \times n_p }\\
\widehat{\boldsymbol{Y}} &= 
\begin{bmatrix}
\widehat{\boldsymbol{d}}^{\hspace{0.02cm}(n_p+1)}, & \widehat{\boldsymbol{d}}^{\hspace{0.02cm}(n_p+2)}, & \cdots, & \hspace{-0.1cm}\widehat{\boldsymbol{d}}^{\hspace{0.02cm}(n_p+n_f)}
\end{bmatrix} \in \mathbb{R}^{n_{\text{dof}}\times n_f}.
\end{aligned}}
\end{equation}

The encoder and decoder models are expressed as $\mathcal{N}_{E,\boldsymbol{\Theta}_E}, \ \mathcal{N}_{D,\boldsymbol{\Theta}_D}$ respectively, and $n_{\text{dof}}$ stands for the total number of degrees of freedom in all displacement solutions at the shared nodes.
We further refer to the overall LSTM encoder-decoder model as the composition ${\mathcal{N}_{\boldsymbol{\Theta}}}$, parametrized by ${\boldsymbol{\Theta}} = \boldsymbol{\Theta}_D \cup \boldsymbol{\Theta}_E$.

In the encoder model, each displacement solution $\boldsymbol{d}^{\hspace{0.02cm} (i)}, i=1,\cdots n_p$ of input sequence $\boldsymbol{X}$ is fed to a LSTM cell (Figure~\ref{lstm cell}) while having hidden and cell states coming from two opposite directions. 
We use expressions $\boldsymbol{h}_{E,j}^{(i)}$ and $\boldsymbol{c}_{E,j}^{(i)}$ to represent the hidden and cell states in the direction $1\to n_p$ (solid arrow) and $\boldsymbol{q}_{E,j}^{(i)}$ and $\boldsymbol{r}_{E,j}^{(i)}$ along $n_p\to1$ (dashed arrow), where $j=1,\cdots,k$ is the layer index.
At the final encoder layer, hidden and cell states from two directions are concatenated separately and provided to the decoder model as initial states $\boldsymbol{h}_{D}^{(0)}$ and $\boldsymbol{c}_{D}^{(0)}$.

The decoder then receives the final item $\boldsymbol{d}^{\hspace{0.02cm} (n_p)}$ in $\boldsymbol{X}$ and recursively produce the predictions $\widehat{\boldsymbol{d}}^{\hspace{0.02cm}(n_p+j)}$, $j=1,\cdots,n_f$ in $\widehat{\boldsymbol{Y}}$.
At each decoding step, the previous prediction will be forwarded to the next step as an input, with a dense neural network bridging the different size between the hidden state and model output.

During training we use a Mean Squared Error (MSE) loss function built from the predicted output $\widehat{\boldsymbol{Y}}$ and true numerical solution $\boldsymbol{Y}$ as
\begin{equation}\label{mse}
\mathcal{L}(\boldsymbol{Y}, \widehat{\boldsymbol{Y}}) = \frac{1}{n_f \cdot n_{\text{dof}}} \sum_{j=1}^{n_f}  \|\boldsymbol{Y}_j -\widehat{\boldsymbol{Y}}_j \|_{2}^2 = \frac{1}{n_f \cdot n_{\text{dof}}} \sum_{j=1}^{n_f} \sum_{i=1}^{n_{\text{dof}}} \rvert Y_{ij}-\widehat{Y}_{ij} \rvert^2 = \frac{1}{n_f\cdot n_{\text{dof}}} \|\boldsymbol{Y}- \widehat{\boldsymbol{Y}}\|_F^2,
\end{equation}
and perform gradient-based updates for the trainable parameters ${\boldsymbol{\Theta}}$.

\subsubsection{Network training and evaluation}\label{sample}
We tailor the training and evaluation of the proposed network to the specific application of interest, i.e., dynamical systems simulated through explicit numerical solution algorithms in time.
In the structural analysis of soft biological tissue, the time step $\Delta t$ is usually in the range $1\times10^{-6} \sim 1\times10^{-3}$ due to the stability condition~\eqref{cfl}.
Such small time step will lead to limited changes between displacement solutions at two successive time steps, and therefore almost identical model input $\boldsymbol{X}$ and true output $\boldsymbol{Y}$. 
%
%
However, for effective training, we would like each of our training sample to contain sufficient information of the underlying dynamics. In other words, the input $\boldsymbol{X}$ and output $\boldsymbol{Y}$ should be sufficiently different for the network to learn a relevant mapping and not just an identity operator, typical of mere steady state conditions.

\begin{figure}[!ht]
    \centering
    \includegraphics[scale=4.9]{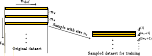}
    \caption{\centering The sample pre-processing step.}
    \label{fig:sampling}
\end{figure}

This is accomplished through a so-called \emph{sample-refill} strategy during the training and evaluation stages, respectively.
First, as illustrated in Figure~\ref{fig:sampling}, instead of using the full dataset, we only pick a displacement solution every $n_s$ steps. 
This \emph{sampled} data is what we group, batch and feed into the deep neural network at training. 
Upon successful training, as described in Figure~\ref{NN-ED}, the network will take $n_p$ displacement solutions in the past and use them to predict $n_f$ future steps.
Because of the pre-processing, the time lag between each predicted solution will still be $n_s$.

\begin{figure*}[!ht]
    \centering
    \includegraphics[scale=6]{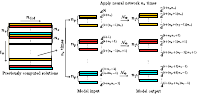}
    \caption{\centering Schematic illustration for the \emph{refill} task. The whole operation starts at step $i$ and requires $n_pn_s$ steps of previously computed solutions.}
    \label{fig:refilling}
\end{figure*}

To fill these gaps, we leverage a \emph{refill} operation during the evaluation stage. Given enough steps computed in the past, we use the model $n_s$ times to produce $n_s\cdot n_f$ predictions. After the first time, the input $\boldsymbol{X}$ is shifted $n_s-1$ times forward to generate the missing displacement predictions at all shared mesh nodes, as illustrated in Figure~\ref{fig:refilling}.
More details of this \emph{refill} task can be found in Algorithm~\ref{alg-syn-avi} below.

\section{A data-driven synchronization-avoiding algorithm}\label{syn-alg}

We propose a data-driven methodology to minimize synchronization in distributed, explicit in time, finite element structural analysis, which starts with data preparation.
This consists in running Algorithm~\ref{alg-para} in parallel, and gathering the sequential displacement solutions for each of the $n_c$ processors.
We than identify all degrees of freedom associated with the shared nodes and form a training dataset using samples spaced by  $n_s$ time steps.
Next, we train $n_c$ independent replicas of the LSTM network illustrated in Figure~\ref{NN-ED}, finally producing a set of optimally trained network models $\mathcal{N}_{\mathbf{\Theta}, [j]}, j=\{1,2,\cdots,n_c\}$. 
We finally apply Algorithm~\ref{alg-syn-avi} where the synchronized displacements at the shared nodes are \emph{modeled} by the network at each processor instead of being communicated across partitions.

The parameter $n_{\text{cri}}$ in Algorithm~\ref{alg-syn-avi} is used to switch between the synchronization process and using displacement predicted by the network for all shared degrees of freedom.
Clearly $n_{\text{cri}}$ is expected to be set as low as possible to attain the most speed-up but it cannot be less than $n_p\cdot n_s+1$ since the model requires the first $n_p\cdot n_s$ steps to start.

The previously mentioned \emph{refill} stage (see Section~\ref{sample}) is realized by indexing every intermediate steps as lists $\mathbf{N}^i_p$ and $\mathbf{N}^i_f$ such that the model inputs are properly shifted for continuous predictions.
Compared with synchronization costs, execution times for evaluating pre-trained network models at every time step are negligible.

\begin{algorithm*}[h!]
\caption{A synchronization-avoiding algorithm for distributed linear elastodynamics.}
\begin{algorithmic}
\State \textbf{Step 1:} \quad Partition the mesh over $n_c$ processors and compute $\Delta t$
\State \textbf{Step 1a:} \hspace{-0.08cm}\ If needed, apply mass scaling following Algorithm~\ref{alg-cms}
\State \textbf{Step 2:} \quad Identify Dirichlet nodes and initial conditions for each processor
\State \textbf{Step 3:} \quad For the generic $j$-th processor, set $t=\Delta t$, $n=1$, $n_{\text{cri}}$, $n_p$, $n_f$, $n_s$
\While{$t\leq T$}
\If{$n\leq n_{\text{cri}}$}
\State \textbf{Apply} Algorithm \ref{alg-para} to compute $\boldsymbol{d}^{\hspace{0.02cm}(n)}$ 
\State $t = t + \Delta t$ 
\State $n = n +1 $
\Else
\For{$i=1,2\cdots,n_s$} \Comment{refill step}
\State $\mathbf{N}^i_p = [i+n-n_p n_s-1 : n_s: i+n-n_s-1]$ \Comment{Index set of NN model input}
\State $\mathbf{N}^i_f = [i+n-1 : n_s: i + n +n_fn_s-n_s-1]$ \Comment{Index set of NN model output}
\State \textbf{Forward} pre-trained NN model: $\boldsymbol{d}^{\hspace{0.02cm}(\mathbf{N}^i_p)}[\text{shared}] \xrightarrow{\mathcal{N}_{\boldsymbol{\Theta},{[j]}}} \widehat{\boldsymbol{d}}^{\hspace{0.02cm}(\mathbf{N}_f^i)}[\text{shared}]$
\EndFor
\State \textbf{Gather} all predictions on shared nodes: 
$\widehat{\boldsymbol{d}}^{\hspace{0.02cm}(\mathbf{N}_f)}[\text{shared}] = \{\widehat{\boldsymbol{d}}^{\hspace{0.02cm}(\mathbf{N}_f^i)}[\text{shared}], i=1,2,\cdots, n_s\}$
\For{$m=n,n+1, \cdots, n+n_sn_f$}
\State \textbf{Apply} Algorithm \ref{alg-para} to compute $\boldsymbol{d}^{\hspace{0.02cm}(m)}$ \textbf{without} ``boxed" steps (synchronization) 
\State \textbf{Update} $\boldsymbol{d}^{\hspace{0.02cm}(m)}$ at shared nodes by the corresponding modeled values in  $\widehat{\boldsymbol{d}}^{\hspace{0.02cm}(\mathbf{N}_f)}[\text{shared}]$

\EndFor
\State \textbf{Impose} Dirichlet boundary conditions to $\boldsymbol{d}^{\hspace{0.02cm}(\mathbf{N}_f)}$
\State $t = t + n_sn_f\Delta t$
\State $n = n+n_sn_f$
\EndIf
\EndWhile
\end{algorithmic}
\label{alg-syn-avi}
\end{algorithm*}

\section{Numerical Examples}\label{chp:numerical}

\subsection{Cantilever model}\label{canti}
\begin{figure}[h!]
\centering
\includegraphics[scale=0.4]{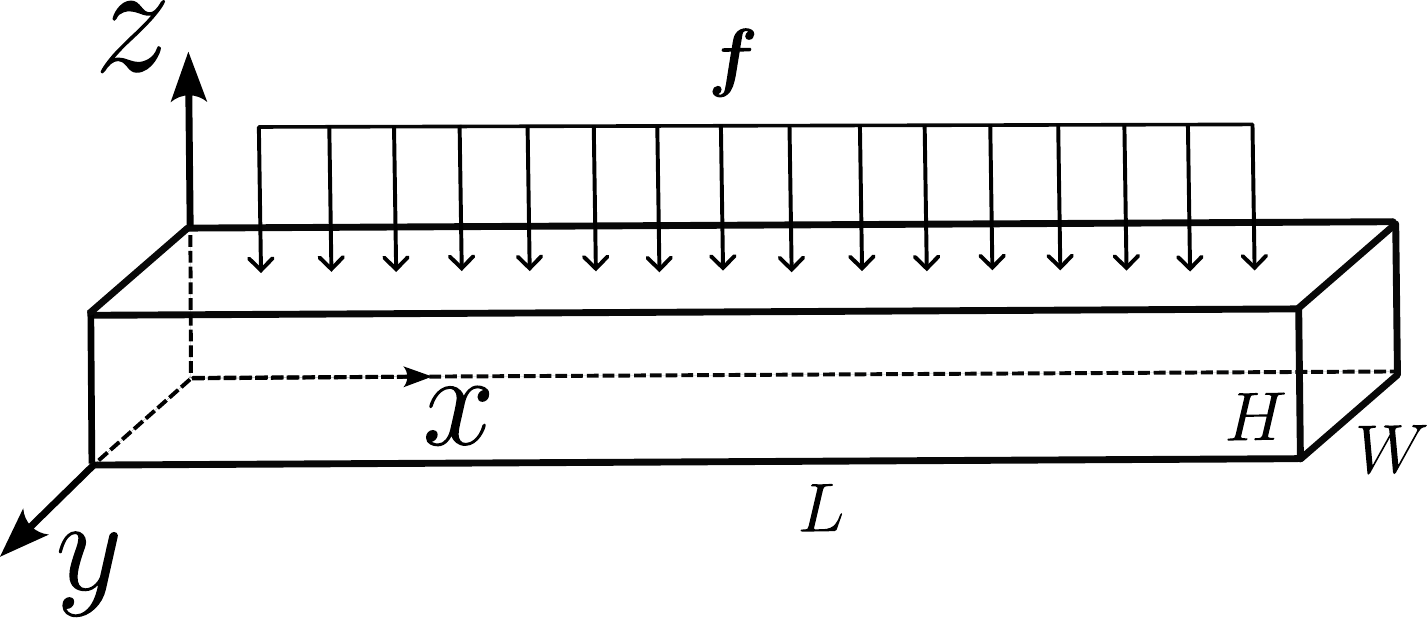}
\caption{Simple cantilever beam model.}
\label{beam}
\end{figure}

To test the proposed computational framework, we use a simple cantilever beam model containing only $110$ vertices and 256 tetrahedral elements.
Specifically, we focus on the under-damped oscillatory regime using a mass proportional damping with factor $\alpha$.
As shown in Figure~\ref{beam}, the cantilever has fully fixed restraints $\boldsymbol{d}\rvert_{x=0} = 0$ at one end, and external load $\boldsymbol{f}$ consists of a ramp of $1s$ (see Remark~\ref{ramp}), followed by a constant distributed body force equal to $f_z$=0.5 dynes/$\text{cm}^3$ in the $z$-direction, i.e., $\boldsymbol{f} = [0, 0,  -f_z]^T$. Further, homogeneous initial conditions are considered here as $\boldsymbol{d}^{\hspace{0.02cm}(0)}=\boldsymbol{v}^{(0)}=\boldsymbol{0}$, and geometric and material model parameters are listed in Table~\ref{cati}.

\begin{figure}[!ht]
\centering
\includegraphics[scale=0.1]{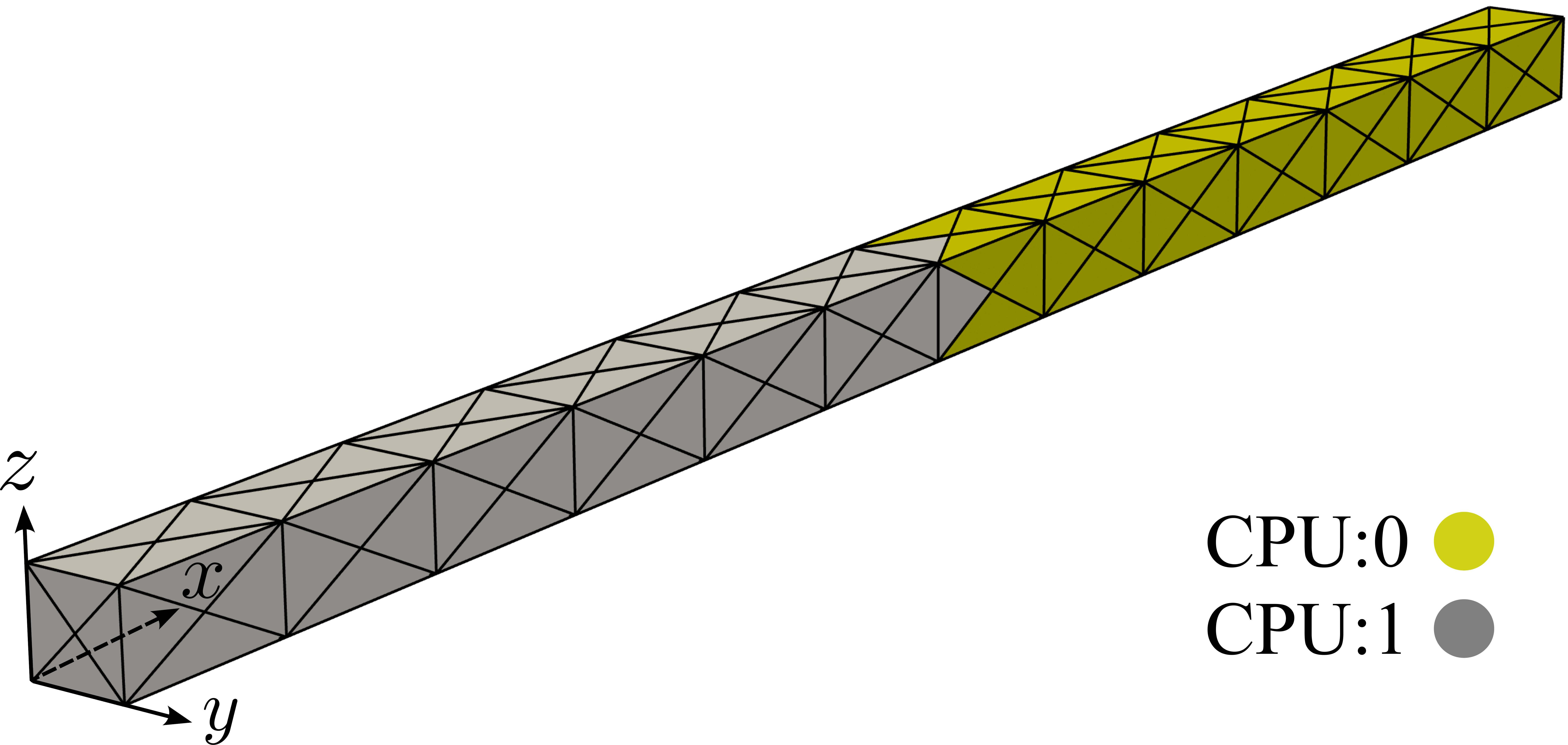}
\caption{\centering Coarse mesh partitioned over 2 CPUs, labelled 0 and 1, respectively.}    
\label{2cpu}
\end{figure}

The time step size is set to $2.48\times 10^{-4}$, and we assume $n_s=80$. The mesh partitioning is realized by \texttt{mgmetis}~\cite{mgmetisChen} based on the \texttt{ParMETIS} library~\cite{METIS}, and parallel computations are managed through the Message Passing Interface (MPI).
Initially we consider a distributed mesh on 2 processors, with 8 shared mesh nodes each. The result of mesh partitioning is shown in Figure~\ref{2cpu}.
\begin{table}[h!]
{\footnotesize
\begin{center}
\begin{tabular}{@{} l l @{}}
\toprule
Length ($L$) & $25$ (cm) \\ 
Width ($W$) & $1$ (cm) \\
Height ($H$)  &$1$ (cm) \\
Young's modulus ($E$) & $1 \times 10^6 \ (\text{dynes/cm}^2)$\\ 
Density ($\rho$) & $1 \ (\text{g/cm}^3)$ \\ 
Poisson's ratio ($\nu$) & 0.3 \\ 
\bottomrule
\end{tabular}
\end{center}}
\caption{Geometric and material parameters for the cantilever beam model.}
\label{cati}
\end{table}

\subsubsection{Hyperparameters and network training} \label{chp:hyper}
In our numerical experiments, we optimize over a number of selected hyperparameter realizations by performing a \emph{grid search} on the mini-batch size $n_B$, the hidden unit size $n_H$ and the initial learning rate $\eta_0$.
In addition, we utilize an exponential learning rate scheduler, where $\eta(x) = \eta_0 \gamma^x$ for a given epoch $x$.
In the search of a good initial learning rate $\eta_0$, we set a minimum learning rate $\eta_{\text{min}} = 5\times 10^{-7}$ and keep the decay rate fixed at $\gamma = 0.9995$. 
As a result, the total number of epochs $n_{\text{epoch}}$ can be calculated as:
\begin{equation}
n_{\text{epoch}} = \lfloor \log_{\gamma}(\frac{\eta_{\text{min}}}{\eta_0}) \rfloor.
\end{equation}
The selected grid of hyperparameter realizations includes $n_B = \text{5, 10, 20, 50}$, $n_H=\text{20, 50, 100}$ and $\eta_0=5\times 10^{-3}$ $(n_{\text{epoch}}  = 18416)$, $5\times 10^{-4}$  $(n_{\text{epoch}}  = 13812)$, $5\times 10^{-5}$  $(n_{\text{epoch}}  = 9208)$, leading to 36 cases in total. 

In terms of how to pick the best combination of $n_B$, $n_H$ and $\eta_0$, we introduce the following criterion rather than simply looking at the loss curves produced by MSE~\eqref{mse} during training
\begin{equation}
\begin{split}
E_{\text{mse}} & = \frac{1}{N\cdot n_f\cdot n_{\text{dof}}}\sum_{j=i}^{i+N\cdot n_f-1} \|\boldsymbol{d}^{\hspace{0.02cm}(j)} - \widehat{\boldsymbol{d}}^{\hspace{0.02cm}(j)}\|_2^2 \\
& = \frac{1}{N\cdot n_f\cdot n_{\text{dof}}}\Bigg( \sum_{j=i}^{i+n_f-1} \|\boldsymbol{d}^{\hspace{0.02cm}(j)} - \widehat{\boldsymbol{d}}^{\hspace{0.02cm}(j)}\|_2^2 + \sum_{j=i+n_f}^{i+2n_f-1} \|\boldsymbol{d}^{\hspace{0.02cm}(j)} - \widehat{\widehat{\boldsymbol{d}}}^{\hspace{0.02cm}(j)}\|_2^2 + \cdots \Bigg),        
\end{split}
\label{msemse}
\end{equation}
that is, after the network is trained, starting from step $i$, we go ahead and use the trained network model $\mathcal{N}_{\boldsymbol{\Theta}}$ for $N$ times. Then, by definition, $E_{\text{mse}}$ quantifies a MSE error of displacement predictions on all shared nodes over $N\cdot n_f$ steps. 
%
Note that during the calculation of $E_{\text{mse}}$, after the first time ($N=1$), we no longer have exact inputs. 
In other words, $\widehat{\widehat{\boldsymbol{d}}}$ is the model output ($N=2$) with model input $\widehat{\boldsymbol{d}}$, predicted from the last step. 
Here we abuse the notation $\widehat{(\cdot)}$ to avoid stacking more ``hats" to predictions.
Trivially,~\eqref{msemse} collapses to~\eqref{mse} when $N=1$.

Utilizing $E_{\text{mse}}$ instead of looking at typical training and validation curves is more consistent with our objective, in the prediction stage, to use previously predicted displacements as inputs to predict displacements at future times.
A standard MSE loss~\eqref{mse} on the other hand, would only account for the performance of a single model application ($n_f$ steps), thus delivering less information about accuracy and stability on a longer time horizon. 
Note that an alternative way to ensure robust long-term predictions is proposed in~\cite{CHEN2022110782,churchill2022deep}, where a \emph{recurrent loss function} similar to~\eqref{msemse} was used in the context of ResNet network training. 
However, their implementation resulted in significantly longer training times.

\begin{figure*}[!ht]
\centering
\begin{subfigure}[b]{0.325\textwidth}
  \centering
  \includegraphics[scale=.48]{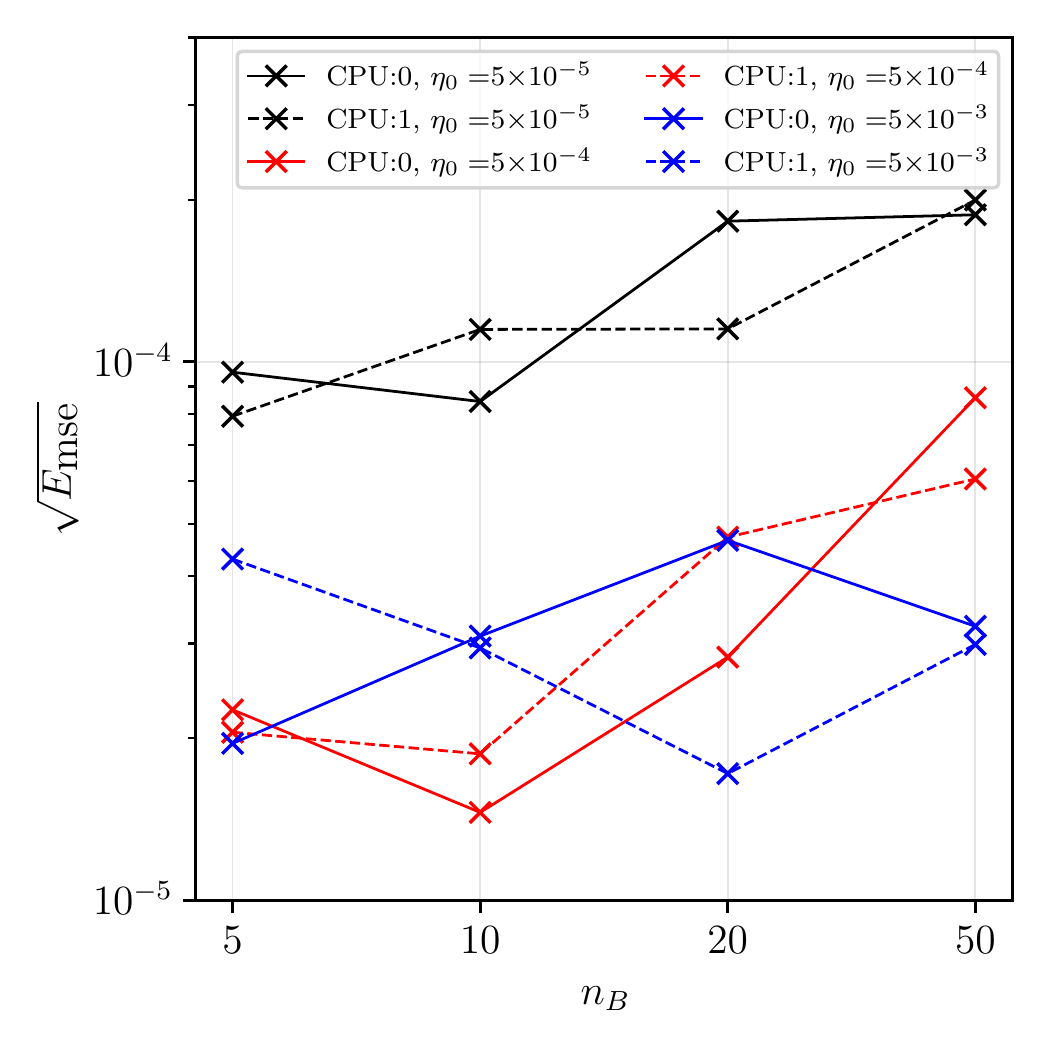}
  \caption{$n_H=20$.}
\end{subfigure}
\begin{subfigure}[b]{0.325\textwidth}
  \centering
  \includegraphics[scale=.48]{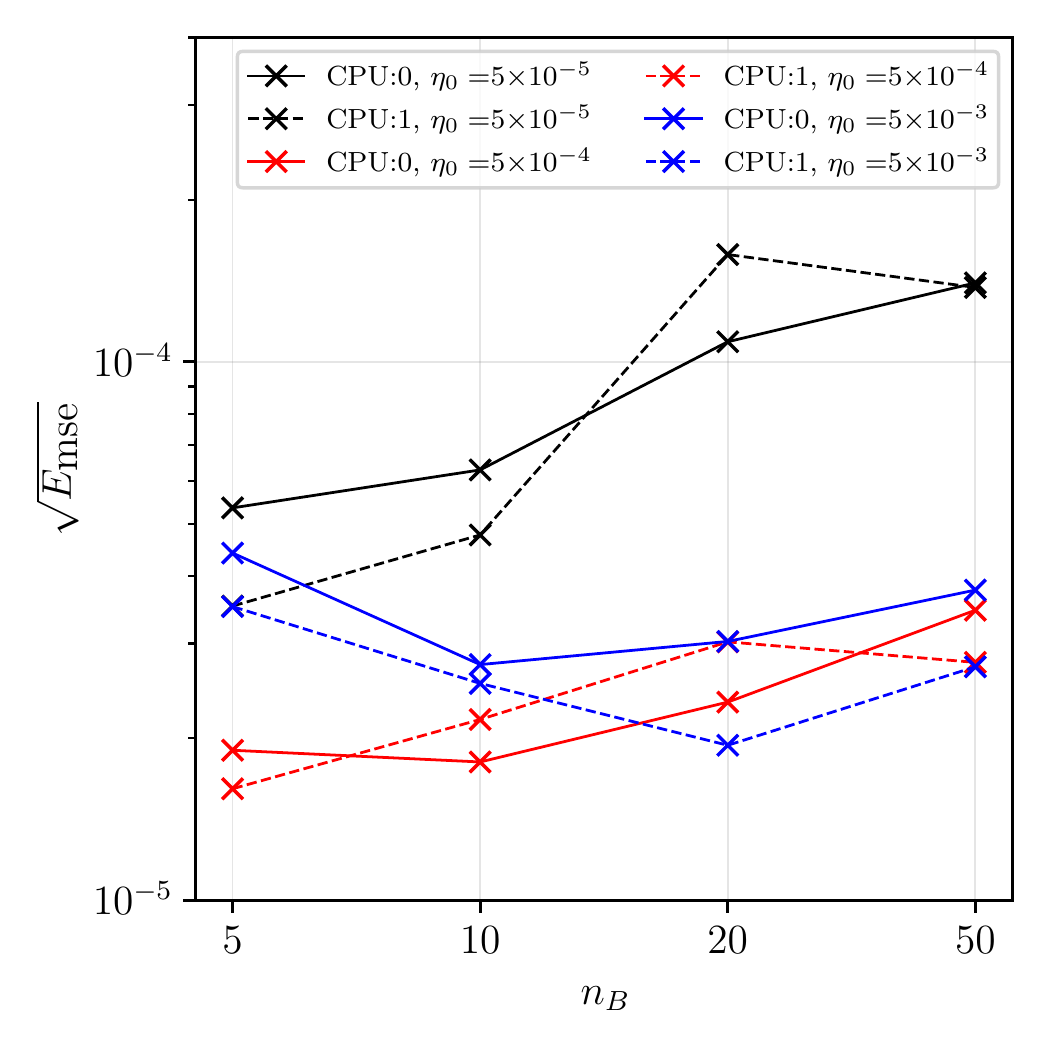}
  \caption{$n_H=50$.}
\end{subfigure}
\begin{subfigure}[b]{0.325\textwidth}
  \centering
  \includegraphics[scale=.48]{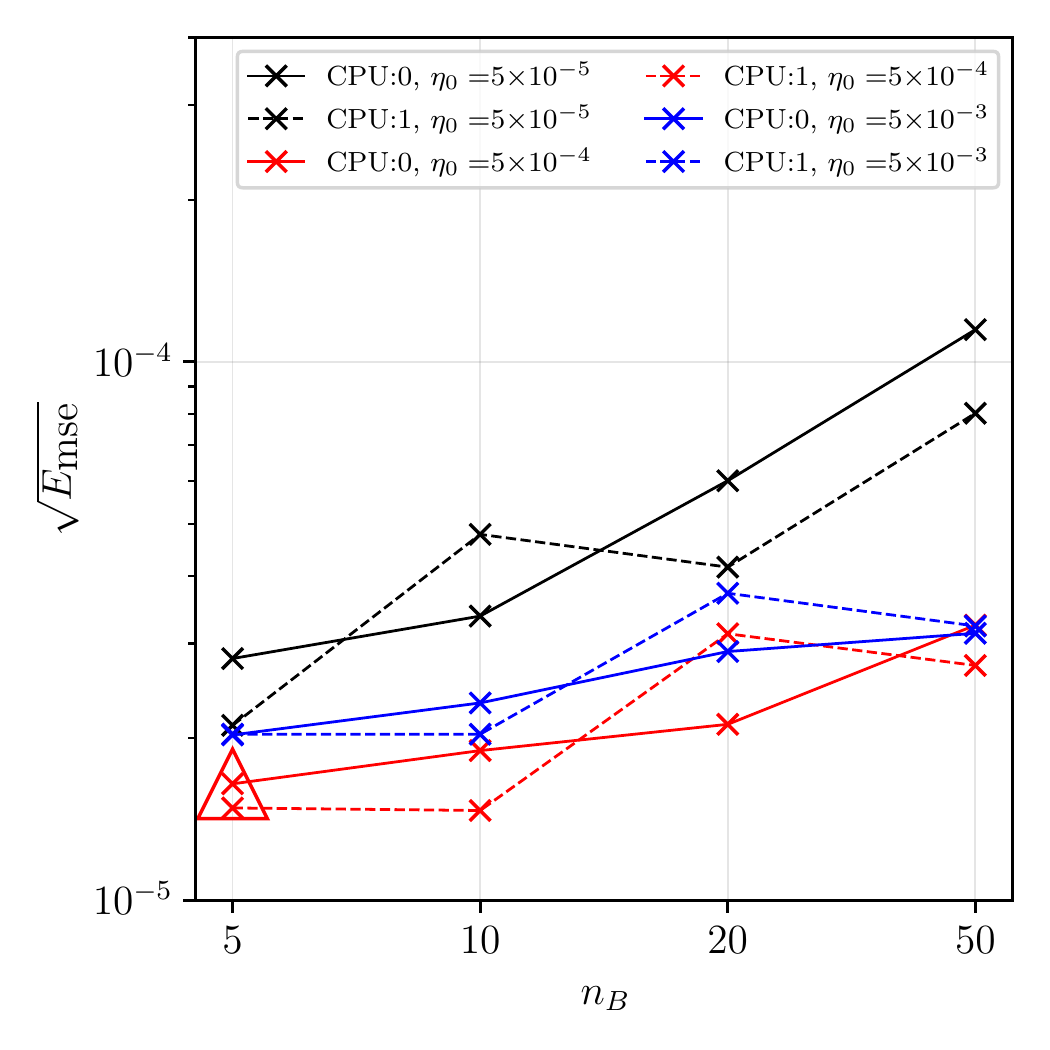}
  \caption{$n_H=100$.}
  \label{best-hyp}
\end{subfigure}
\caption{\centering Values of $\sqrt{E_{\text{mse}}}$ from grid search on mini-batch size $n_B$, hidden unit size $n_H$ and initial learning rate $\eta_0$. The optimal hyperparameter combination is selected as $n_B=5$, $n_H=100$, $\eta_0=5\times 10^{-4}$, marked by ``{\color{red}$\bigtriangleup$}".}\label{FIG:GRID}
\end{figure*}

In practice, we set $N=60$ and calculate the square root of $E_{\text{mse}}$ as a measure of accuracy per degree of freedom and per prediction.
All the results are reported in Figure~\ref{FIG:GRID}, where the lowest learning rate $\eta_0=5\times 10^{-5}$ provides the worst accuracy.
In addition, we train a number of networks equal to the number of mesh partitions (two in this case, see Figure~\ref{2cpu}), so we can quantify the variability in the prediction accuracy produced by multiple network replicas, and favor models producing robust predictions across processors.
This observation penalizes networks with a small number of hidden units ($n_H=20$), characterized by large inter-processor variability. %
The optimal hyperparameter combination (marked by ``{\color{red}$\bigtriangleup$}" in Figure~\ref{best-hyp}) is finally selected as $\mathbf{n_B=5, n_H=100, \eta_0=5\times 10^{-4}}$, since it offers the best compromise between a moderate training cost, a sufficient accuracy and limited model discrepancy.
It is worth noting that one could further refine this choice by expanding the number of hyperparameter combinations in the grid search. For example, we expect better accuracy for even smaller mini-batch size $n_B$ when $\eta_0=5\times 10^{-4}$ via the observation of a roughly monotonic behavior. 
However, the selected accuracy metric $\sqrt{E_{\text{mse}}}$ is already close to $10^{-5}$, regarded as satisfactory.

We also study the effects of the sequence lengths $n_p$, $n_f$ and the training set size $n_{ts}$ in section~\ref{npnf}. 
The parameters $n_p$, $n_f$ specify the amount of past and future displacement observations used in training, whereas the training set size $n_{ts}$ controls the total amount of pre-computed numerical solutions fed into the network, i.e., temporally speaking, we use $n_{ts}$ (and later sampled by $n_s$) of all finite element solutions at each partition to train our surrogate models in parallel.
In experiments of Figure~\ref{FIG:GRID}, these parameters are fixed as $n_p=n_f=20$, $n_{ts}=50\%$.

Training is performed through the~\texttt{Adam} optimizer~\cite{kingma2014adam}, shuffling the order and choice of the mini-batches at each epoch.

The sample size $n_s$ mentioned in Section~\ref{sample} is also regarded as a hyperparameter that may be dependent on the time step size $\Delta t$, problem type, etc. In our experiments, we found that  sub-optimal choices of $n_s$ can lead to severe over-fitting in the early stage of training, hence a few preliminary training tests can help with the selection of $n_s$.

\subsubsection{Offline and online data-driven model evaluation}\label{offon}

\begin{figure}[!ht]
\centering
\includegraphics[scale=0.59]{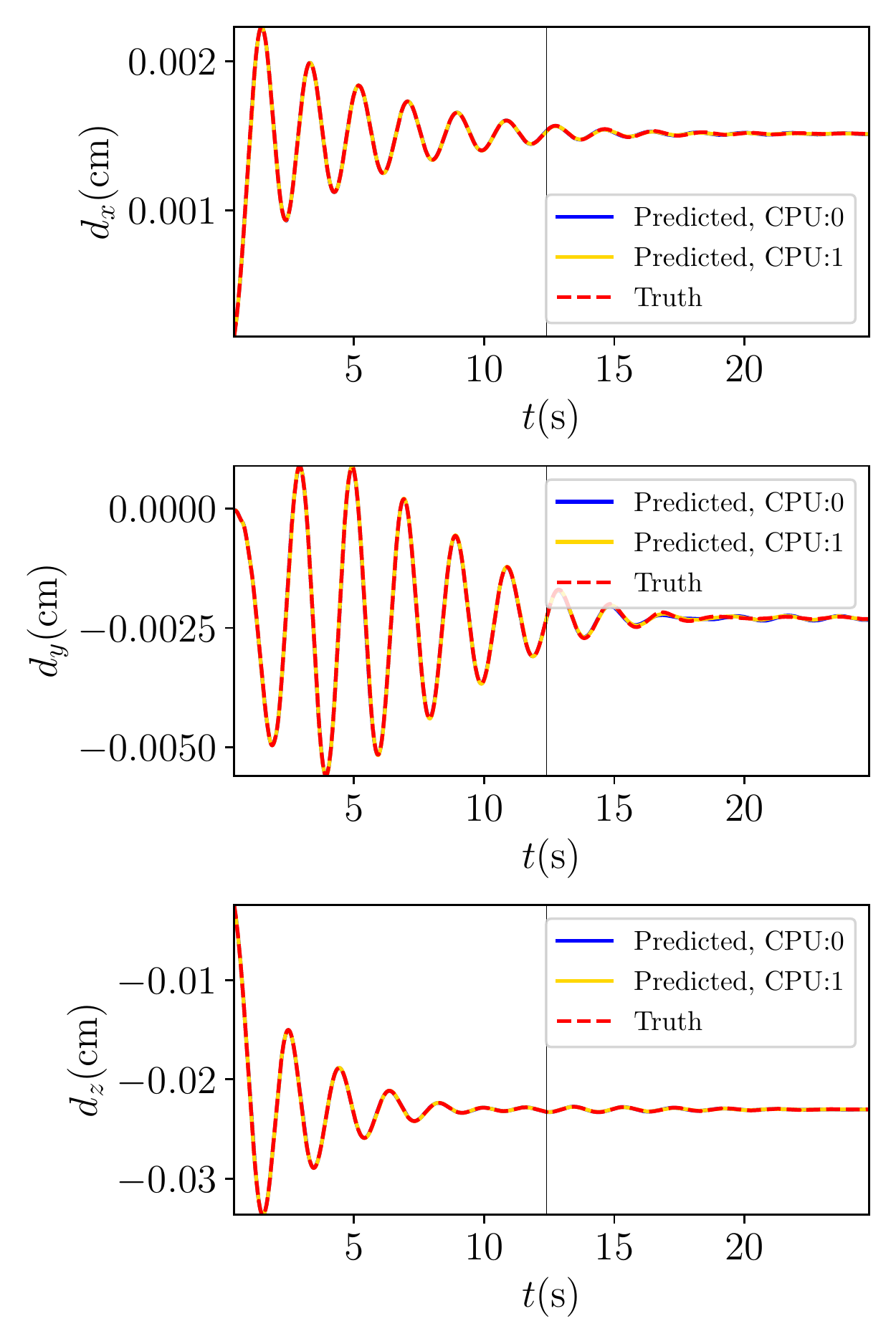}
\caption{\centering Offline shared node displacement predictions for cantilever model. Predicted steps: 98400. Node coordinate: (11.54, 0, 1).}
\label{fig:offline}
\end{figure}

We proceed to define two ways to use our data-driven model. 
First, as discussed in Section~\ref{syn-alg}, we train several network models in parallel based on the shared degrees of freedom at each partition. Next, we inspect whether displacement evolution on each set of shared nodes are ideally learned, which we refer as the \emph{offline} prediction.
Offline prediction performance using the hyperparameters determined in Section~\ref{chp:hyper} is shown in Figure~\ref{fig:offline}, along with the finite element solutions, denoted as the \emph{truth}.
Besides a satisfactory accuracy, two separately trained network models on the same set of shared degrees of freedom show very good agreement, with hardly noticeable discrepancies after 15 seconds on the lateral $x$ and $y$ components.
A vertical line in each subplot of Figure~\ref{fig:offline} separates a region on the left where $1/n_s$ of the simulated data is used for training, from a \emph{pure prediction} region on the right, the system's response is modelled based on the learned dynamics.

%
In addition, Figure~\ref{fig:offline} shows that both the damped dynamics and the convergence to the steady state are accurately learned. 

The offline prediction discussed above serves as a prerequisite for an \emph{online} prediction task which constitutes the basis for the synchronization-avoiding strategy shown in Algorithm~\ref{alg-syn-avi}.
When predictions are performed \emph{online}, we start to combine the pre-trained neural network surrogates with the distributed finite element solver.
This inevitably causes the error of data-driven model to propagate from share nodes to other nodes during each explicit update, due to a non-diagonal stiffness matrix $\mathbf{K}$.

\begin{figure*}[!ht]
    \centering
    \includegraphics[scale=0.45]{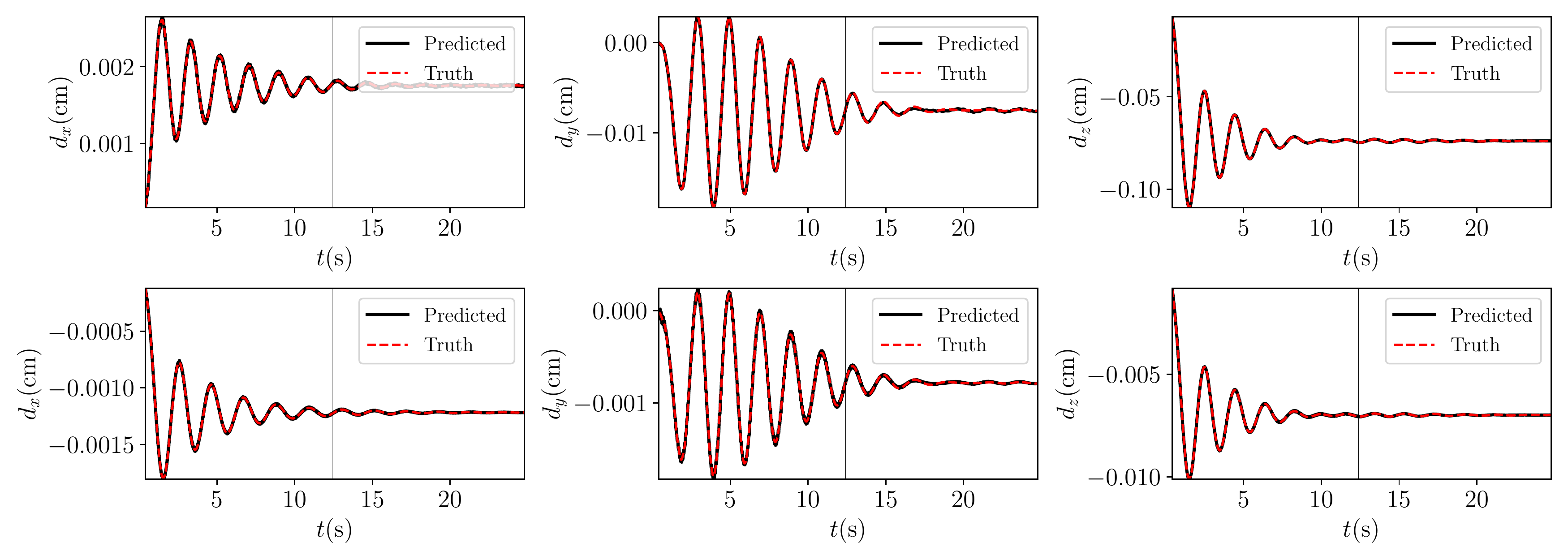}
    \caption{\centering Online displacement predictions of the cantilever model. CPU label: 0,1. Training is based on shared degrees of freedom at each partition, but plotted nodes are not shared. Predicted steps: 98400. Top row: CPU:0, node (25.0,0.0,1.0). Bottom row: CPU:1, node (5.77,0.0,1.0). }
    \label{fig:online}
\end{figure*}

\begin{figure}[!ht]
    \centering
    \includegraphics[scale=0.3]{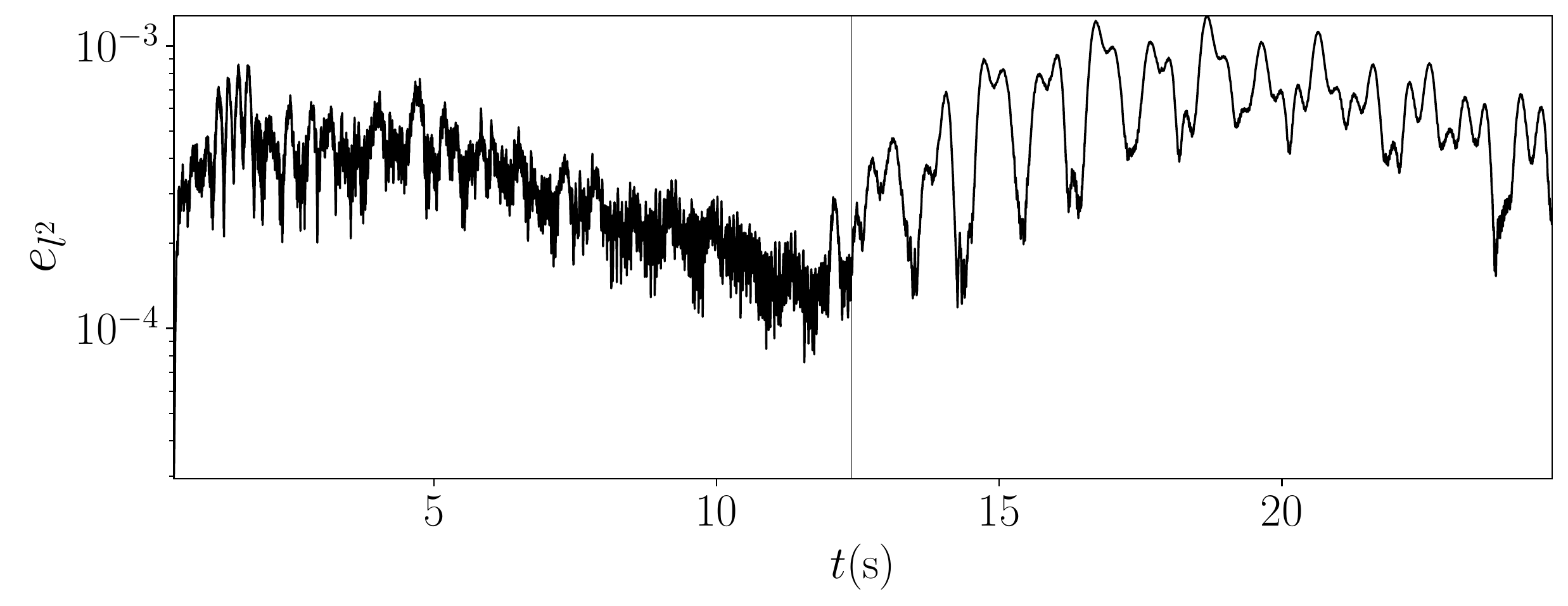}
    \caption{\centering Evolution of $l^2$ error in online predictions. Predicted steps: 98400.}
    \label{fig:subset l2}
\end{figure}

In Figure~\ref{fig:online} we plot the \emph{online} predicted displacement dynamics at two non-shared nodes, one for each partition. The predictions appear accurate and stable such that the previously mentioned error propagated from shared nodes to non-shared nodes does not grow unbounded. A $l^2$ error is shown in Figure~\ref{fig:subset l2} with its definition at time $t$ as
\begin{equation}
\small{
\mathlarger{e}^{(t)}_{l^2} = 
\|\boldsymbol{d}^{\hspace{0.02cm}(t)} - \widehat{\boldsymbol{d}}^{\hspace{0.02cm}(t)}\|_2 
}.
\label{l2e}
\end{equation}
From Figure~\ref{fig:subset l2}, we observe an initial increase in the error. As time evolves, $\mathlarger{e}_{l^2}$ gradually reduces before the vertical bar due to the increasingly smaller oscillation amplitude. 
After the vertical bar, $\mathlarger{e}_{l^2}$ increases again, but eventually gradually drops, suggesting a stable prediction of the steady state.

In the current cantilever problem, predictions in the lateral directions ($x,y$) would be more susceptible to get polluted by an error of the magnitude shown in Figure\ref{fig:subset l2}, as we can see from the following sections. 
However, this should be of less concern since the error remains bounded and the dynamics is governed by the dominant $z$ component.

\subsubsection{Network performance tests}\label{npnf}

\begin{figure*}[!ht]
    \centering
    \includegraphics[scale=0.37]{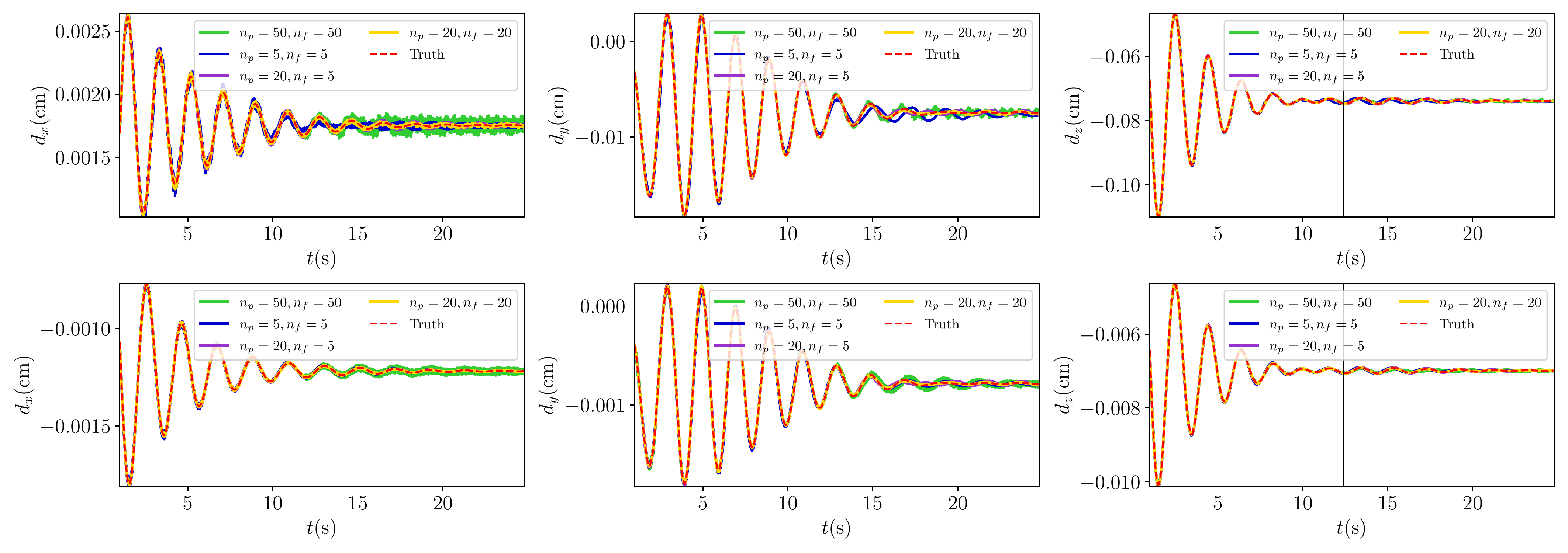}
    \caption{\centering Predicted dynamics of the cantilever model trained with different input-output sequence length ($n_p, n_f$). CPU label: 0,1. Training is based on shared degrees of freedom of each partition and plotted nodes are not shared. Top row: CPU:0, node (25.0,0.0,1.0). Bottom row: CPU:1, node (5.77,0.0,1.0).}
    \label{fig:npnf}
\end{figure*}

\noindent{\bf Input-output sequence length} - We investigate the effects produced on the network accuracy by changing the parameters $n_p$ and $n_f$, indicating the number of time steps included in the network input and output, respectively.
We consider three symmetric cases with an increasing number of steps $n_p=n_f=(5,20,50)$ and an asymmetric case with $n_p=20$ and $n_f=5$. As shown in Figure~\ref{fig:npnf}, the best results are obtained for an intermediate number of steps either symmetric or asymmetric while an excessive number of steps seems to reduce the flexibility of the network predictions. We believe this is due to a less complicated model, since learning a much longer input-output dependence as $n_p=n_f=50$ would demand more hidden units, deeper encoder etc. Therefore, without making unnecessary model refinement, a choice of $n_p=n_f=20$ should suffice for the current dynamical system.

Finally, it is also worth mentioning that $n_p, n_f$ also affect the overall speedup of our data-driven framework, since less sequence length means more model usage. This is an another reason why we prefer $n_p=n_f=20$ over $n_p=n_f=5$ or $n_p=20, n_f=5$, although their resulting accuracy are comparable.  

\vspace{3pt}

\begin{figure*}[!ht]
    \centering
    \includegraphics[scale=0.37]{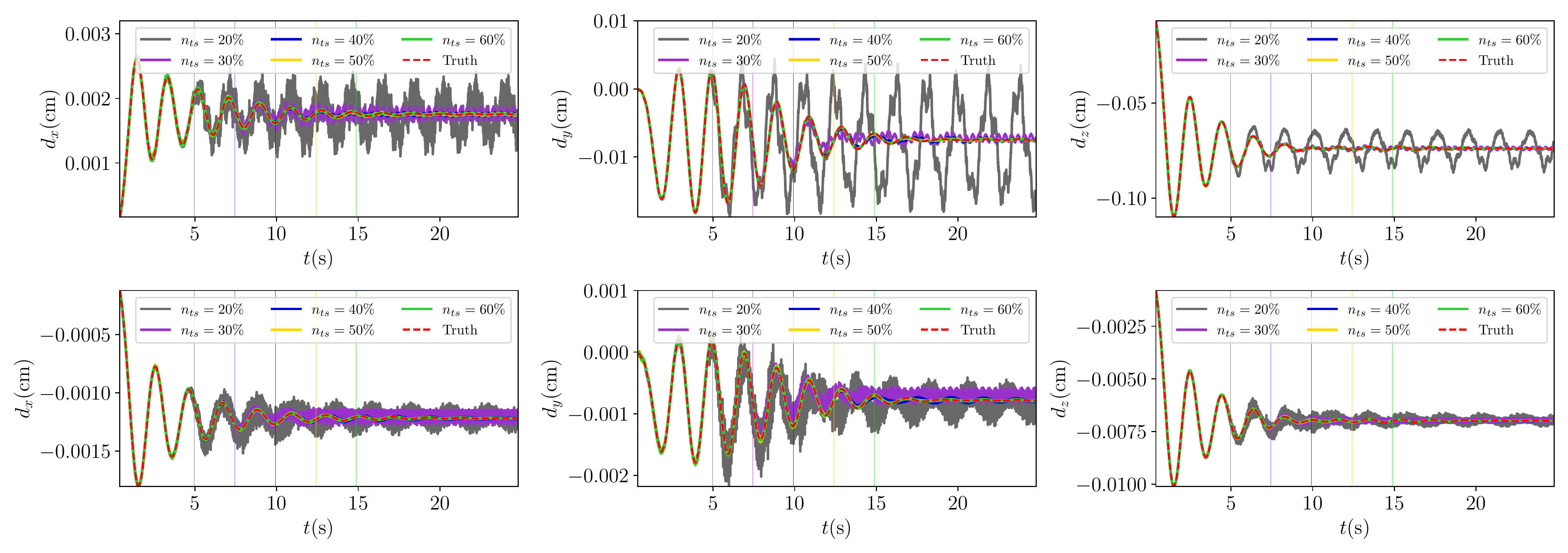}
    \caption{\centering Predicted dynamics of the cantilever model trained with different training data size ($n_{ts}$). CPU label: 0,1. Training is based on shared degrees of freedom of each partition and plotted nodes are not shared. Predicted steps: 98400. Top row: CPU:0, node (25.0,0.0,1.0). Bottom row: CPU:1, node (5.77,0.0,1.0).}
    \label{fig:nts}
\end{figure*}

\noindent{\bf Training set size}\label{nst} - In this section we perturb the hyperparameter $n_{ts}$, to see how the total number of training examples affect the accuracy of the network predictions. As expected (see Figure~\ref{fig:nts}), accuracy improves for an increasing size of the training dataset and approximately three periods of training data are needed to accurately learn the damped oscillatory dynamics in the $z$ direction. 
%

\vspace{3pt}

\begin{figure}[!ht]
    \centering
    \includegraphics[scale=0.1]{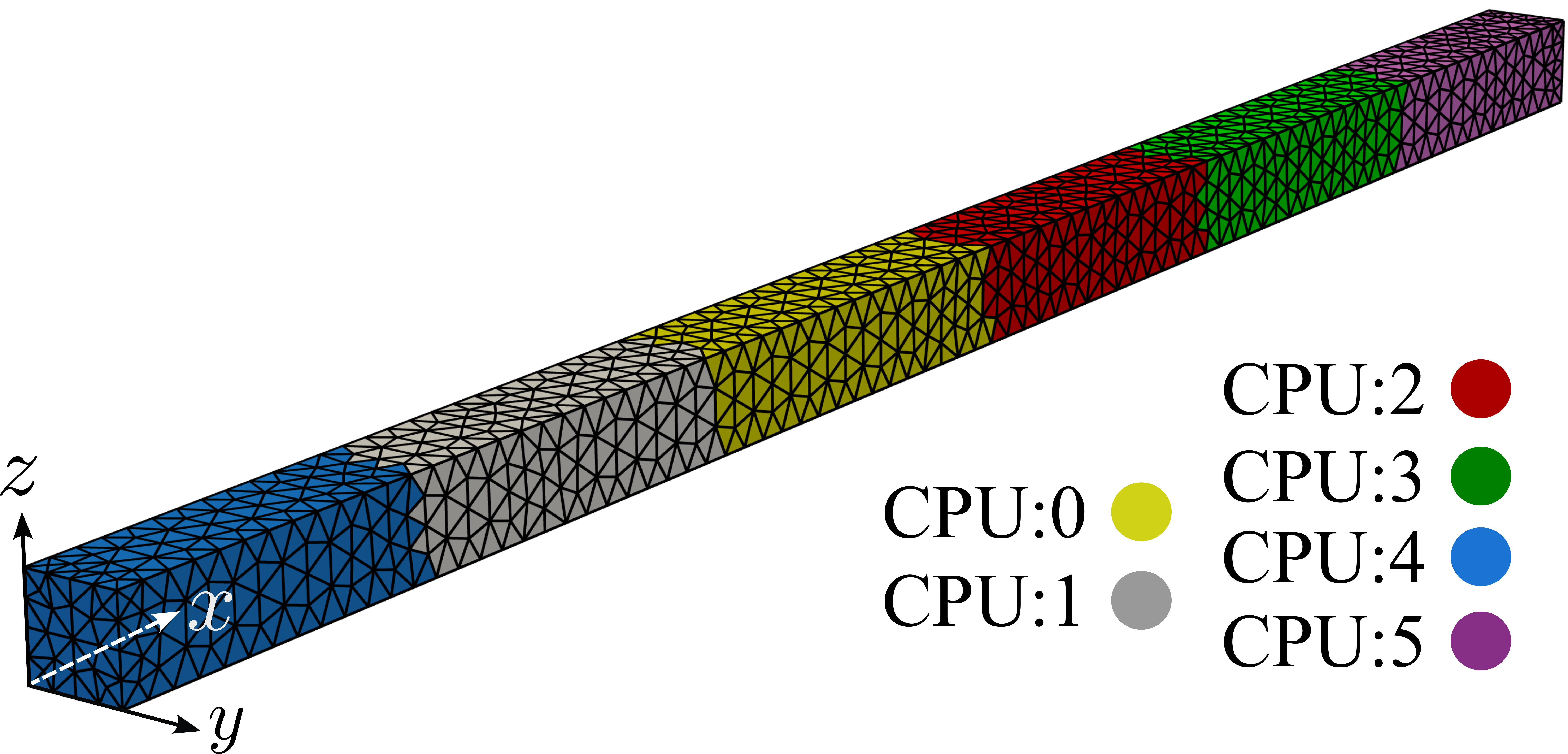}
    \caption{\centering Refined mesh partitioned over 6 CPUs.}
    \label{6cpu}
\end{figure}

\noindent{\bf Refined mesh with additional processors} - Next, we extend our data-driven framework to a refined cantilever model partitioned over 6 CPUs and shown in Figure~\ref{6cpu}. The mesh contains 4615 tetrahedral elements, we select $\Delta t = 6.2 \times 10^{-5}$ and we use the previously discussed optimal combination of hyperparameters $n_B=5$, $n_H=100$, $\eta_0=5\times 10^{-4}$, $n_p=n_f=20$, $n_{ts}=0.5$. 
We tune the sample size $n_s$ to 350, roughly proportional to the ratio of time step size between the two mesh resolutions.
Each model learns the dynamics of approximately 105 degrees of freedom on average (35 shared nodes).
Even though some differences are observed in the lateral $y$ direction for all models, the principal dynamics in the $z$-direction is accurately learned.
\begin{figure*}[!ht]
\centering
     \begin{subfigure}[b]{0.325\textwidth}
        \centering
        \includegraphics[scale=.48]{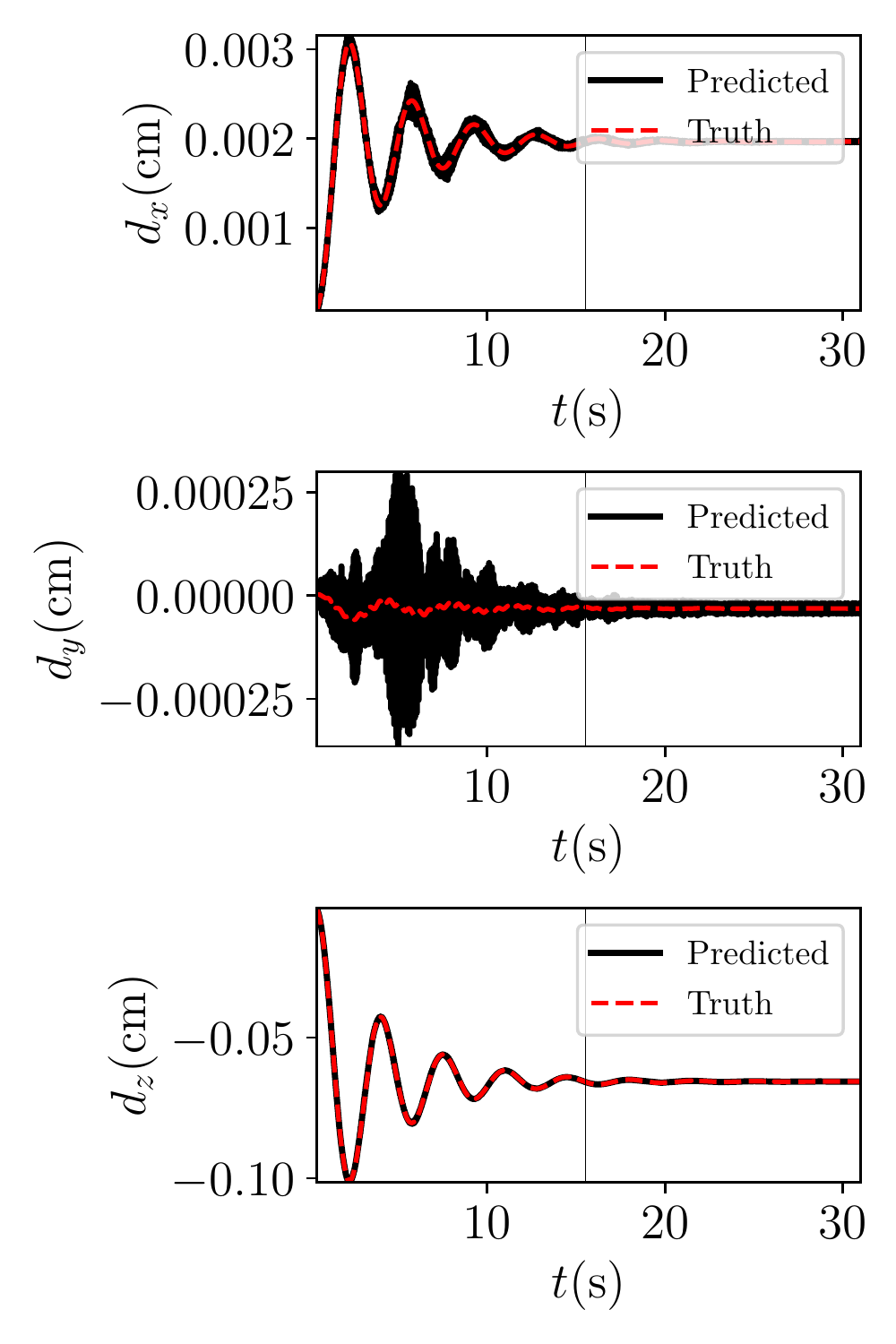}
        \caption{CPU:0, node: $(10.85,0,0.69)$.}
    \end{subfigure}
    \begin{subfigure}[b]{0.325\textwidth}
        \centering
        \includegraphics[scale=.48]{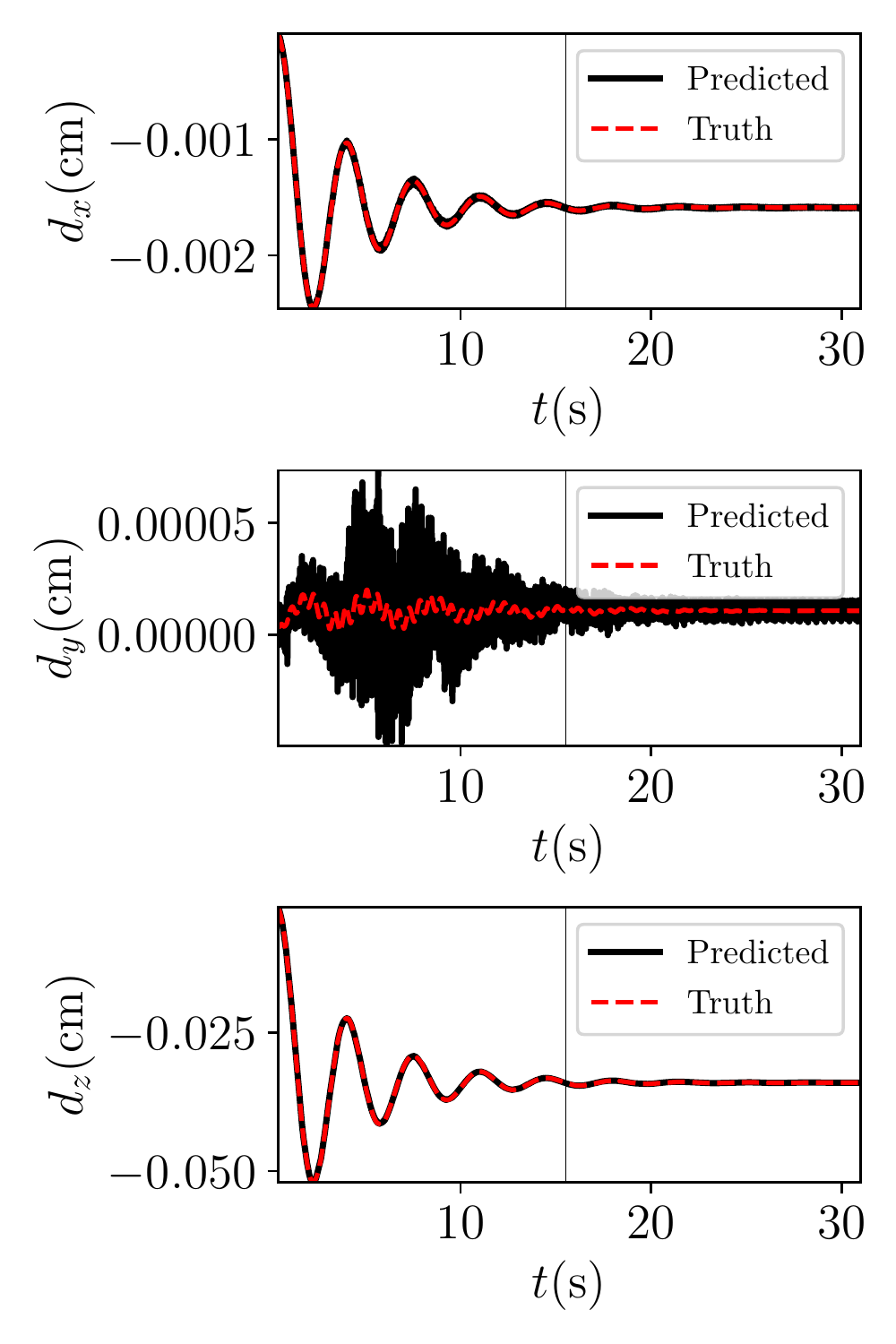}
        \caption{CPU:1, node: $(7.45,0.69,0.31)$.}
    \end{subfigure}
    \begin{subfigure}[b]{0.325\textwidth}
        \centering
        \includegraphics[scale=.48]{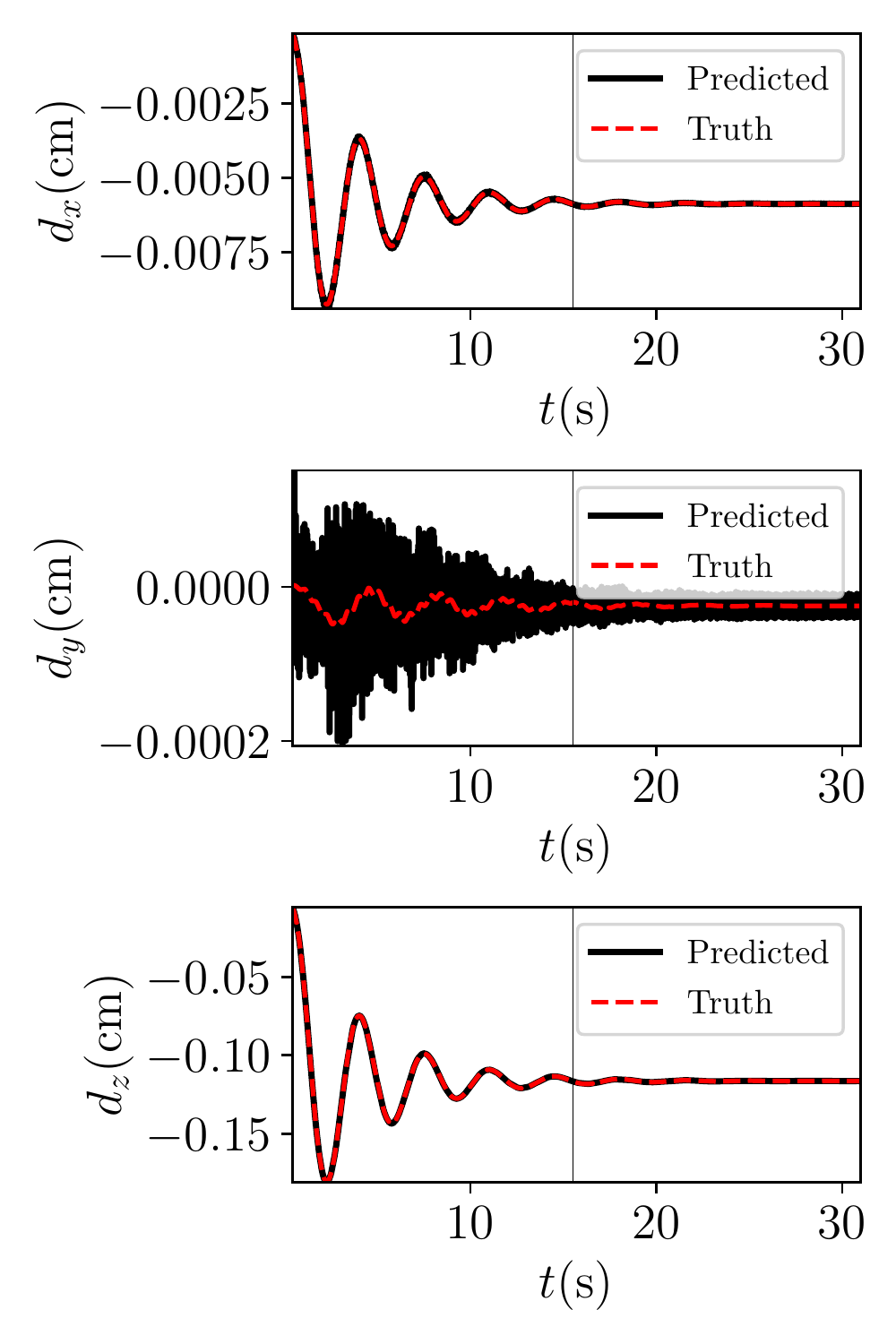}
        \caption{CPU:2, node: $(15.44,0.65,0)$.}
    \end{subfigure}\\
     \begin{subfigure}[b]{0.325\textwidth}
        \centering
        \includegraphics[scale=.48]{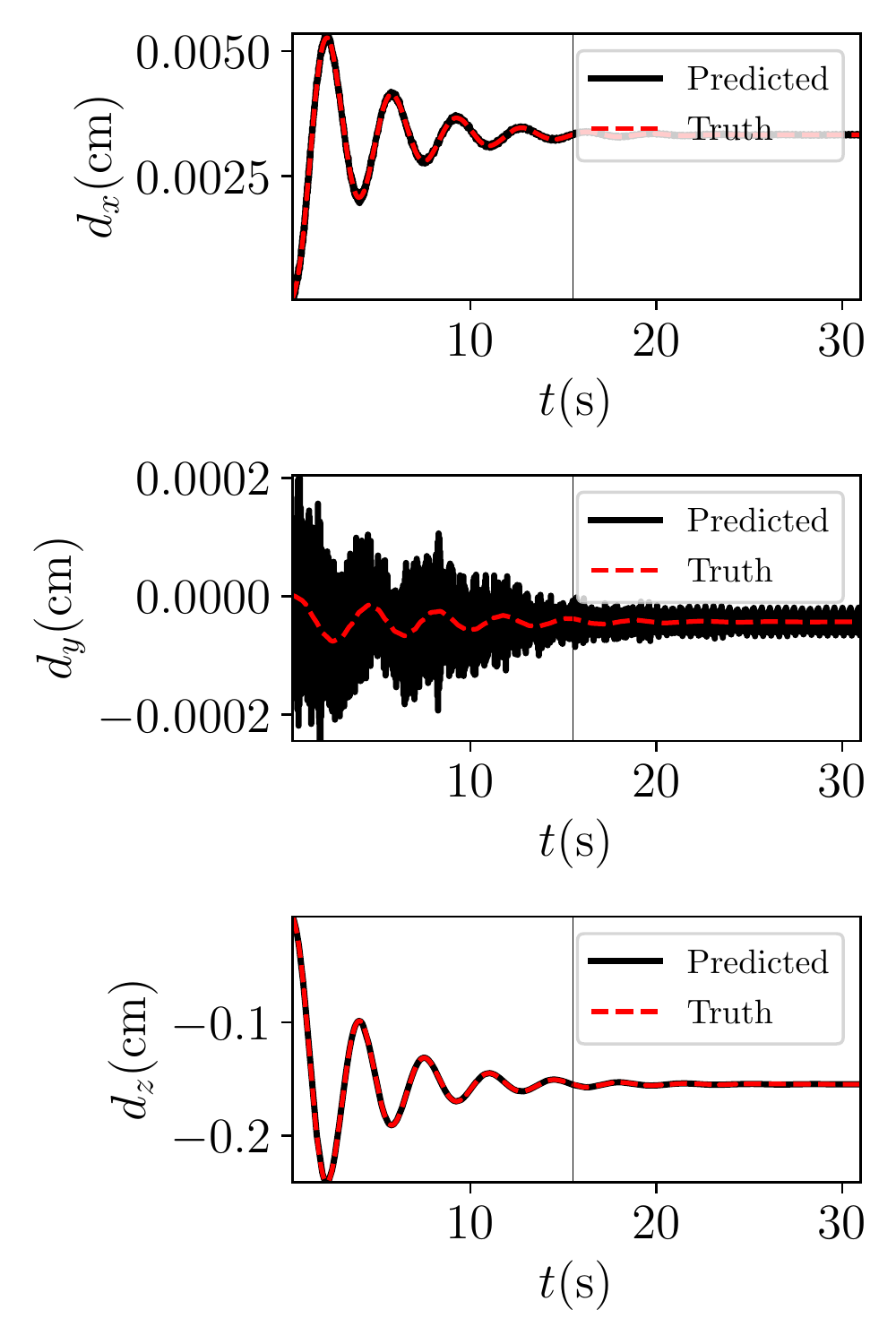}
        \caption{CPU:3, node: $(18.6,1,0.77)$.}
    \end{subfigure}
    \begin{subfigure}[b]{0.325\textwidth}
        \centering
        \includegraphics[scale=.48]{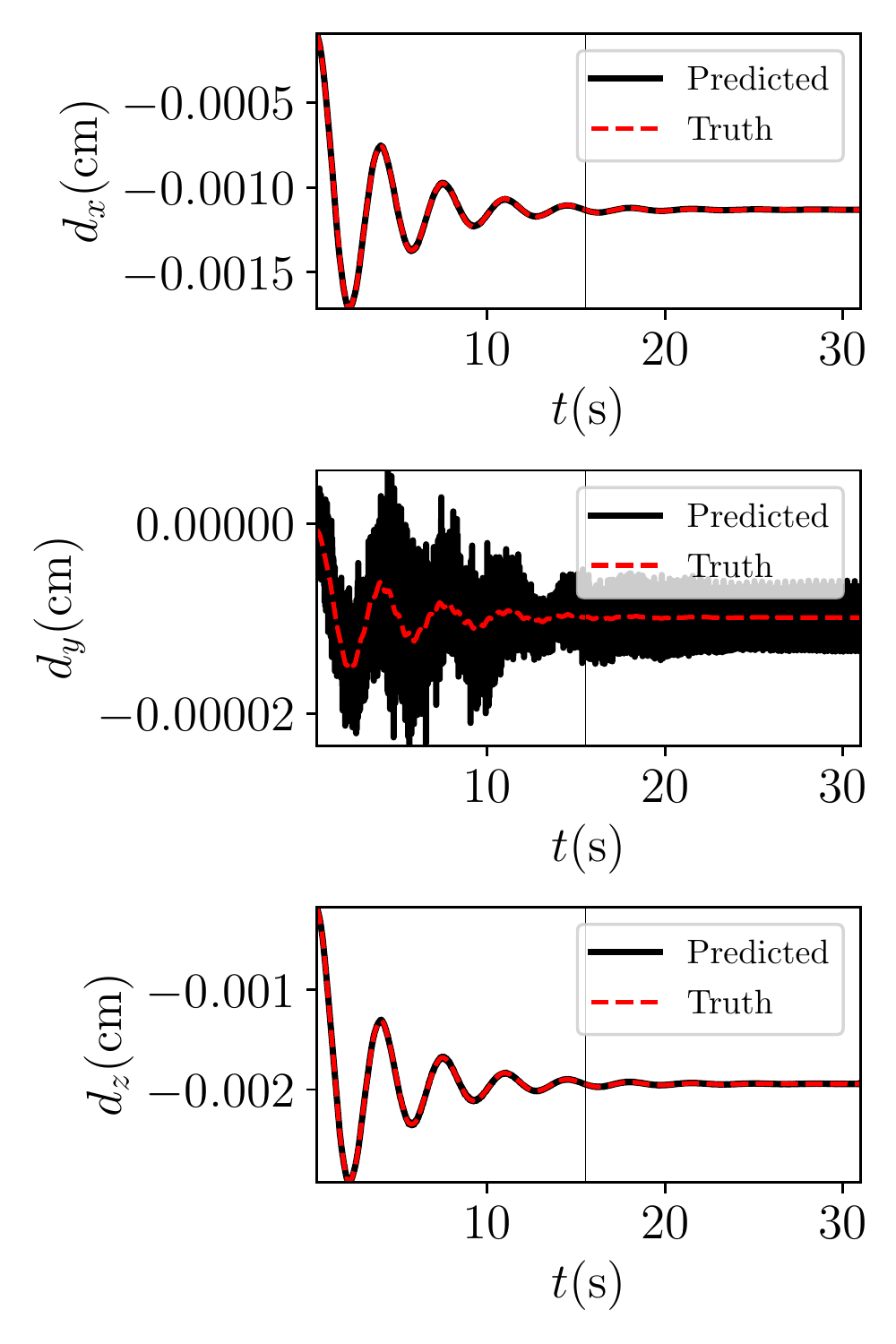}
        \caption{CPU:4 node: $(1.64,0.5,0)$.}
    \end{subfigure}
    \begin{subfigure}[b]{0.325\textwidth}
        \centering
        \includegraphics[scale=.48]{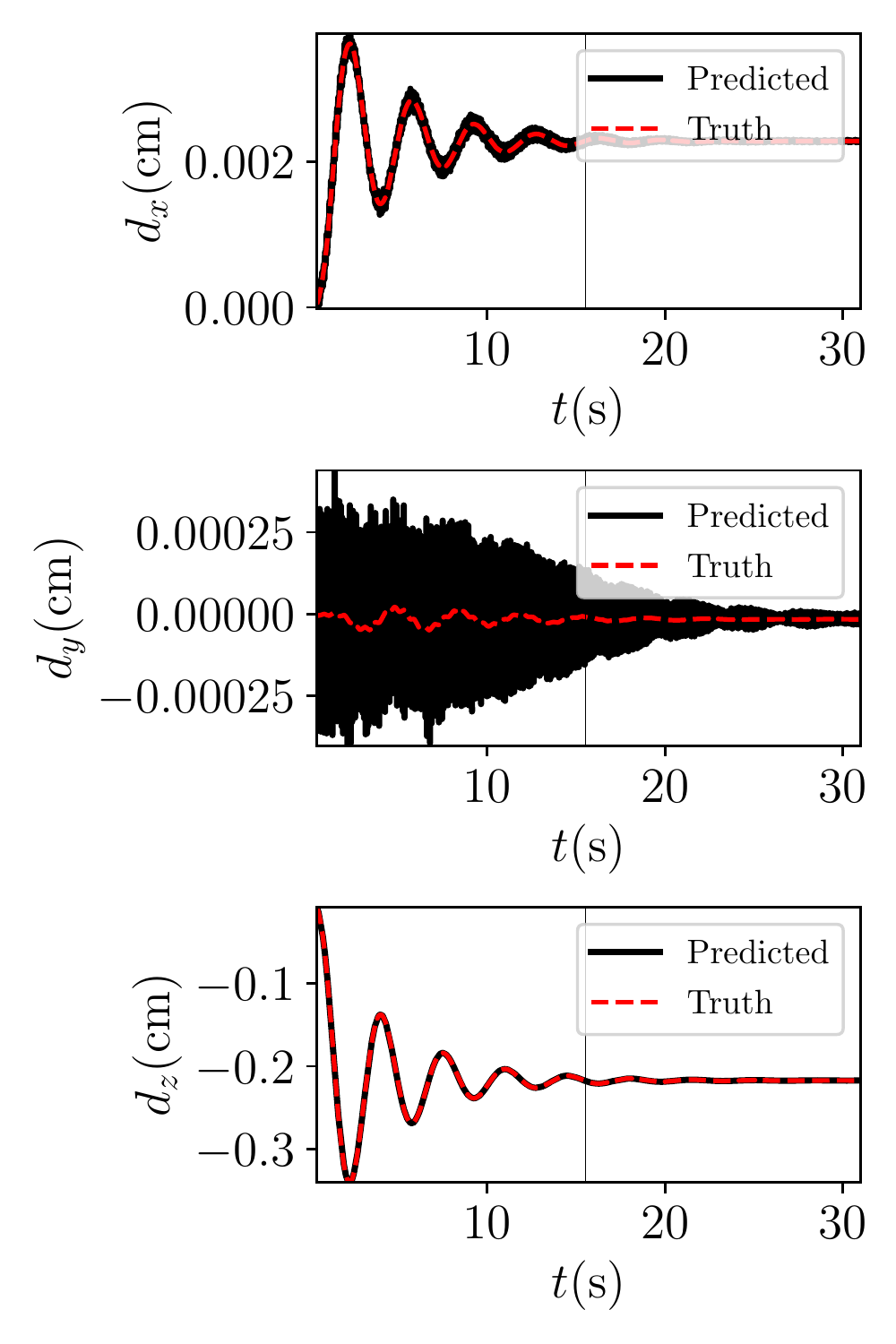}
        \caption{CPU:5, node: $(23.66,1,0.68)$.}
    \end{subfigure}
    \caption{\centering Predicted dynamics of the refined cantilever model. CPU label is from 0 to 5. Training is based on shared degrees of freedom of each partition and the plotted nodes are not shared. Predicted steps: 493000. }\label{6cpus-fig-dis}
\end{figure*}

\noindent{\bf Test with skewed forcing} - We add an additional external loading component in the $y$-direction, i.e., $f_y=f_z=0.5$ dynes/cm$^3$, resulting in extra non-zeros in the update for the displacement solutions~\eqref{lumped}, and therefore with additional complexity imposed to the network training. 

\begin{figure*}[!ht]
    \centering
    \includegraphics[scale=0.45]{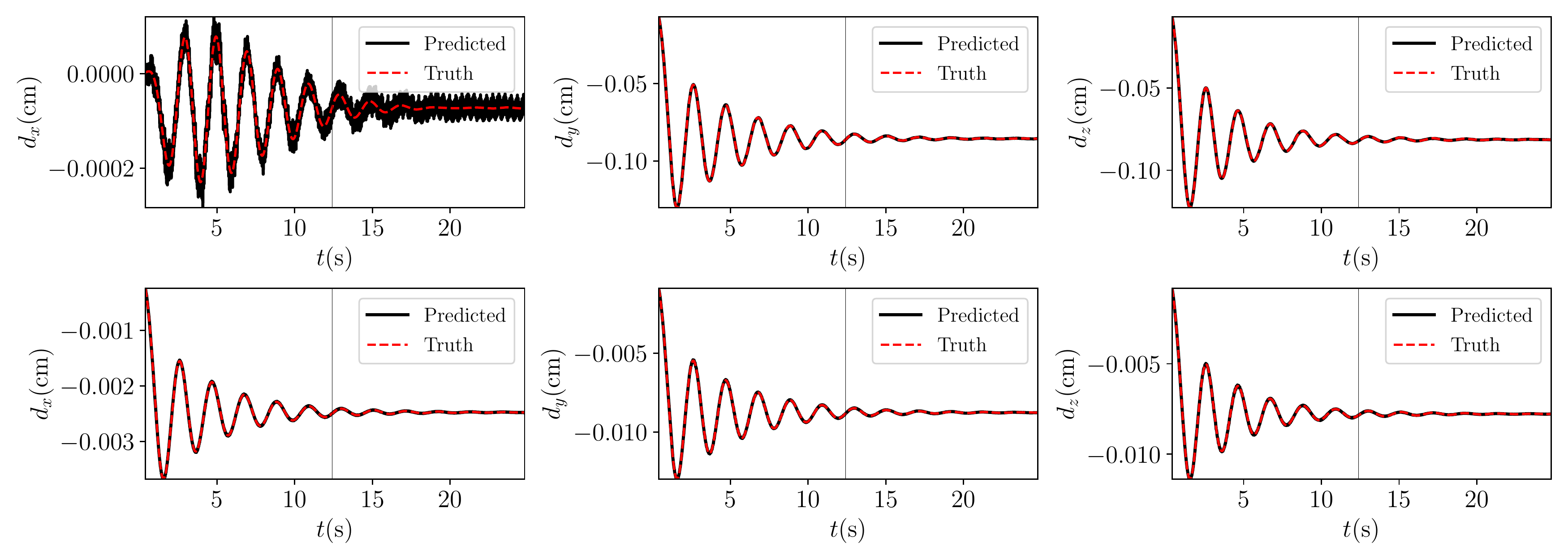}
    \caption{\centering Predicted dynamics of the cantilever model with skewed loading. CPU label: 0,1. Training is based on shared degrees of freedom of each partition and plotted nodes are not shared. Predicted steps: 98400. Top row: CPU:0, node (25.0,0.0,1.0). Bottom row: CPU:1, node (5.77,0.0,1.0). }
    \label{fig:skewed}
\end{figure*}

Results are shown in Figure~\ref{fig:skewed} where the network successfully learns the correct dynamics in the $y,z$ plane. 

\begin{figure*}[!ht]
    \centering
    \includegraphics[scale=0.45]{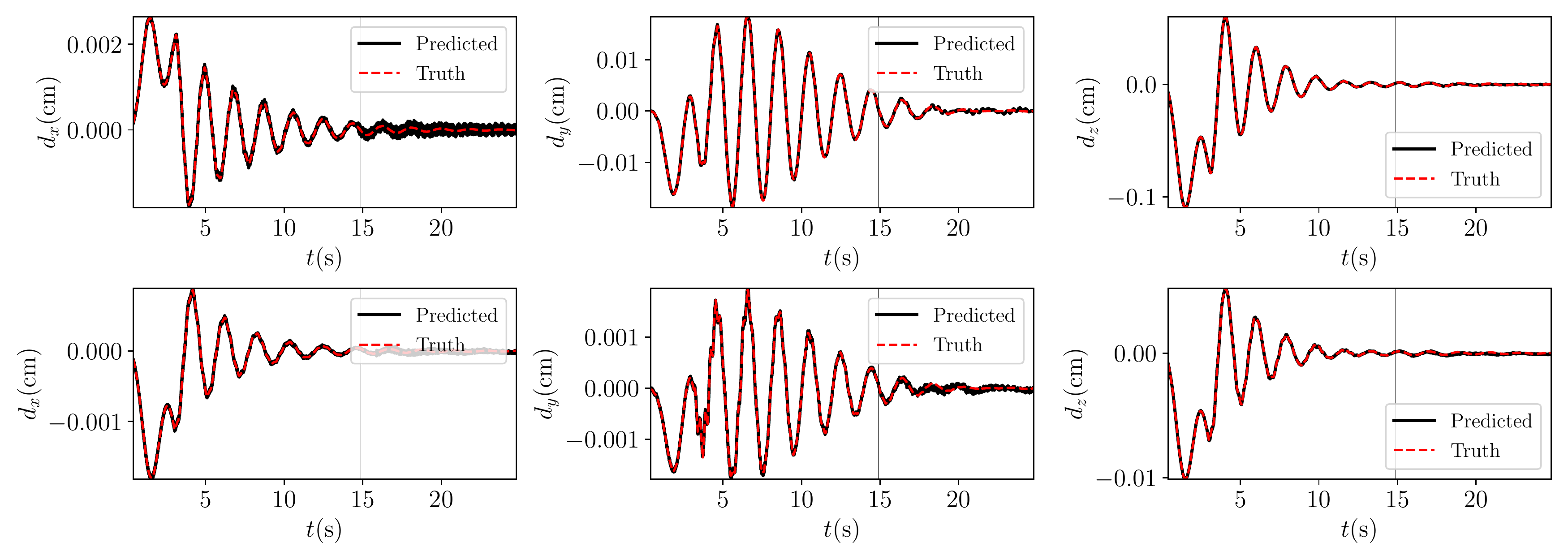}
    \caption{\centering Predicted dynamics of the cantilever model with discontinuous loading. CPU label: 0,1. Training is based on shared degrees of freedom of each partition and plotted nodes are not shared. Predicted steps: 98400. Top row: CPU:0, node (25.0,0.0,1.0). Bottom row: CPU:1, node (5.77,0.0,1.0). }
    \label{fig:dis}
\end{figure*}

\noindent{\bf Test on a discontinuous loading} - Next, we switch the external loading back to $z$-direction but consider a discontinuous load with respect to time: $\boldsymbol{f}(t)= [0, 0, -f_z\mathbb{I}_{t<3}(t)]^T$, where the indicator function $\mathbb{I}_{t<3}(t)$ defined as
\begin{equation}
\small{
\mathbb{I}_{t<3}(t) = 
\begin{cases}
1 \qquad t<3s\\
0 \qquad t\geq 3s.
\end{cases}
}
\end{equation}
As shown in Figure~\ref{fig:dis}, having part of the training data associated with a forced rather than free dynamic response does not negatively affect the ability of the network to reach a steady state with zero displacements.

\vspace{3pt}

\begin{figure*}[!ht]
\centering
\includegraphics[scale=0.37]{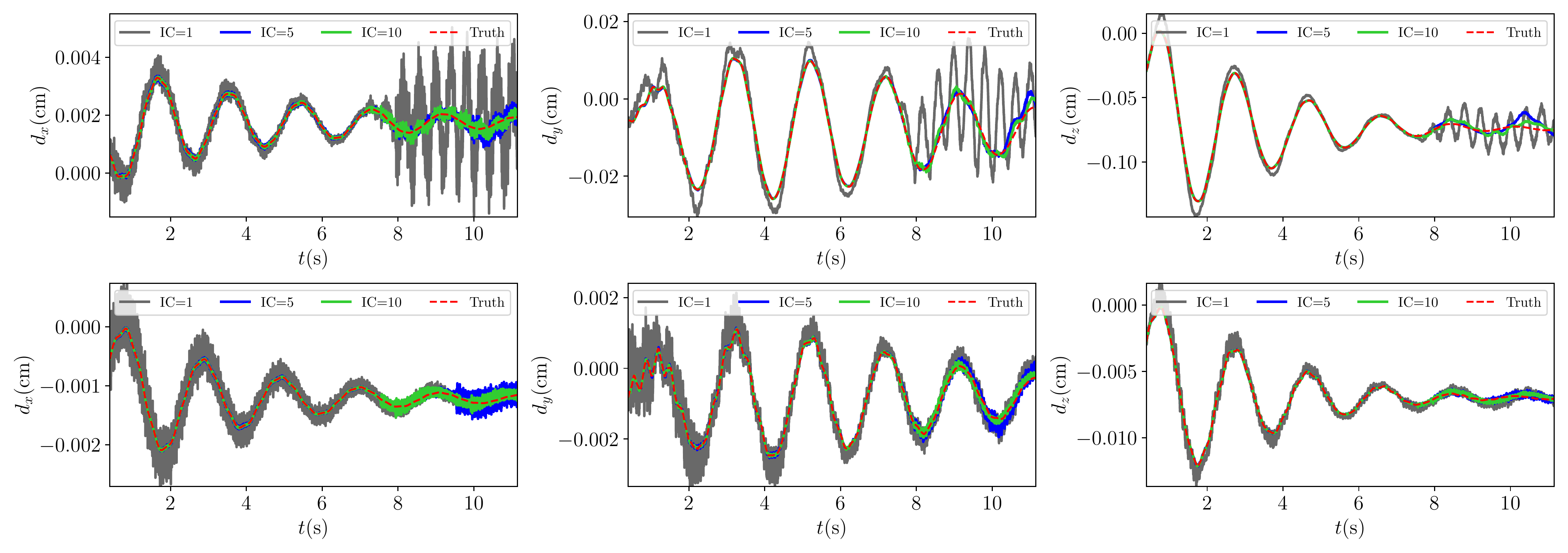}
\caption{\centering Predicted dynamics of the cantilever model trained with multiple initial conditions. CPU label: 0,1. Training is based on shared degrees of freedom of each partition and plotted nodes are not shared. Predicted steps: 43400. Top row: CPU:0, node (25.0,0.0,1.0). Bottom row: CPU:1, node (5.77,0.0,1.0). The tested initial condition is not in the training dataset.}
\label{fig:multiple-IC}
\end{figure*}

\noindent{\bf General initial conditions}\label{chp:mIC} - For more general applications, we wish our proposed LSTM network to learn and predict the evolution of a \emph{class} of dynamical systems rather than a very specific case.
To do so, we first propose a data-driven model trained on a collection of displacement solutions, generated from different initial conditions (IC) $\boldsymbol{d}^{\hspace{0.02cm} (0)}$ and check if this model is able to evolve the correct dynamics from an initial condition unseen at training. 
Note that this strategy has also been used in~\cite{he2016deep, CHEN2022110782, WU2020109307} for training general residual networks approximating dynamical systems.
To generate training data, we random perturb the steady solution $\bar{\boldsymbol{d}}$ up to 25\% and use it as the initial displacement, i.e.
\begin{equation}
\small{ 
\boldsymbol{d}^{\hspace{0.02cm} (0)} = (1 + u) \bar{\boldsymbol{d}}
},
\end{equation}
where $u$ is a uniformly distributed random variable $ u\sim \mathcal{U}(-0.25,0.25)$.
Figure~\ref{fig:multiple-IC} shows how the accuracy of the proposed approach is affected by the number of initial conditions included in the training dataset. For networks trained with 10 sets of initial conditions, the dynamics in the global $z$-direction is captured with satisfactory accuracy by both network replicas.

\vspace{3pt}

\noindent{\bf General external forcing}\label{chp:loadings} - Next, we seek to achieve generalization with respect to the external forcing.
This task is intrinsically more complicated than varying the initial conditions, as a time-dependent load $\boldsymbol{f}$ can affect the dynamic system response (see, e.g., \emph{forced} vibrations~\cite{clough1993dynamics}). 
Similarly to the previous section, we expand the training data set by adding displacement ensembles generated by our distributed finite element solver through the application of multiple external loads. A parametric family of loading conditions is obtained by introducing a uniform random variable to the $z$-component of $\boldsymbol{f}$ 
\begin{equation}
    \small{
    \boldsymbol{f} = [0,0,-\alpha_f]^T; \quad \alpha_f \sim\mathcal{U}(\alpha_{f,\min},\alpha_{f,\max}),
    }
\end{equation}
where $\alpha_{f,\min}=0.3,\alpha_{f,\max}=0.7$ are the selected prior bounds.

However, simply increasing the size of the dataset is not sufficient in this case to produce accurate predictions. We therefore modify the network architecture using a \emph{conditional} decoder~\cite{2022-conditional, pmlr-v80-garnelo18a} (e.g. see Figure~\ref{conditional}) , where the loading condition is concatenated to the last item in the input sequence, i.e., $\boldsymbol{d}^{\hspace{0.02cm} (n_p)}$. By this extra structure, the network is able to learn how to conditionally decode in the training phase, by the provided loading information. It is also worth noting that we find by stacking more identical $\alpha_f$ values to $\boldsymbol{d}^{\hspace{0.02cm} (n_p)}$ can slightly improve the accuracy. In our experiment, we concatenate 12 copies of $\alpha_f$ to $\boldsymbol{d}^{\hspace{0.02cm} (n_p)}$.

Results produced by this \emph{conditional} encoder-decoder LSTM network are illustrated in Figure~\ref{fig:multiple-F}, where training with 10 different external loads is sufficient to achieve accurate predictions for an external load not seen at training.
Finally, note how the current conditional structure can be trivially extended to the previous test case on multiple initial conditions and we expect better performance than Figure~\ref{fig:multiple-IC}, especially for long-time behaviour of small lateral (i.e., $x$ or $y$) displacements.
\begin{figure}[!ht]
    \centering
    \scalebox{.7}{\tikzset{every picture/.style={line width=0.75pt}} 

\begin{tikzpicture}[x=0.75pt,y=0.75pt,yscale=-1,xscale=1]

\draw [line width=0.75]    (280.17,32) -- (280.32,54) ;
\draw [shift={(280.33,56)}, rotate = 269.6] [color={rgb, 255:red, 0; green, 0; blue, 0 }  ][line width=0.75]    (8.74,-2.63) .. controls (5.56,-1.12) and (2.65,-0.24) .. (0,0) .. controls (2.65,0.24) and (5.56,1.12) .. (8.74,2.63)   ;
\draw  (433.78,58.59) -- (483,58.59) -- (483,90.95) -- (433.78,90.95) -- cycle ;
\draw [line width=0.75]    (320.61,74.58) -- (343.5,74.97) ;
\draw [shift={(345.5,75)}, rotate = 180.96] [color={rgb, 255:red, 0; green, 0; blue, 0 }  ][line width=0.75]    (8.74,-2.63) .. controls (5.56,-1.12) and (2.65,-0.24) .. (0,0) .. controls (2.65,0.24) and (5.56,1.12) .. (8.74,2.63)   ;
\draw  (346.33,61.63) -- (408.13,61.63) -- (408.13,88.71) -- (346.33,88.71) -- cycle ;
\draw [line width=0.75]    (408.14,74.58) -- (432.36,74.97) ;
\draw [shift={(434.36,75)}, rotate = 180.92] [color={rgb, 255:red, 0; green, 0; blue, 0 }  ][line width=0.75]    (8.74,-2.63) .. controls (5.56,-1.12) and (2.65,-0.24) .. (0,0) .. controls (2.65,0.24) and (5.56,1.12) .. (8.74,2.63)   ;
\draw  (243.74,55.84) -- (319.94,55.84) -- (319.94,94.82) -- (243.74,94.82) -- cycle ;
\draw [line width=0.75]    (189.17,31) -- (189.17,72) ;
\draw [shift={(189.17,74)}, rotate = 270] [color={rgb, 255:red, 0; green, 0; blue, 0 }  ][line width=0.75]    (8.74,-2.63) .. controls (5.56,-1.12) and (2.65,-0.24) .. (0,0) .. controls (2.65,0.24) and (5.56,1.12) .. (8.74,2.63)   ;
\draw [line width=0.75]    (189.17,74) -- (242.5,74) ;
\draw [shift={(244.5,74)}, rotate = 180] [color={rgb, 255:red, 0; green, 0; blue, 0 }  ][line width=0.75]    (8.74,-2.63) .. controls (5.56,-1.12) and (2.65,-0.24) .. (0,0) .. controls (2.65,0.24) and (5.56,1.12) .. (8.74,2.63)   ;
\draw [line width=0.75]    (279.86,95.21) -- (279.86,111.21) ;
\draw [shift={(279.86,113.21)}, rotate = 270] [color={rgb, 255:red, 0; green, 0; blue, 0 }  ][line width=0.75]    (8.74,-2.63) .. controls (5.56,-1.12) and (2.65,-0.24) .. (0,0) .. controls (2.65,0.24) and (5.56,1.12) .. (8.74,2.63)   ;
\draw  (168.8,1.09) -- (209.76,1.09) -- (209.76,31.5) -- (168.8,31.5) -- cycle ;
\draw [line width=0.75]    (460,91.5) -- (460,107.5) ;
\draw [shift={(460,109.5)}, rotate = 270] [color={rgb, 255:red, 0; green, 0; blue, 0 }  ][line width=0.75]    (8.74,-2.63) .. controls (5.56,-1.12) and (2.65,-0.24) .. (0,0) .. controls (2.65,0.24) and (5.56,1.12) .. (8.74,2.63)   ;
\draw   (105.83,96.42) .. controls (105.83,94.25) and (107.59,92.5) .. (109.75,92.5) -- (122.42,92.5) .. controls (124.58,92.5) and (126.33,94.25) .. (126.33,96.42) -- (126.33,108.17) .. controls (126.33,110.33) and (124.58,112.08) .. (122.42,112.08) -- (109.75,112.08) .. controls (107.59,112.08) and (105.83,110.33) .. (105.83,108.17) -- cycle ;
\draw    (126,102) -- (187.34,74.81) ;
\draw [shift={(189.17,74)}, rotate = 156.09] [color={rgb, 255:red, 0; green, 0; blue, 0 }  ][line width=0.75]    (10.93,-3.29) .. controls (6.95,-1.4) and (3.31,-0.3) .. (0,0) .. controls (3.31,0.3) and (6.95,1.4) .. (10.93,3.29)   ;

\draw (437,65.58) node [anchor=north west][inner sep=0.75pt]  [font=\small]  {$\widehat{\boldsymbol{d}}^{\hspace{0.02cm} (n_p+1)}$};
\draw (256,65.5) node [anchor=north west][inner sep=0.75pt]  [font=\small]  {$\boldsymbol{h}_{D}^{(1)} ,\ \boldsymbol{c}_{D}^{(1)}$};
\draw (357,68) node [anchor=north west][inner sep=0.75pt]   [align=left] {Dense};
\draw (173,8.5) node [anchor=north west][inner sep=0.75pt]  [font=\small]  {$\boldsymbol{d}^{\hspace{0.02cm}(n_p)}$};
\draw (276.5,109) node [anchor=north west][inner sep=0.75pt]    {$\vdots $};
\draw (276.5,6) node [anchor=north west][inner sep=0.75pt]    {$\vdots $};
\draw (231,12) node [anchor=north west][inner sep=0.75pt]    {$\cdots $};

\draw (456.5,106) node [anchor=north west][inner sep=0.75pt]    {$\vdots $};
\draw (127,103.89) node [anchor=north west][inner sep=0.75pt]  [font=\footnotesize,rotate=-335.34] [align=left] {concatenate};
\draw (108,97) node [anchor=north west][inner sep=0.75pt]  [font=\small]  {$\alpha_f$};

\draw [line width=0.75]    (210,16) -- (230.5,16) ;
\draw [shift={(230,16)}, rotate = 180] [color={rgb, 255:red, 0; green, 0; blue, 0 }  ][line width=0.75]    (8.74,-2.63) .. controls (5.56,-1.12) and (2.65,-0.24) .. (0,0) .. controls (2.65,0.24) and (5.56,1.12) .. (8.74,2.63)   ;

\end{tikzpicture}}
    \caption{Modified conditional decoder}
    \label{conditional}
\end{figure}
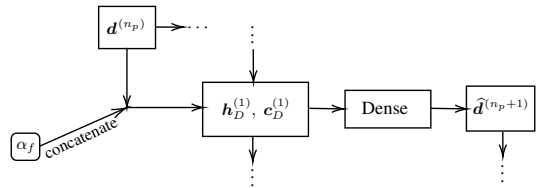
\begin{figure*}[!ht]
    \centering
    \includegraphics[scale=0.335]{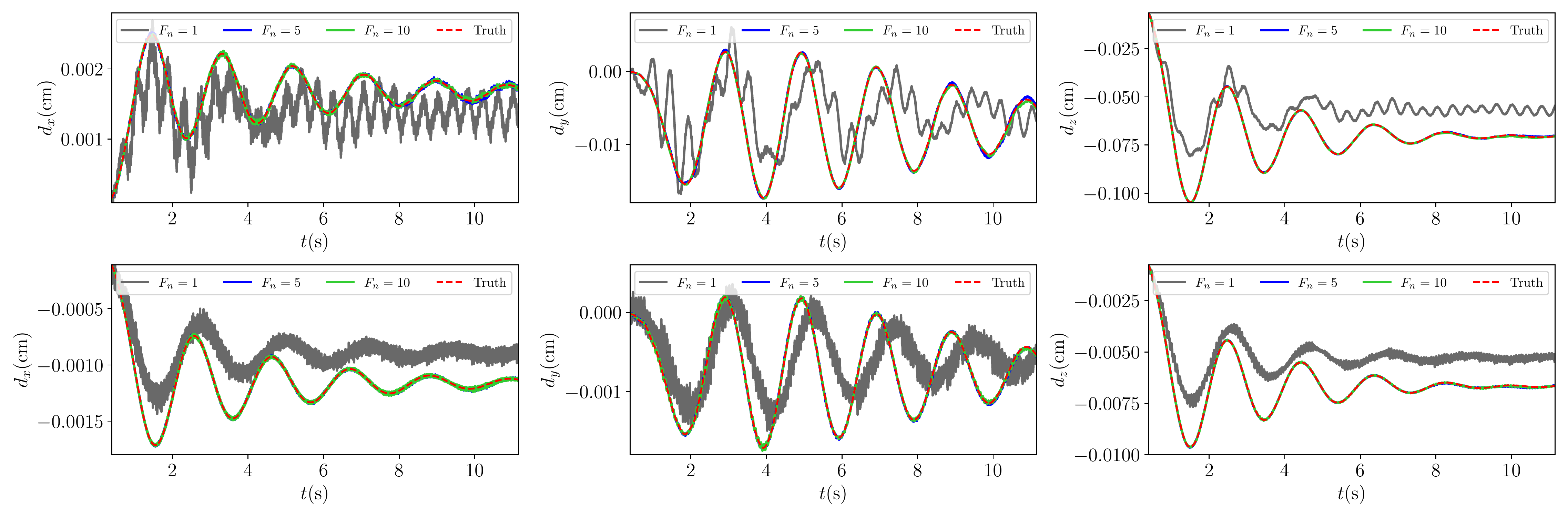}
    \caption{\centering Predicted dynamics of the cantilever model trained with multiple external loadings. CPU label: 0,1. Training is based on shared degrees of freedom of each partition and plotted nodes are not shared. Predicted steps: 43400. Top row: CPU:0, node (25.0,0.0,1.0). Bottom row: CPU:1, node (5.77,0.0,1.0). The tested external loading is not in the training dataset.}
    \label{fig:multiple-F}
\end{figure*}

\vspace{3pt}

\noindent{\bf Full system modeling} - While tests in the previous sections focus on predicting the system dynamics at a small number of shared locations, we would like to see how the accuracy of the proposed network is affected for an increasing number of such locations. Thus, we use the proposed LSTM network to predict the dynamic response for a coarse discretization of the entire cantilever beam.
The hyperparameters are selected as $n_B=5$, $n_H=100$, $\eta_0=5\times 10^{-4}$, $n_p=n_f=20$, $n_s=80$, $n_{ts}=0.5$ and the external loading is $\boldsymbol{f} = [0,0,-f_z]^T$. 
Clearly in this case, once we start to use the trained network model, no more finite element computations are required, leading to a substantial reduction in the computational effort.

\begin{figure*}[!ht]
\centering
     \begin{subfigure}[b]{0.325\textwidth}
        \centering
        \includegraphics[scale=.48]{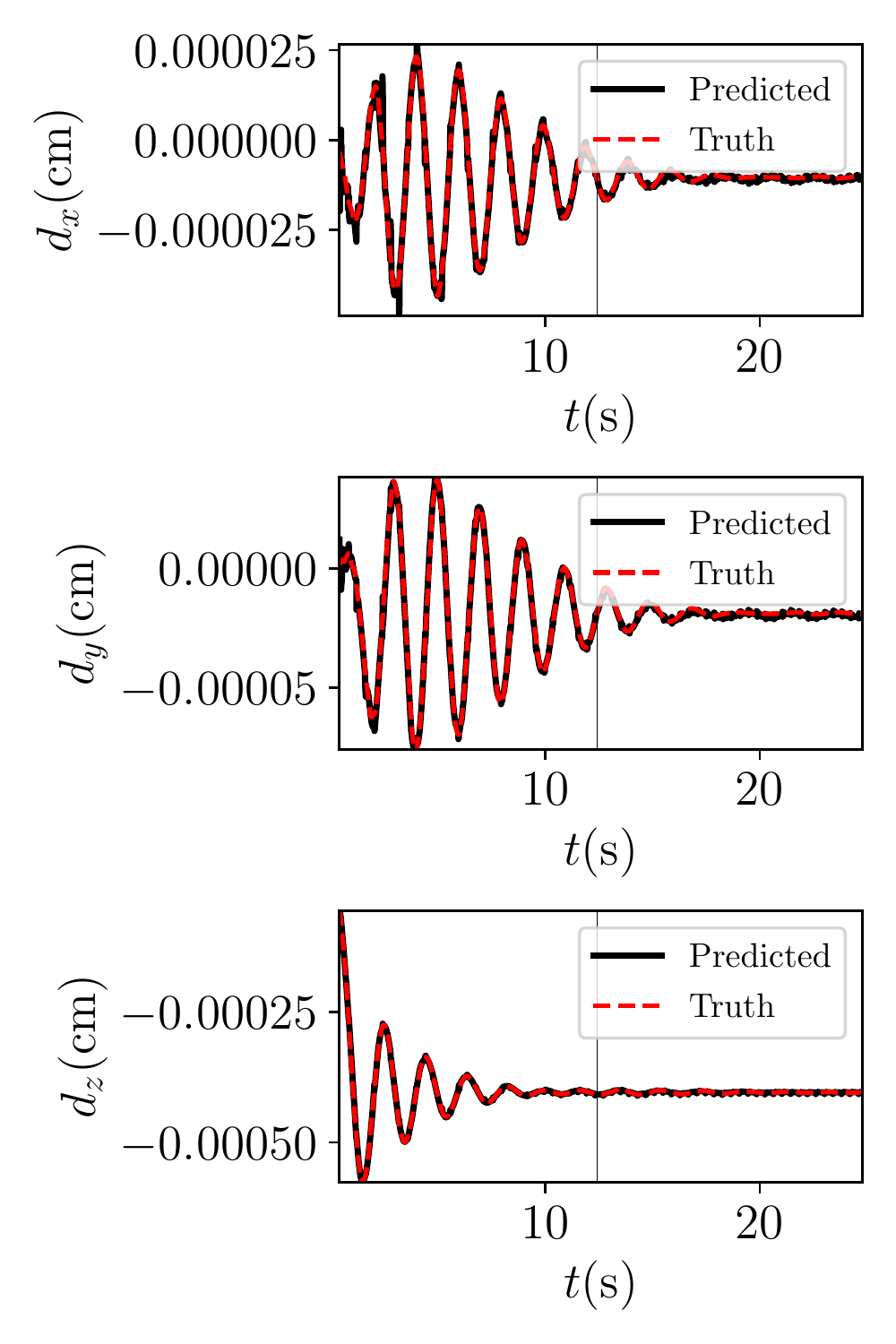}
        \caption{Node: $(0.96,1,0.5)$.}
    \end{subfigure}
    \begin{subfigure}[b]{0.325\textwidth}
        \centering
        \includegraphics[scale=.48]{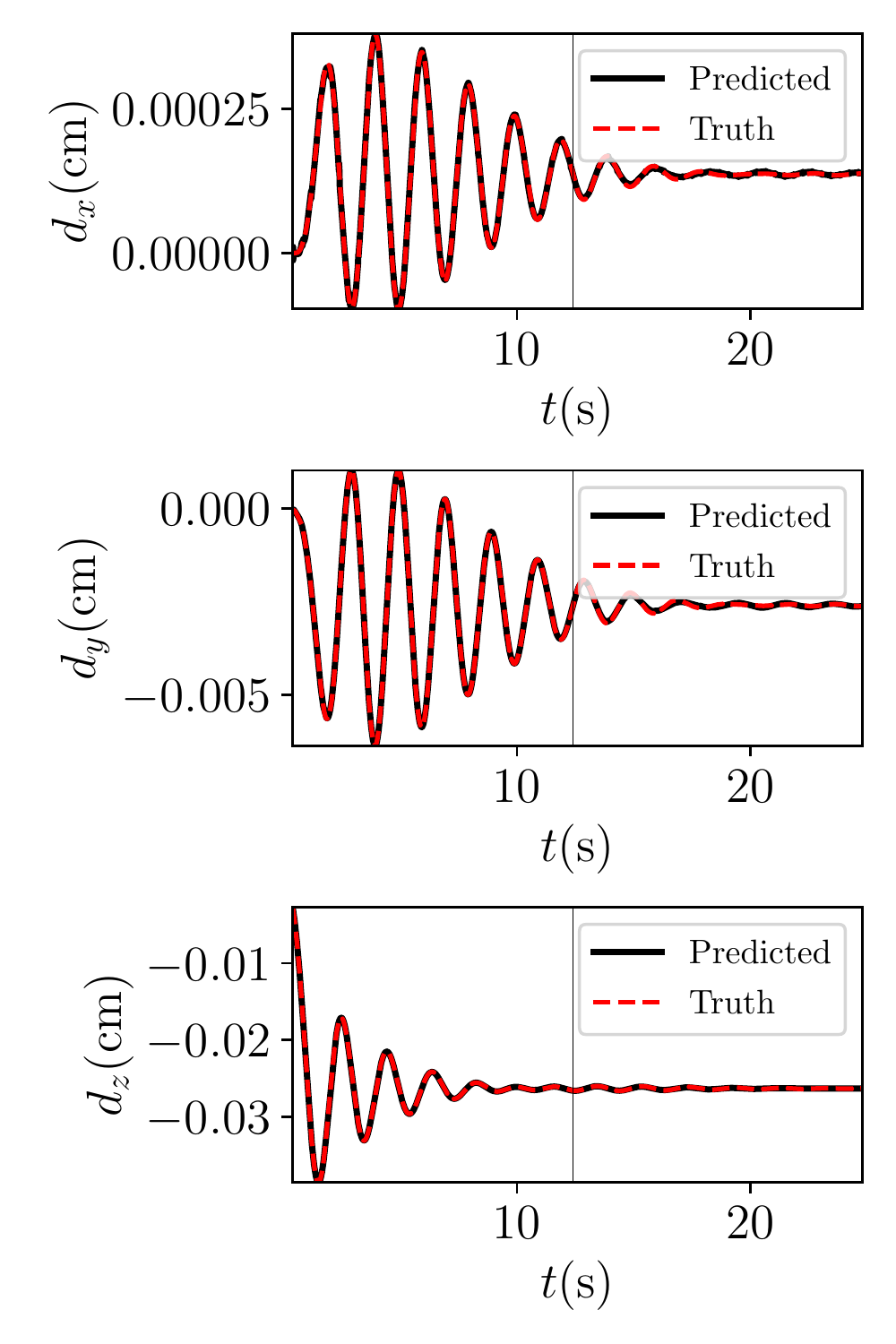}
        \caption{Node: $(12.5,1,0.5)$.}
    \end{subfigure}
    \begin{subfigure}[b]{0.325\textwidth}
        \centering
        \includegraphics[scale=.48]{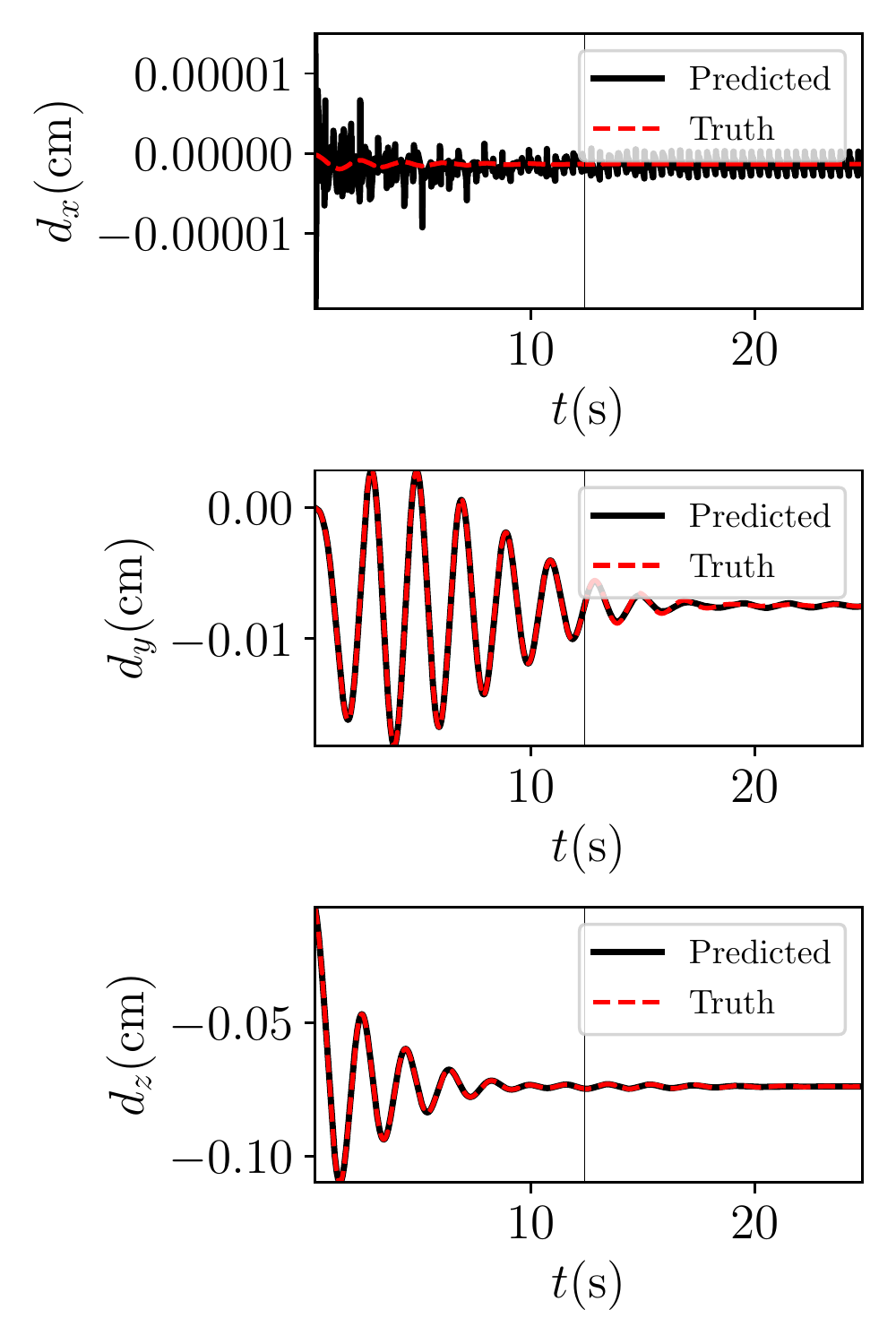}
        \caption{Node: $(25,0.5,0.5)$.}
    \end{subfigure}
    \caption{\centering Predicted dynamics at three distinctive nodes of the cantilever model. Training is based on all degrees of freedom. Predicted steps: 98400.}\label{serial-fig-dis}
\end{figure*}

\begin{figure}[!ht]
\centering
\begin{subfigure}[b]{0.4\textwidth}
  \includegraphics[scale=.087]{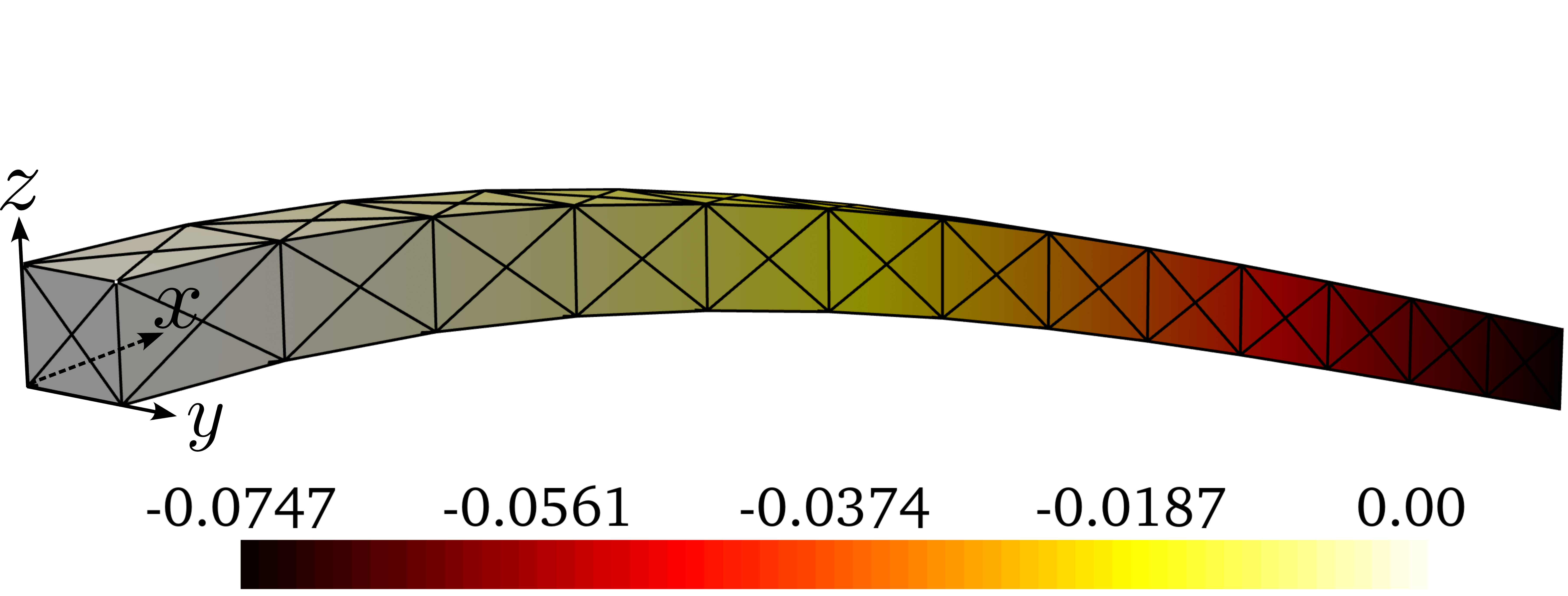}
  \caption{Exact [cm].}
\end{subfigure}
\begin{subfigure}[b]{0.4\textwidth}
  \includegraphics[scale=.087]{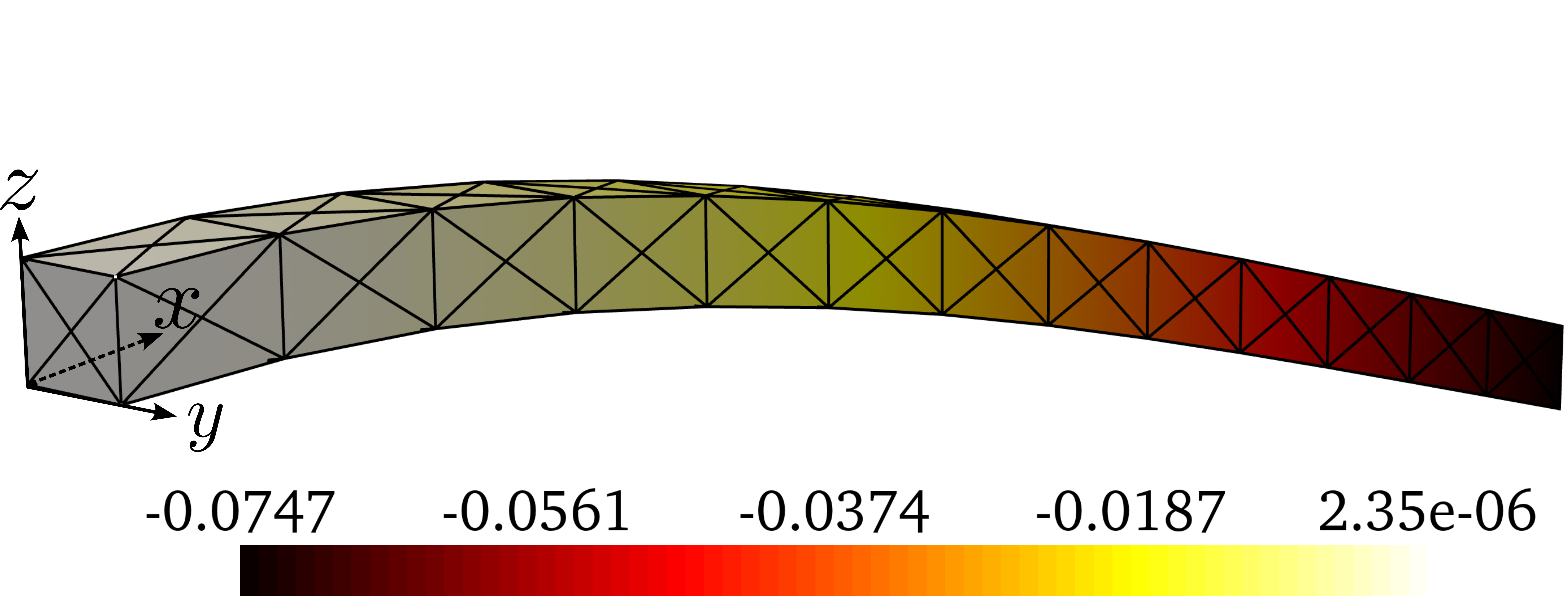}
  \caption{Predicted [cm].}
\end{subfigure}
\caption{\centering Network prediction for full cantilever model. Comparison of exact and predicted $z$-displacement solution at $t\sim 12.4s$.}
\label{fig:wrappedz}
\end{figure}

We first show the evolution of displacement predictions at three distinctive locations in the mesh, i.e., near the clamped end, in the middle and at the tip. Figure~\ref{serial-fig-dis} shows a satisfactory accuracy in all cases, except for the $x$-displacement at the tip, but, in this case, the displacement is practically zero and therefore the absolute error still acceptable. 
We also show a comparison of the exact and predicted $z$-displacement contours in Figure~\ref{fig:wrappedz}. In such a case, the surrogate model has no knowledge on how to strongly enforce a Dirichlet boundary condition at the clamped end. However, the predicted displacement of $2.35\times 10^{-6}$ is sufficiently small.

\begin{figure}[!ht]
    \centering
    \includegraphics[scale=0.3]{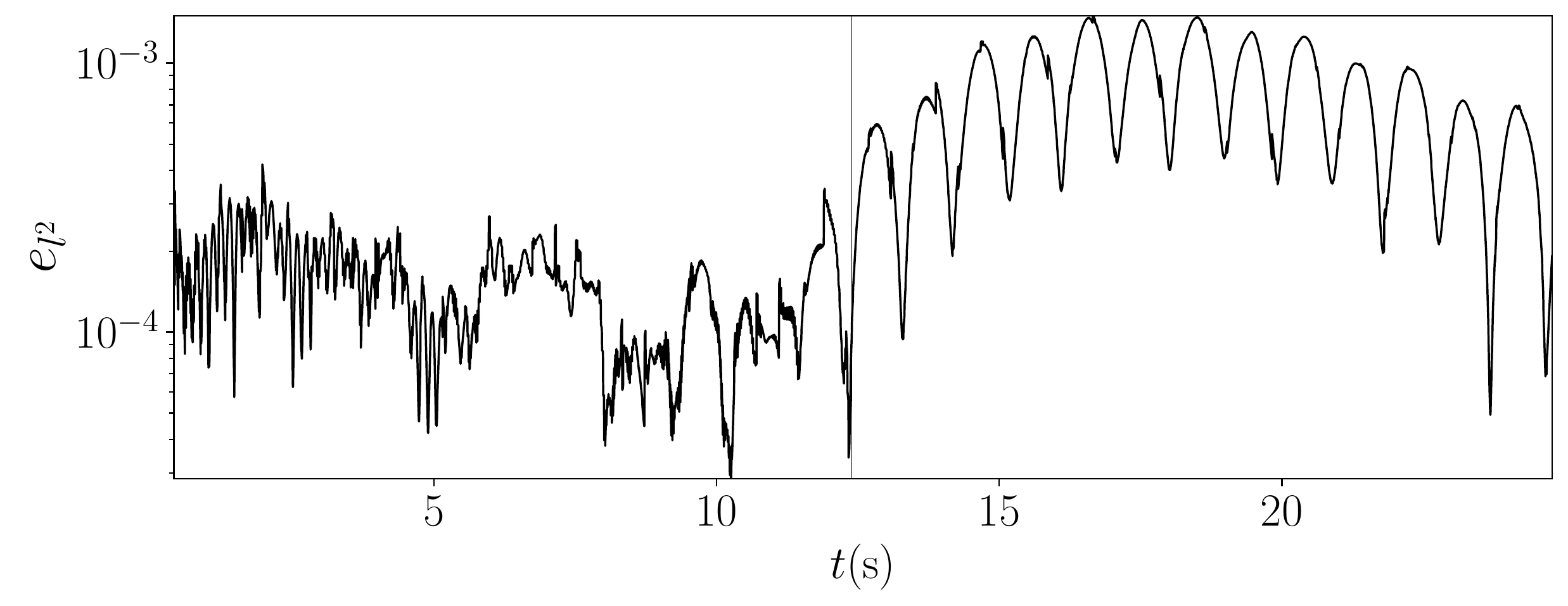}
    \caption{\centering Time history of the $l^2$ error for the predicted solutions of the full cantilever model. The number of predicted steps is equal to 98400.}
    \label{fig:full l2}
\end{figure}

Finally, Figure~\ref{fig:full l2} shows the evolution of the $l^2$ error using Equation~\eqref{l2e}, where we observe a trend similar to that reported in Figure~\ref{fig:subset l2}.

\subsection{Coronary model}\label{cardio}
\begin{figure*}[!ht]
\centering
\begin{subfigure}[b]{0.49\textwidth}
        \centering
        \includegraphics[scale=.13]{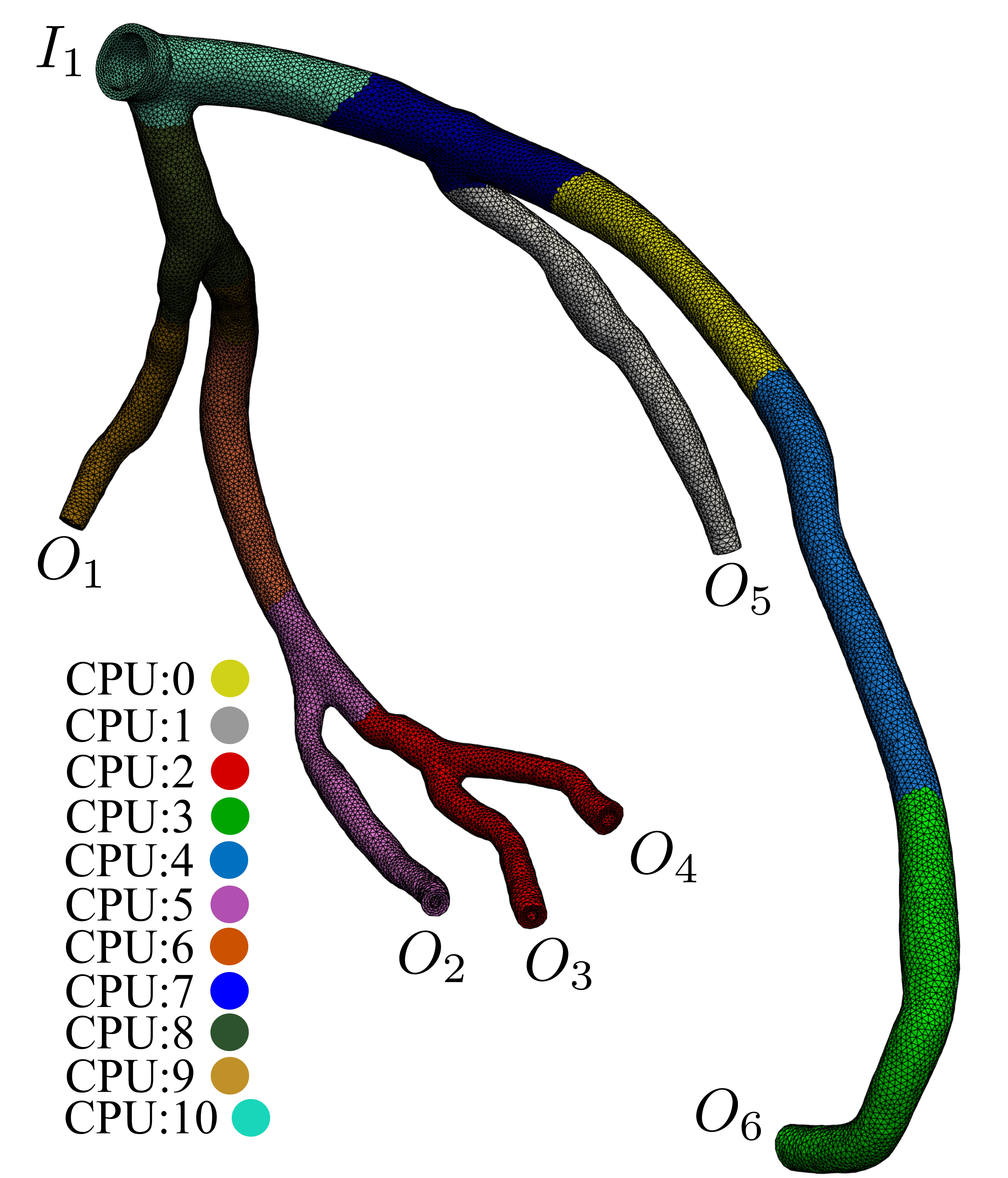}
        \caption{\centering Geometry of the left coronary artery model. The mesh is partitioned over 11 CPUs, labelled 0 to 10.}
        \label{coronary}
    \end{subfigure}
     \begin{subfigure}[b]{0.49\textwidth}
        \centering
        \includegraphics[scale=.32]{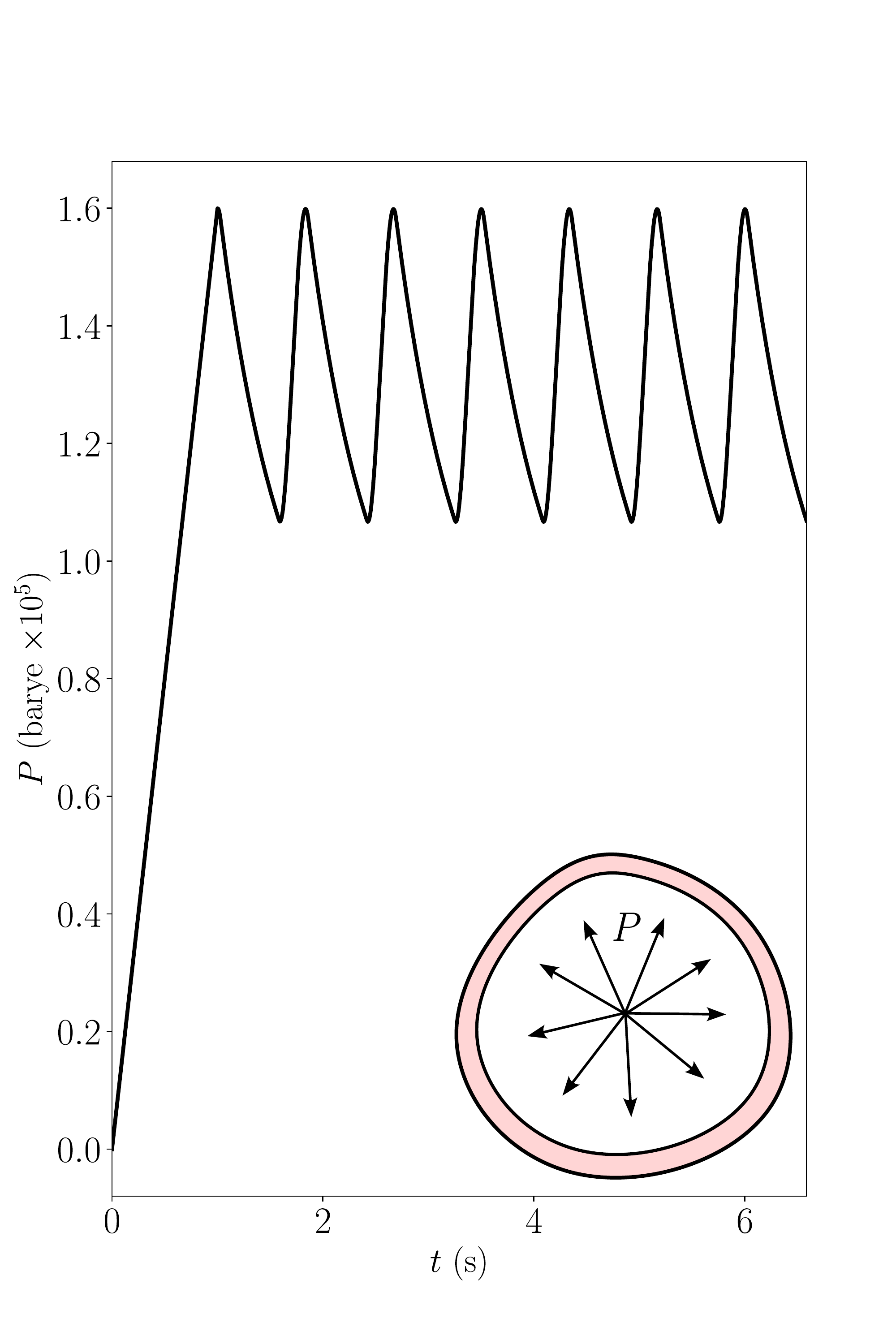}
        \caption{\centering Pressure waveform acting on the vessel lumen.}
        \label{waveform}
    \end{subfigure}
    \caption{\centering Geometry, mesh and loading condition for the coronary artery model.}
\end{figure*}

%
In this section, we extend the proposed computational framework to a realistic cardiovascular simulation. We adopt a patient-specific human left coronary artery model, which was investigated in previous FSI and UQ studies~\cite{seo2019performance,seo2020effects}. 
The model dynamics is driven by a pulsatile pressure acting on the vessel lumen, whose periodic waveform is illustrated in Figure~\ref{waveform}. 
The pressure is gradually applied to the model through a linear ramp active during the first $1$ second of the simulation ({\it{cf.}} Remark~\ref{ramp}, $t_{\text{end}}=1 s$).  It fluctuates from a systolic maximum of $1.6\times10^5$ baryes (120 mmHg) to a diastolic minimum of $1.067\times10^5$ baryes (80 mmHg) with a period of approximately 0.833 s (72 beats per minute), as per the normal systemic pressure and heart rate at rest of a healthy adult.
Additionally, we consider an elastic modulus of $E = 6.26\times 10^6$ $\text{dynes/cm}^2$, a density of the vascular tissue of $\rho=1 \text{g/cm}^3$~\cite{kim2010patient} and a Poisson ratio equal to $\nu = 0.4$ which approximates incompressibility conditions.

\begin{figure}[!ht]
    \centering
    \includegraphics[scale=0.31]{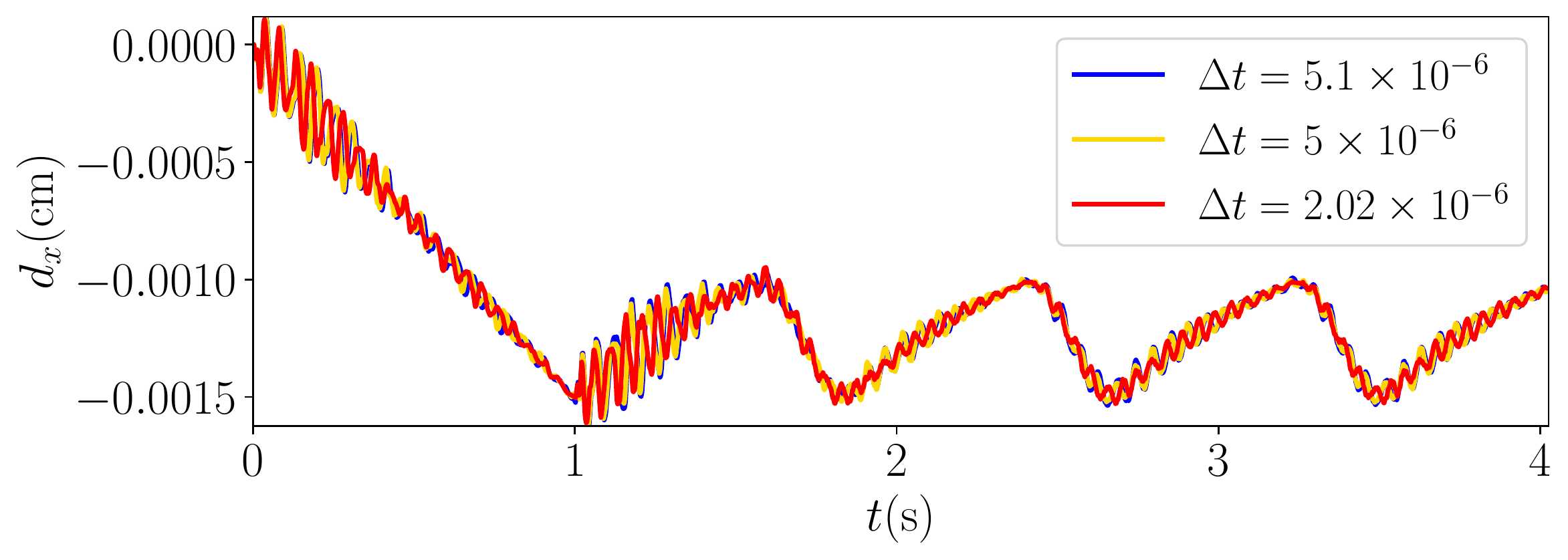}
    \caption{\centering Comparison of $x$-component displacement dynamics with and without artificial mass scaling. Plotted node: (-13.7, 6.4, -9.6) }
    \label{fig:mass-scaling}
\end{figure}

Homogeneous Dirichlet boundary conditions are strongly enforced at the main inlet ($I_1$) and $6$ outlets in the bottom ($O_1,\cdots,O_6$), as shown in Figure~\ref{coronary}. 
The model is discretized using 373,435 tetrahedral elements with 250,659 final degrees of freedom and is partitioned over 11 cores, where each core shares about 262 nodes with its neighbors.
The original time step size calculated by equation~\eqref{cfl} is about $2.02\times 10^{-6}$, and we have increased it up to $5\times 10^{-6}$ via artificial mass scaling (e.g. Algorithm~\ref{alg-cms}), saving 60\% of the computational time, but leading to a sensible increase in the total mass of the model equal to 21.13\%. However, we verified in Figure~\ref{fig:mass-scaling} that this increase corresponded to a marginal effect on the model dynamics. We also include a comparison test in Figure~\ref{fig:mass-scaling} using the original time step size $2.02\times 10^{-6}$, current size $5\times 10^{-6}$ and a further slightly-increased size $ 5.1\times 10^{-6}$, which corresponds to 24.7\% of mass increment.
%
%

Note that a model of the coronary circulation offers an ideal benchmark for the proposed approach, since the small size of the coronary arteries leads to a lower bound in the explicit time step with respect to other anatomical regions.

\begin{figure*}[!ht]
\centering
     \begin{subfigure}[b]{0.325\textwidth}
        \centering
        \includegraphics[scale=.48]{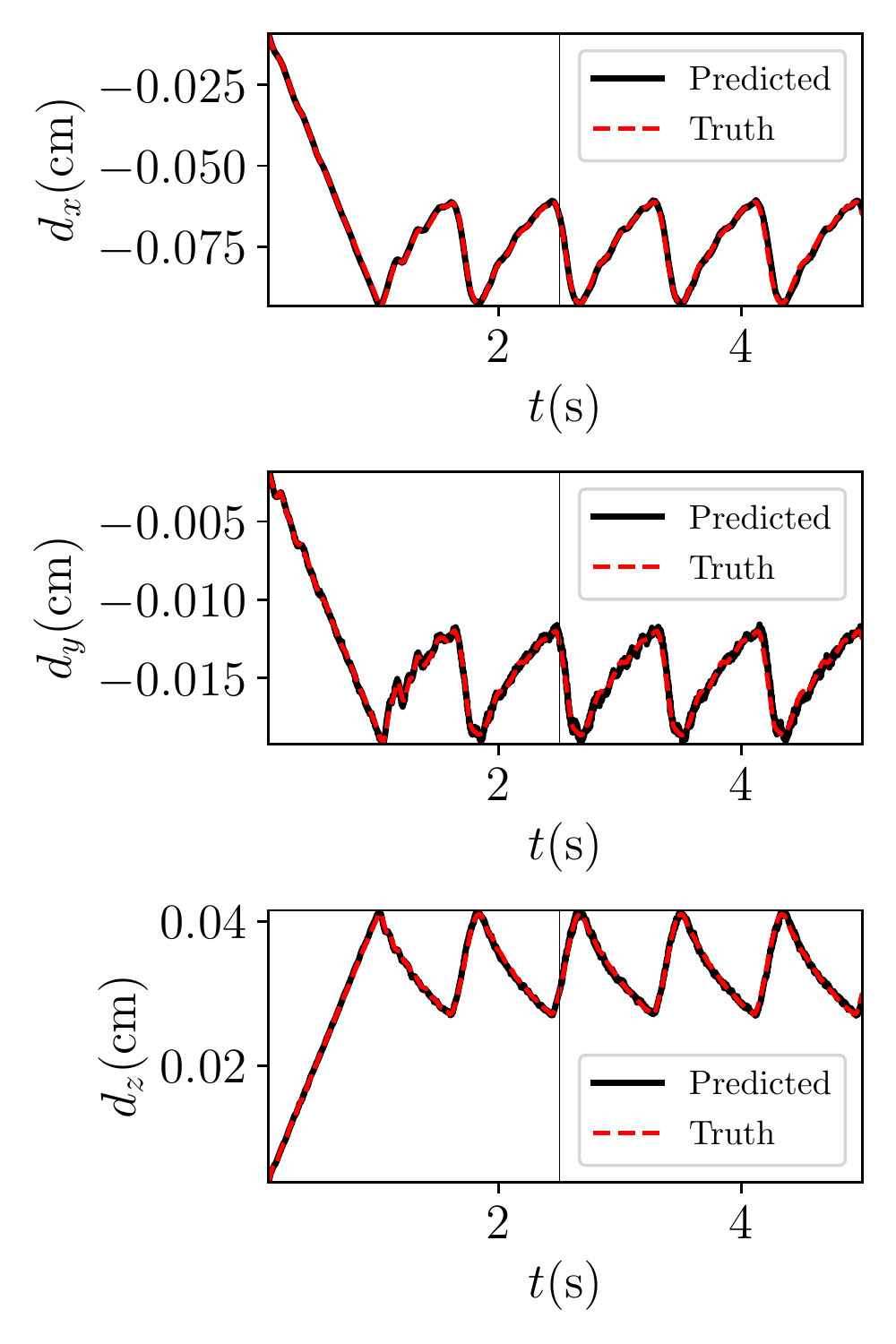}
        \caption{CPU:0, node: (-11.9,10.4,-5.4).}
    \end{subfigure}
        \begin{subfigure}[b]{0.325\textwidth}
        \centering
        \includegraphics[scale=.48]{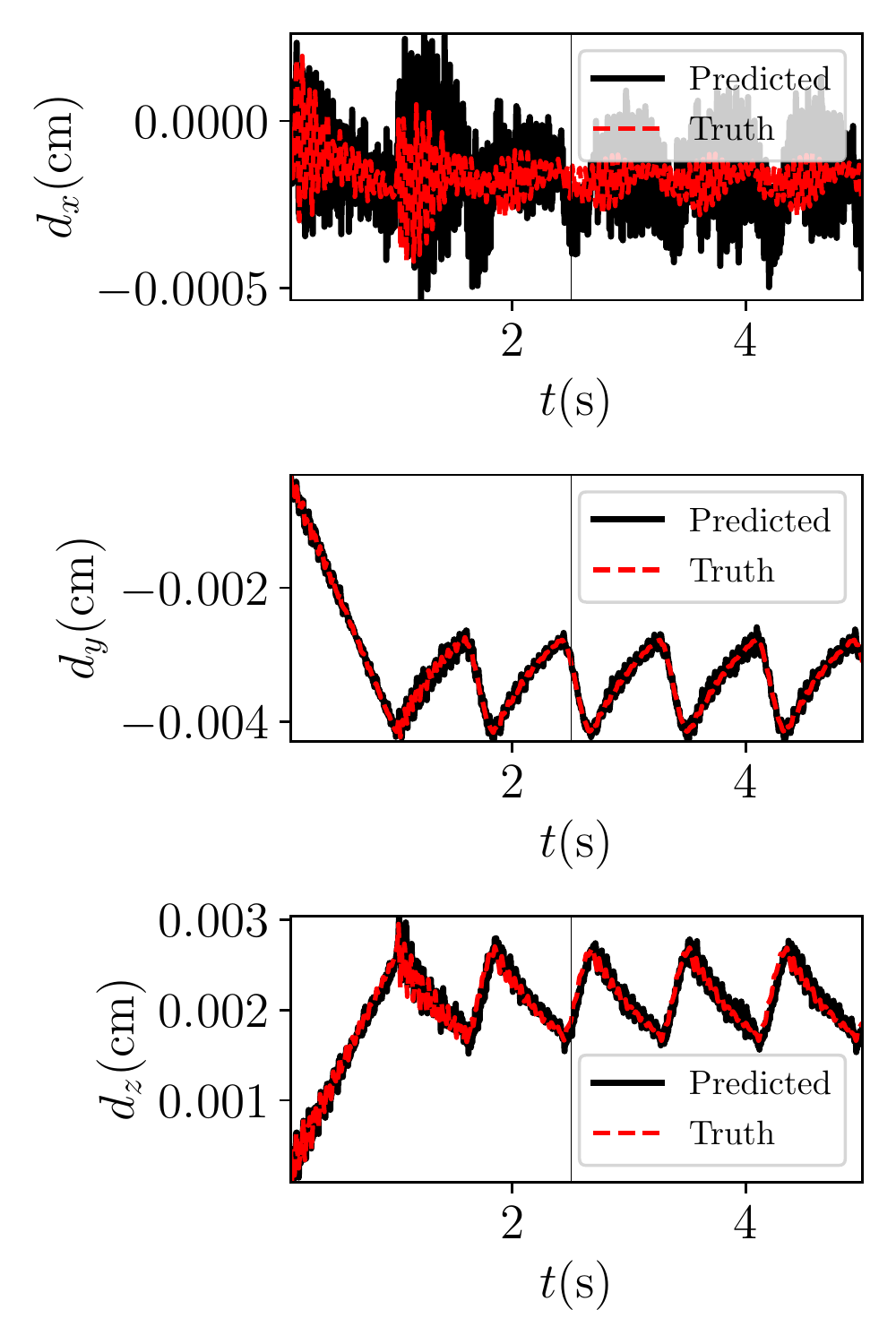}
        \caption{CPU:2, node: (-13.6,6.2,-9.7).}
    \end{subfigure}
        \begin{subfigure}[b]{0.325\textwidth}
        \centering
        \includegraphics[scale=.48]{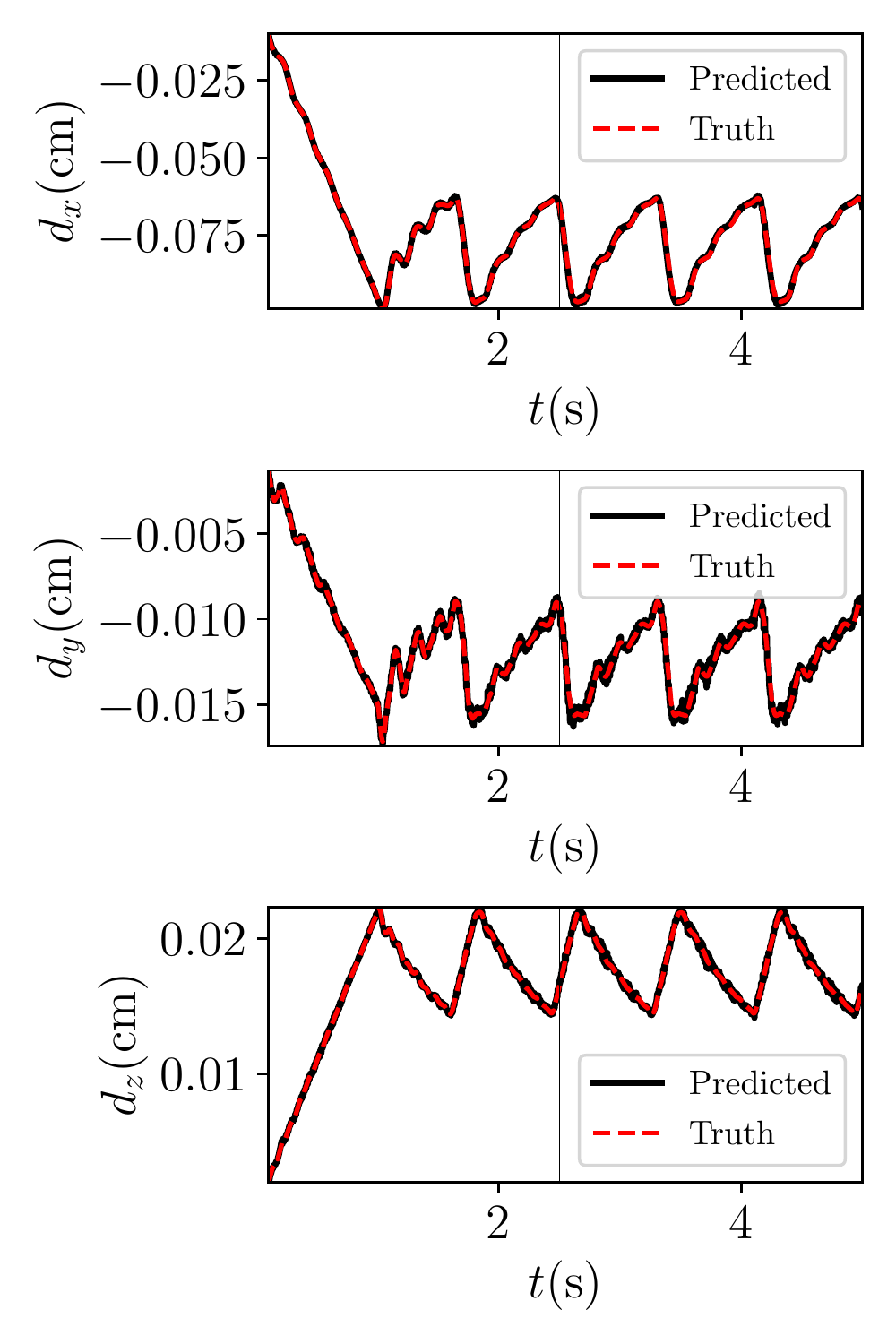}
        \caption{CPU:4, node:(-12.3,13.0,-9.1).}
    \end{subfigure}\\
    \begin{subfigure}[b]{0.325\textwidth}
        \centering
        \includegraphics[scale=.48]{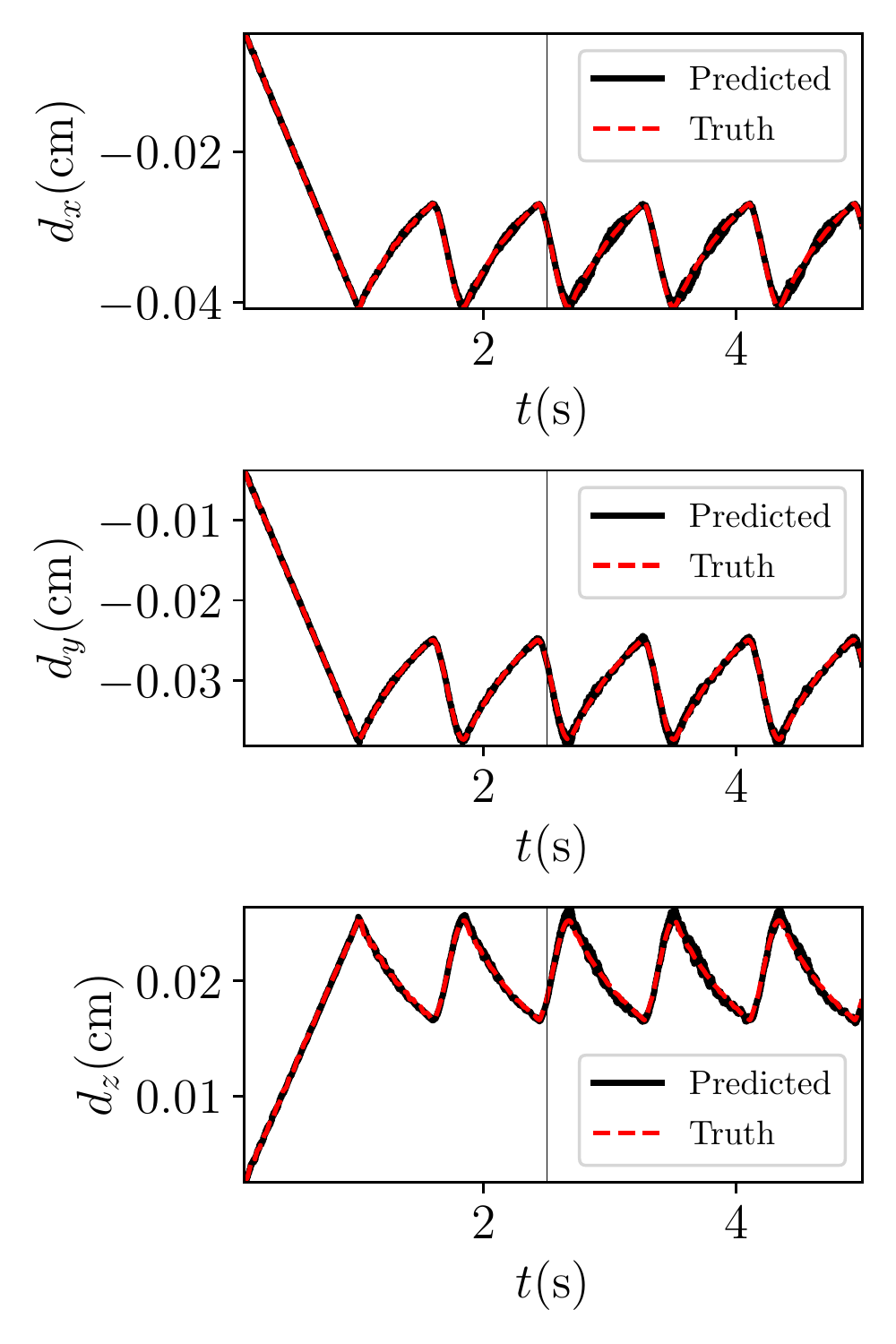}
        \caption{CPU:6, node:(-12.3,4.7,-7.2).}
    \end{subfigure}
    \begin{subfigure}[b]{0.325\textwidth}
        \centering
        \includegraphics[scale=.48]{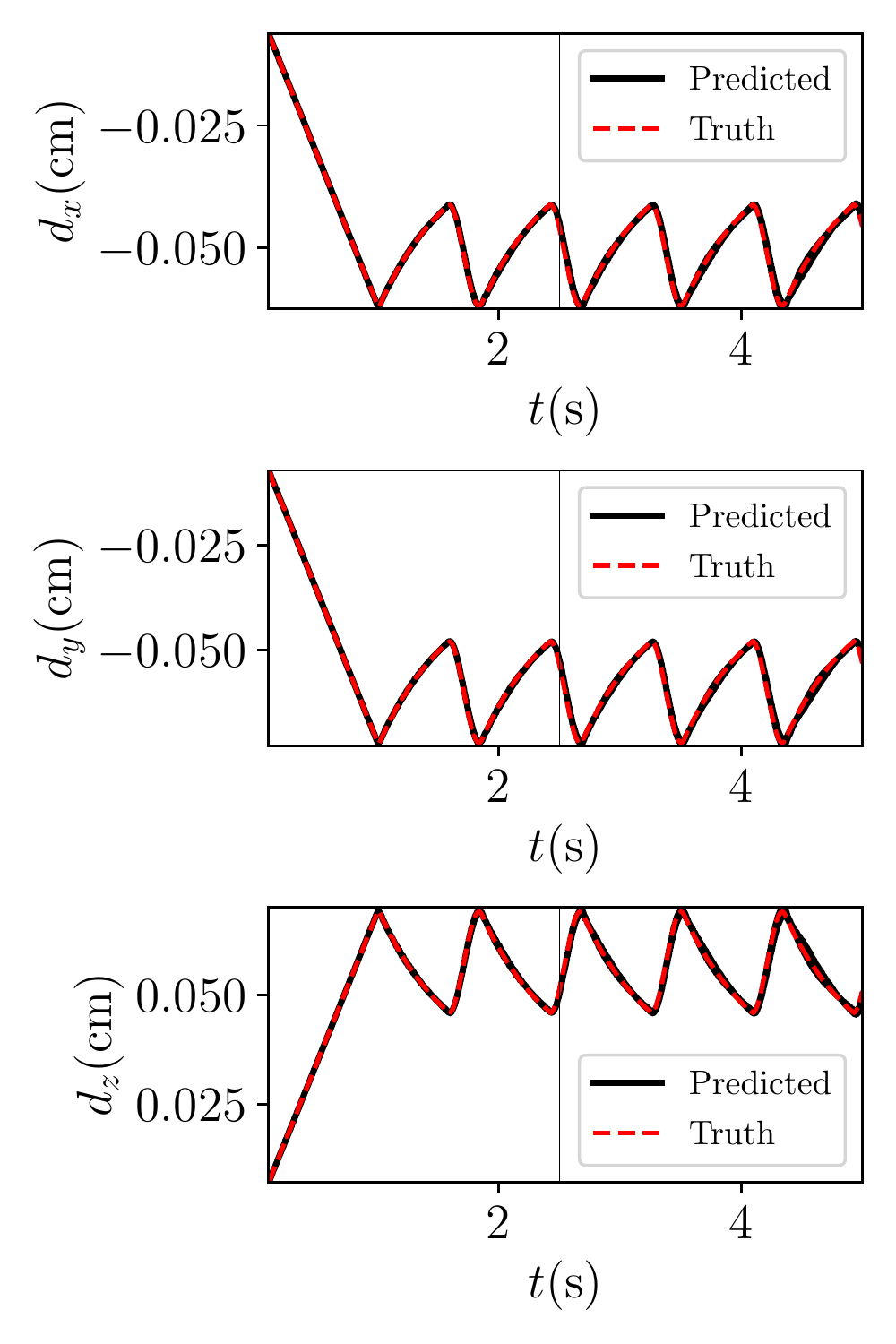}
        \caption{CPU:8, node:(-10.9,6.3,-4.8)}
    \end{subfigure}
    \begin{subfigure}[b]{0.325\textwidth}
        \centering
        \includegraphics[scale=.48]{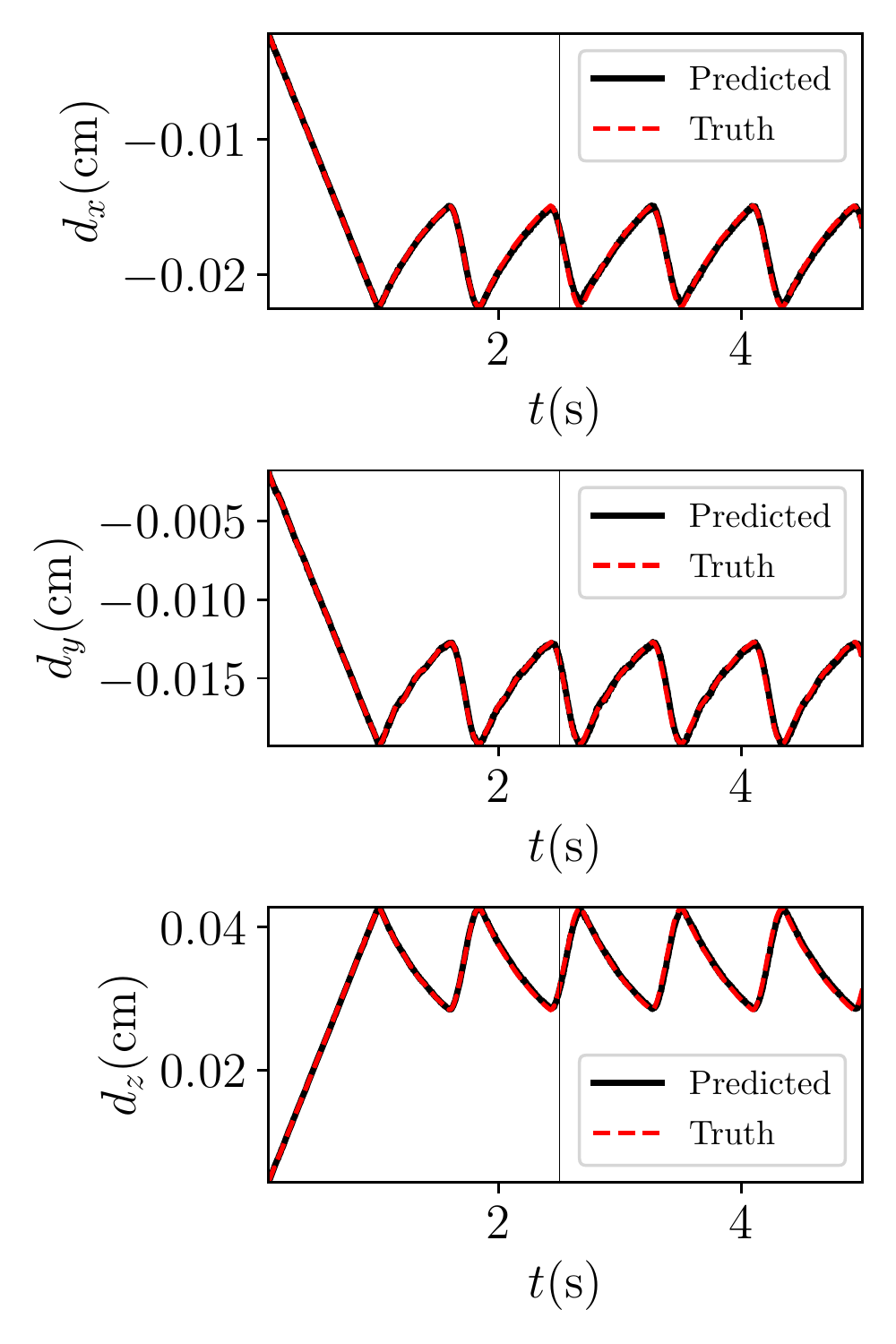}
        \caption{CPU:10, node:(-10.7,7.5,-3.6).}
    \end{subfigure}
    \caption{\centering Predicted dynamics of the human left coronary artery model. CPU labels are 0,2,4,6,8,10. Training is based on shared degrees of freedom of each partition and plotted nodes are not shared. Predicted steps: 980000.}\label{11cpus-fig-dis}
\end{figure*}

To test our methodology, we run Algorithm~\ref{alg-syn-avi} using the optimal hyperparamter combination as discussed in the previous sections, with an increased sample size $n_s = 1000$, and train 11 separate deep neural network surrogates.
Figure~\ref{11cpus-fig-dis} shows the predicted displacement dynamics at a non-shared node of a few partitions and confirms that the displacement evolution of the system is sufficiently learned even from the limited available data. Note that the discontinuity brought by the linear ramp condition is also correctly learned.
Larger relative errors can be observed for nodes with small displacement amplitudes but the dominant displacement components appear to be accurately modelled.

\section{Error control}\label{chp:ec}
%
In this section, we discuss metrics for error estimation that are naturally provided by the proposed approach.
As mentioned above, provided the discretization error from the Galerkin method is neglected, then the only source of error comes from the predicted dynamics at the shared nodes.
If we consider, for example, the coronary model in Figure~\ref{coronary}, there are exactly two, separately trained, deep neural networks associated with each shared node, that ideally should provide identical predictions.
However, in practice, these predictions may differ, providing a means for estimating the approximation error in Algorithm~\ref{alg-syn-avi}.

\begin{figure}[!ht]
    \centering
    \includegraphics[scale=0.37]{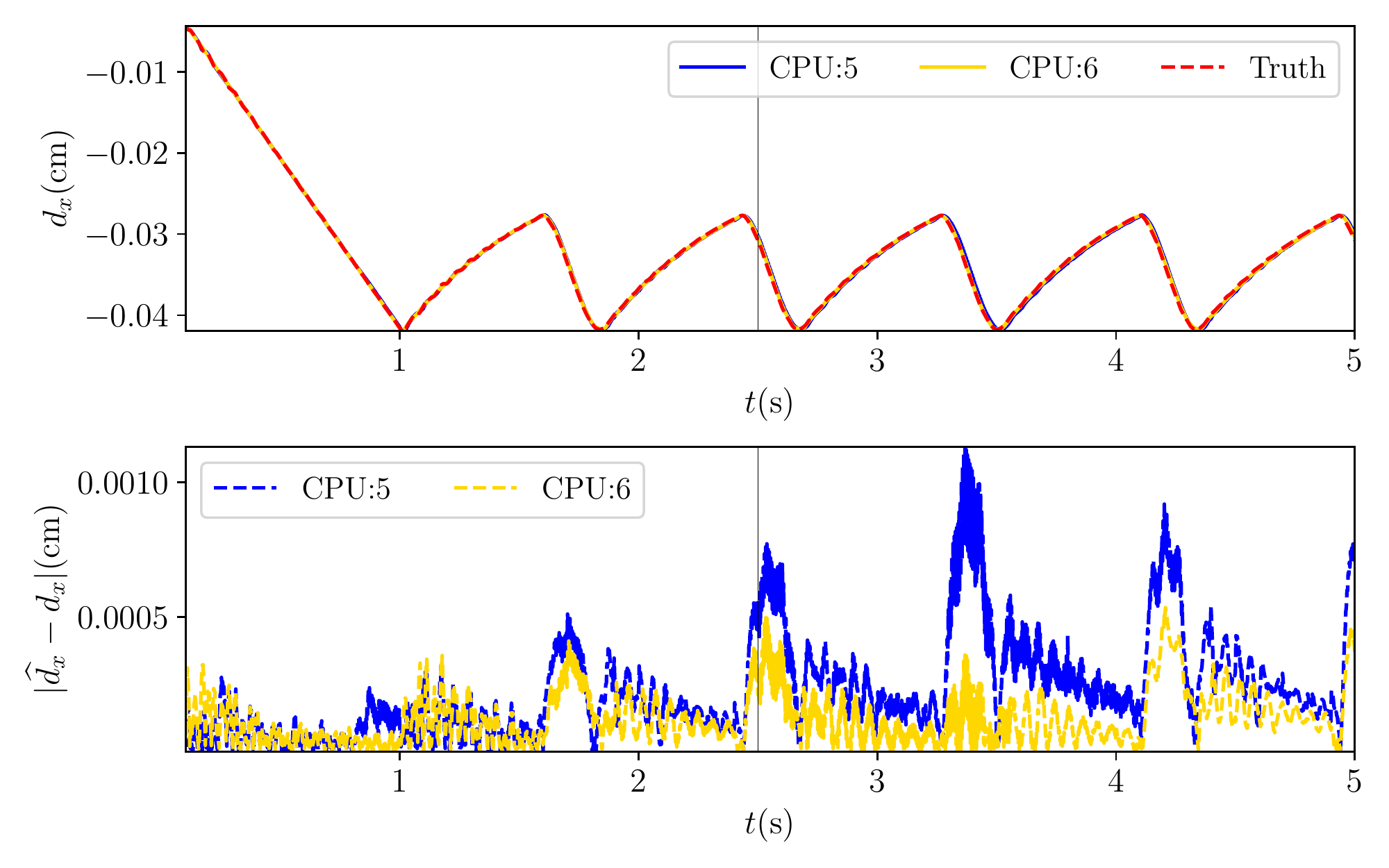}
    \caption{\centering Offline predicted $d_x$ displacement and time history of absolute error between two neighbor processors for coronary model. The plotted node (-12.8, 4.9, -7.5) is shared by CPU5 and CPU6.}
    \label{fig:error-single}
\end{figure}

In Figure~\ref{fig:error-single}, we illustrate the prediction variability for the same shared node produced by networks associated with two different partitions, following \emph{offline} evaluation, as discussed in Section~\ref{offon}.
It is shown that CPU6 consistently produces errors smaller than CPU5.

We then introduce the following criterion to quantify such variability \emph{globally}, i.e., over all the shared nodes
\begin{equation}
    \mathlarger{es}^{(t)}_{l^2}= \frac{1}{3\cdot N_a} \sum_{i=1}^{N_a}\|\widehat{\boldsymbol{d}}^{(t)}_{(i)}-\boldsymbol{d}^{(t)}_{(i)}\|_2.
    \label{eq:es}
\end{equation}
Therefore, $\mathlarger{es}^{(t)}_{l^2}$ represents a global average $l^2$ error per shared degree of freedom at time $t$. Moreover, since there are exactly two processors associated with each shared node, we denote their arithmetic average as $\bar{\mathlarger{es}}^{(t)}_{l^2}$. We also calculate the difference between two predictions for the same shared node by replacing the exact solution $\boldsymbol{d}^{(t)}_{(i)}$ in equation~\eqref{eq:es} with the solution predicted by the second network. The resulting quantity is denoted as: $\widehat{\mathlarger{es}}^{(t)}_{l^2}$.
\begin{figure}[!ht]
    \centering
    \includegraphics[scale=0.32]{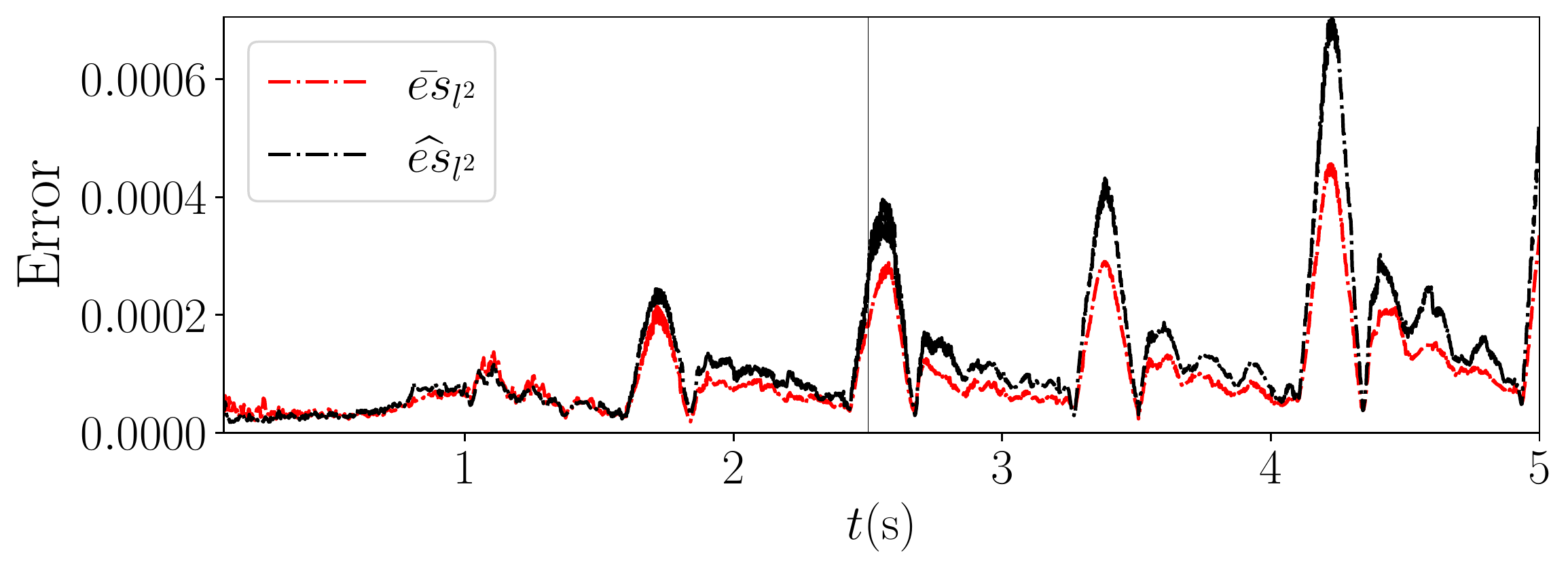}
    \caption{\centering Coronary model. Evolution of space-averaged $l^2$ error per shared degree of freedom.}
    \label{fig:error-global}
\end{figure}
As shown in both Figure~\ref{fig:error-single} and Figure~\ref{fig:error-global}, the error increases around regions of high curvature in the displacement response. 
It can also be observed that the discrepancy between two predictions at the same shared node is highly correlated with the displacement error and is greater in most cases. Moreover, it is easy to compute and therefore particularly appropriate for error monitoring.

\begin{figure}[!ht]
    \centering
    \includegraphics[scale=0.34]{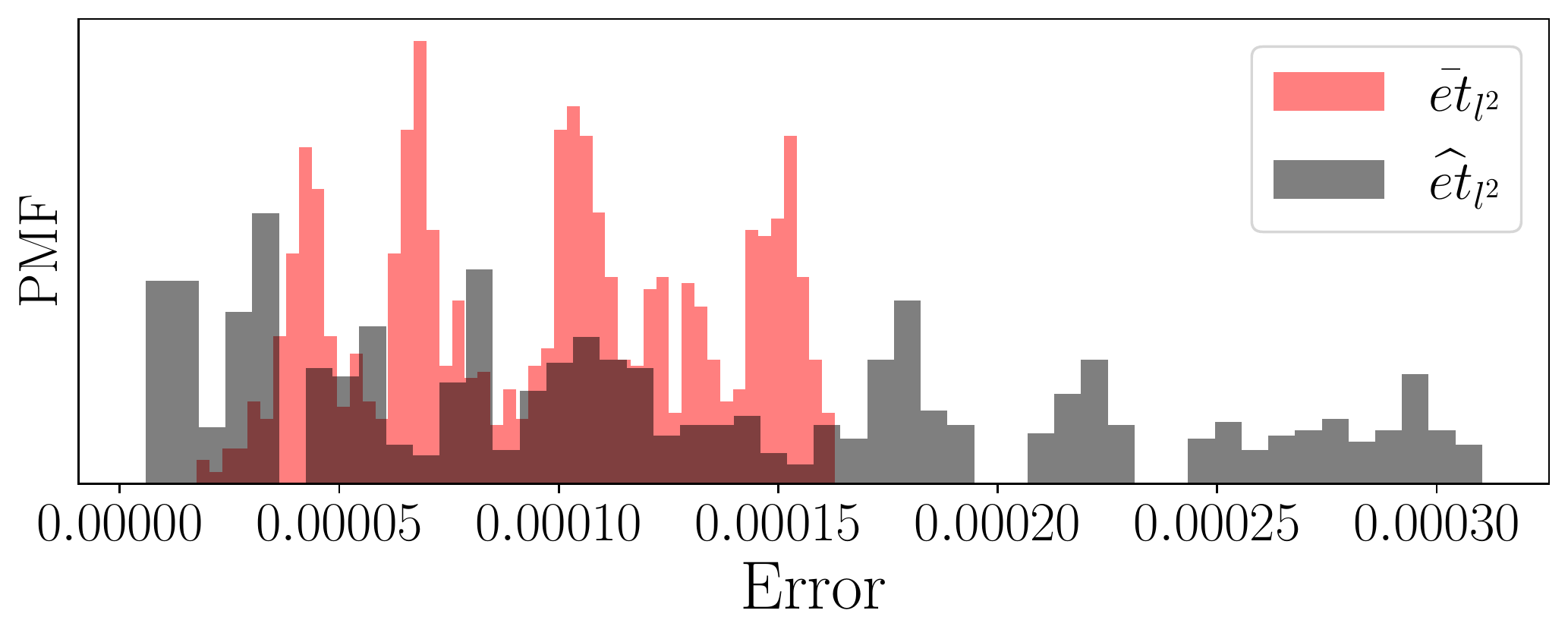}
    \caption{\centering Coronary model. PMF of time-averaged $l^2$ error per shared degree of freedom.}
    \label{fig:error-histo}
\end{figure}

We finally provide histograms in Figure~\ref{fig:error-histo} to show the temporal average of such variability. We further define
\begin{equation}
    \mathlarger{et}_{(j),l^2}= \frac{1}{3\cdot n_T} \sum_{i=1}^{n_T}\|\widehat{\boldsymbol{d}}_{(j)}^{(i)}-\boldsymbol{d}_{(j)}^{(i)}\|_2,
\end{equation}
to quantify this error at shared node $j$, where $n_T$ is the total number of predicted time steps. Again, we put a bar on top of $\mathlarger{et}_{(j),l^2}$ to indicate an average of two generic processors that share node $j$, and denote $\widehat{\mathlarger{et}}_{(j),l^2}$ for the difference in two predictions, at the same shared node $j$.

From Figure~\ref{fig:error-histo} and consistent with our discussion above, the time-averaged model differences provide an upper bound for the approximation error at the shared nodes.
\section{Performance analysis}\label{chp:perf}
%
As presented in Algorithm~\ref{alg-para}, at each time iteration of the proposed distributed solver, the main computational tasks are (1) the evaluation of element-level quantities $\mathbf{K}_e$, $\boldsymbol{f}_e^{(n),\text{ext}}$, (2) matrix-vector product for displacement update and (3) displacement synchronization at the shared nodes.
Application of the proposed synchronization-avoiding Algorithm~\ref{alg-syn-avi} allows one to avoid most of the synchronization cost by leveraging pre-trained data-driven surrogates.
To better quantify the cost of the above operations, we introduce several quantities in Table~\ref{table: perf-ts}.

The superscript $\bar{(\cdot)}$ in Table~\ref{table: perf-ts} means \emph{max-average}, i.e., we first run the structural solver for a number of steps to get the \emph{average} costs per step.
Then, we locate the processor with the \emph{maximum} total cost $\bar{t}_t$ and analyze its $\bar{t}_s$, $\bar{r}_s$, etc.
This approach allows one to account for the lack of load balance between partitions. The speedup factor $\bar{\zeta}$ is simply the ratio between $\bar{t}_t$ and $\bar{t}_t^{\prime}$, where $(\cdot)^{\prime}$ denotes whether the data-driven surrogate is used to reduce the synchronization costs.

We are interested in two scenarios. The first considers a fixed number of cores and several models of increasing size. The second considers a fixed-size mesh partitioned over an increasing number of cores. We further analyze these cases with or without pre-assembling element-level quantities (see Remark~\ref{rmk:pre}).
All the tests below are carried out using the cantilever model with different levels of refinement.
\begin{table}[h!]
{\footnotesize
\begin{center}
\begin{tabular}{@{} l l @{}}
\toprule
Total time& $\bar{t}_t$ (s)\\ 
Time for element quantity evaluation & $\bar{t}_e$ (s)\\
Time for synchronization & $\bar{t}_s$ (s)\\ 
Time for matrix-vector product & $\bar{t}_m$ (s)\\ 
Time for applying data-driven model & $\bar{t}_d$ (s)\\
Ratio of element quantity formation cost & $\bar{r}_e (\%)$
\\
Ratio of synchronization cost & $\bar{r}_s (\%)$
\\
Ratio of matrix-vector product cost & $\bar{r}_m (\%)$
\\
Ratio of data-driven model cost & $\bar{r}_d (\%)$
\\
Number of shared nodes & $N_a$ \\
Number of processor used           & $n_c$  \\
Speedup factor & $\bar{\zeta}$
\\
\bottomrule
\end{tabular}
\end{center}}
\caption{Definition of quantities for performance analysis.}
\label{table: perf-ts}
\end{table}

\subsection{Test with pre-assembly of element-level quantities}
For linear structural problems in the small strain regime, the matrices $\mathbf{M}, \mathbf{K}$ in problem~\eqref{lumped} are constants in time and therefore can be assembled only once before the time loop, stored and re-used.
In such a case, the cost of displacement synchronization dominates over the relatively inexpensive matrix-vector product, making the proposed approach particularly appealing.

\vspace{3pt}

\begin{figure}[!ht]
    \centering
    \includegraphics[scale=0.6]{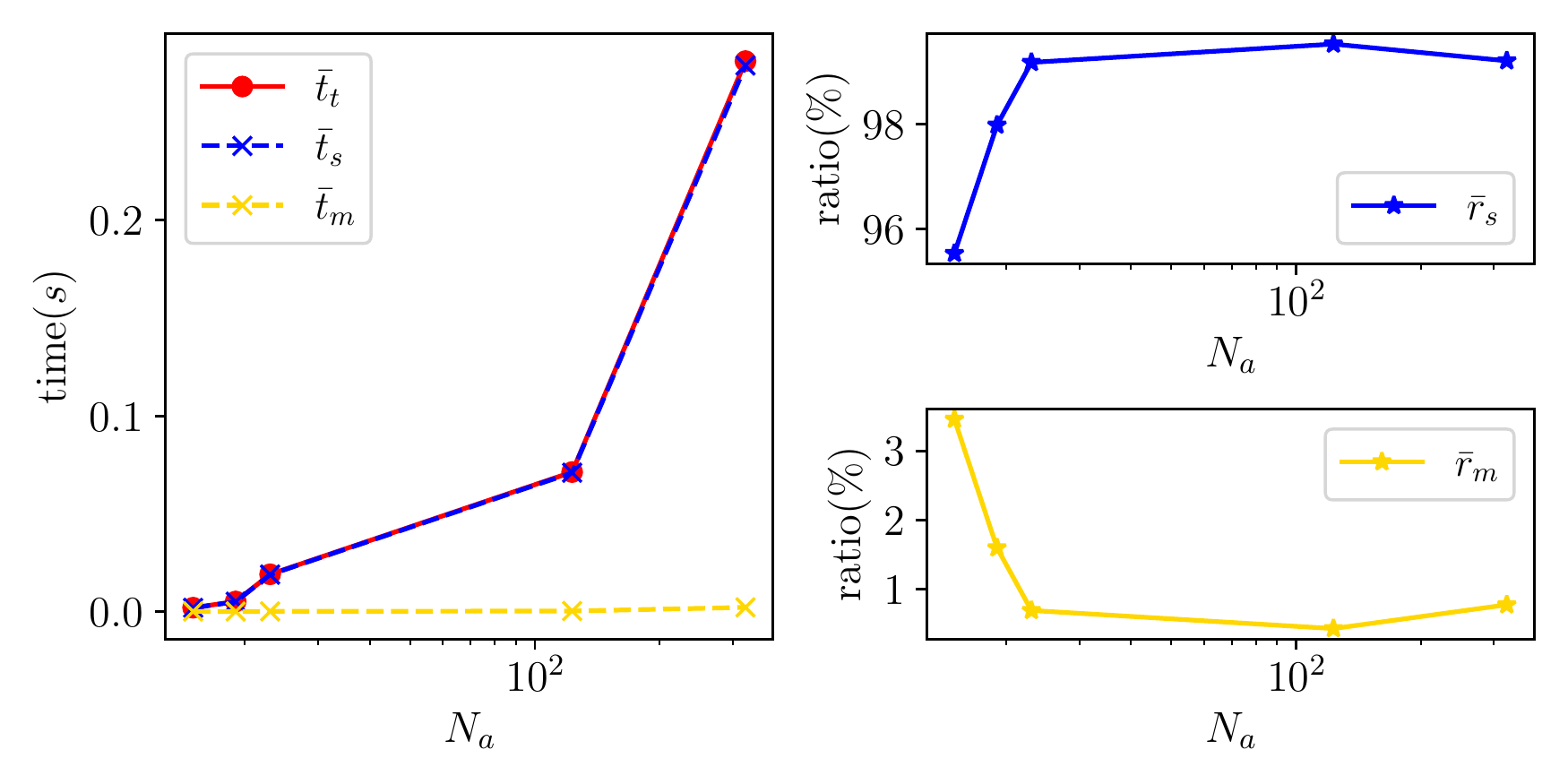}
    \caption{\centering Timing statistics with pre-assembly and without data-driven model. The number of partitioning is fixed at $n_c=8$. Left: average cost per step  vs. number of shared nodes. Right: average ratio per step vs. number of shared nodes.}
    \label{fig:w-pre-8cpu-syn}
\end{figure}

\begin{figure}[!ht]
    \centering
    \includegraphics[scale=0.6]{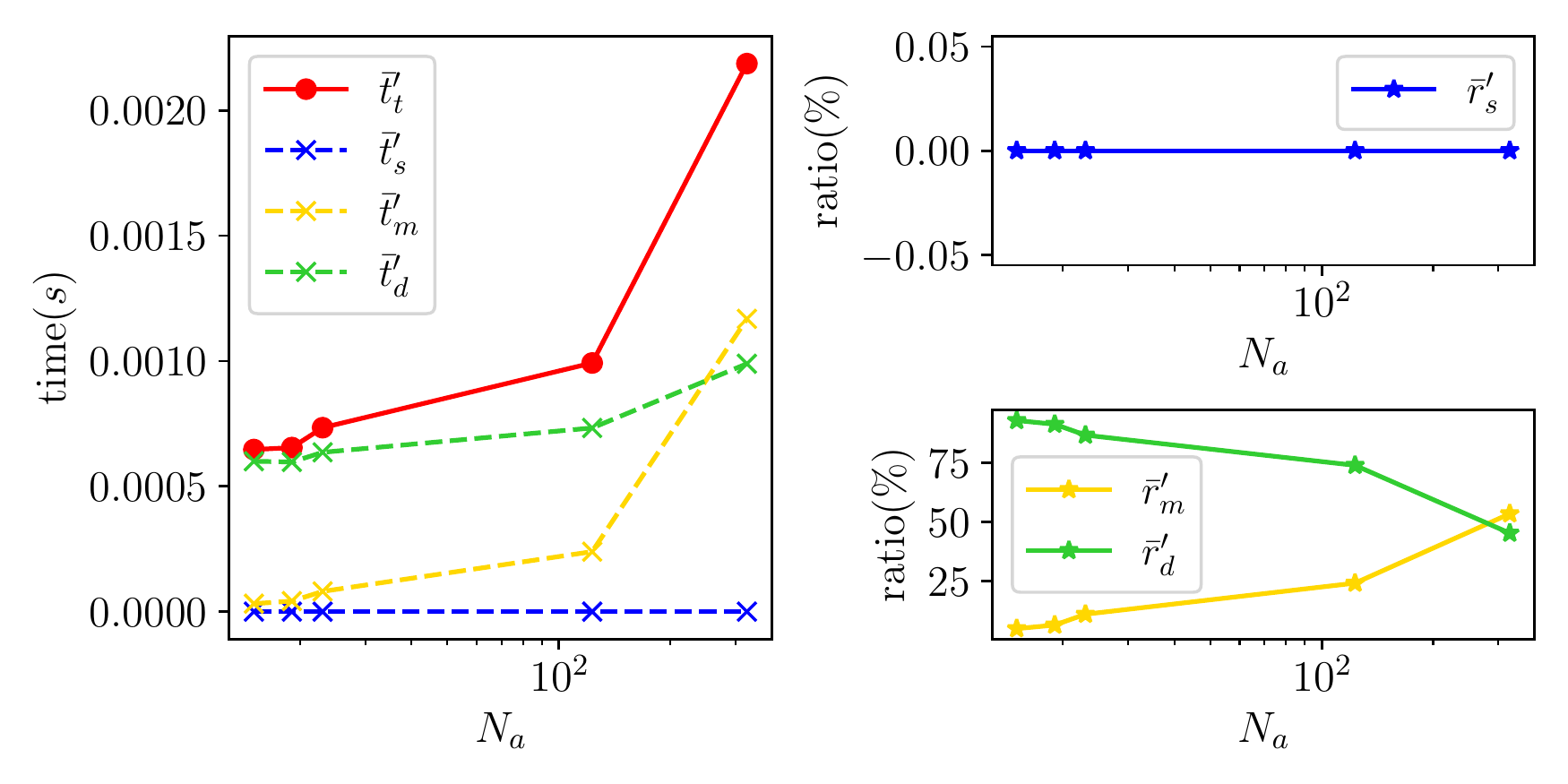}
    \caption{\centering Timing statistics with pre-assembly and with data-driven model. The number of partitioning is fixed at $n_c=8$. Left: average cost per step vs. number of shared nodes. Right: average ratio per step vs. number of shared nodes.}
    \label{fig:w-pre-8cpu-dnn}
\end{figure}

\begin{figure}[!ht]
    \centering
    \includegraphics[scale=0.6]{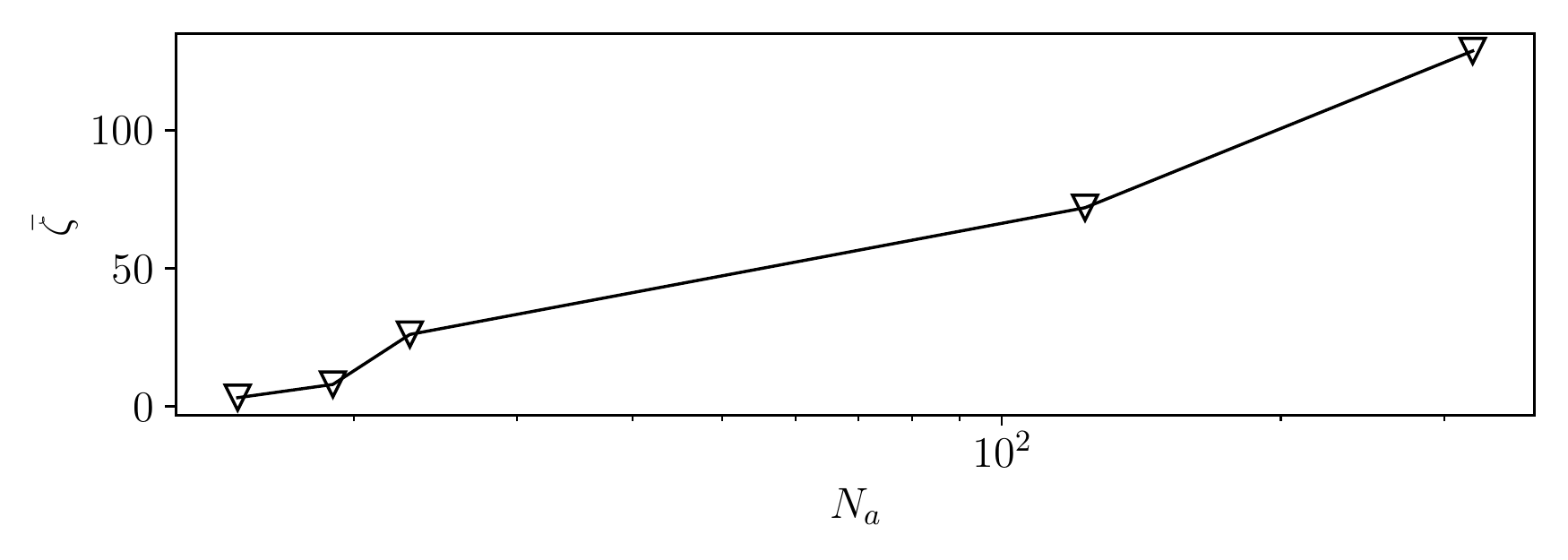}
    \caption{\centering Data driven model speed-up with pre-assembly for $n_c=8$.}
    \label{fig:w-pre-8cpu-sp}
\end{figure}

\noindent{\bf Fixed number of partitions} - We consider a series of pre-assembled explicit structural simulations with increasing mesh size, solved by an 8-core machine. 
Since the number of shared nodes for each partition increases with the mesh size, the amount of communication also increases. 
This is clearly shown for all cases in Figure~\ref{fig:w-pre-8cpu-syn}, as $\bar{t}_s$ grows rapidly with respect to $N_a$ and takes more than 95\% of the total cost.
In such a case our data-driven synchronization algorithm is particularly effective, resulting in a significant speed-up as shown in Figures~\ref{fig:w-pre-8cpu-dnn} and~\ref{fig:w-pre-8cpu-sp}, since $\bar{t}'_d$ is orders of magnitude less expensive than $\bar{t}_s$.

Generally speaking, as shown in Figure~\ref{NN-ED}, by fixing the network depth (number of encoder layers), the sequence length $n_p$/$n_f$, and the number of hidden units $n_H$, the model cost $\bar{t}_d^{\prime}$ only depends on the size of input $\boldsymbol{d}$, which essentially scales with $N_a$. 
This is evident in Figure~\ref{fig:w-pre-8cpu-dnn}, but we can also observe how the cost of the matrix-vector product gradually exceeds the network evaluation cost.

\begin{figure}[!ht]
    \centering
    \includegraphics[scale=0.6]{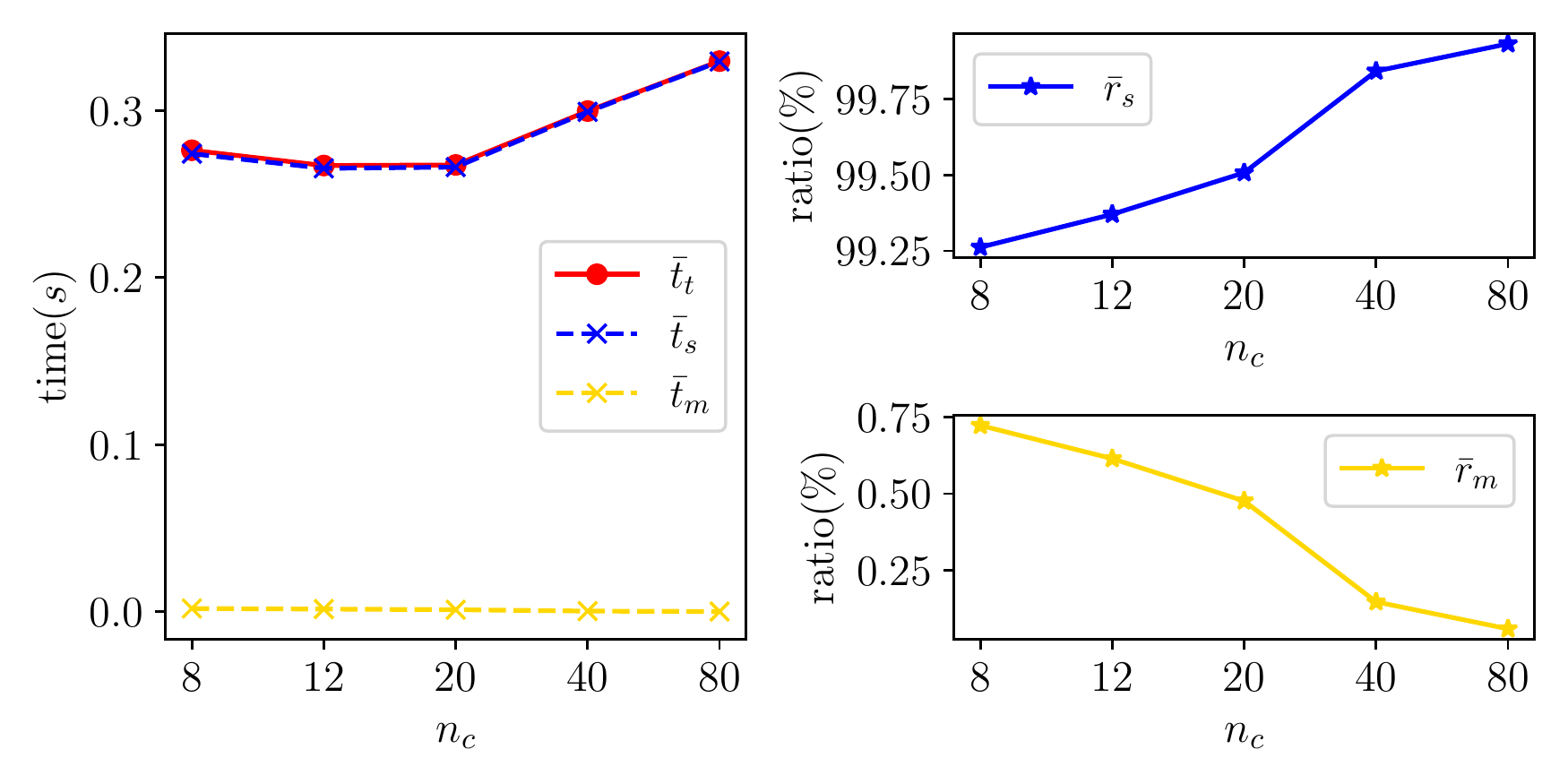}
    \caption{\centering Timing statistics of the test with pre-assembly and without data-driven model. The mesh resolution is fixed. Left: average cost per step  vs. number of partitioning. Right: average ratio per step vs. number of partitioning.}
    \label{fig:w-pre-mesh-syn}
\end{figure}

\noindent{\bf Fixed mesh size} - In this section, we fix the mesh size and distribute it over an increasing number of processors. 
In Figure~\ref{fig:w-pre-mesh-syn}, the synchronization cost remains approximately stable for up to 20 processors but increases sensibly with 40 and 80 processors. This relates to the different cost of communication in computational environments with shared memory rather than distributed memory. In other words, we need 2 and 4 24-core machines to realize the final two cases characterized by 40 and 80 processors, respectively.

\begin{figure}[!ht]
    \centering
    \includegraphics[scale=0.6]{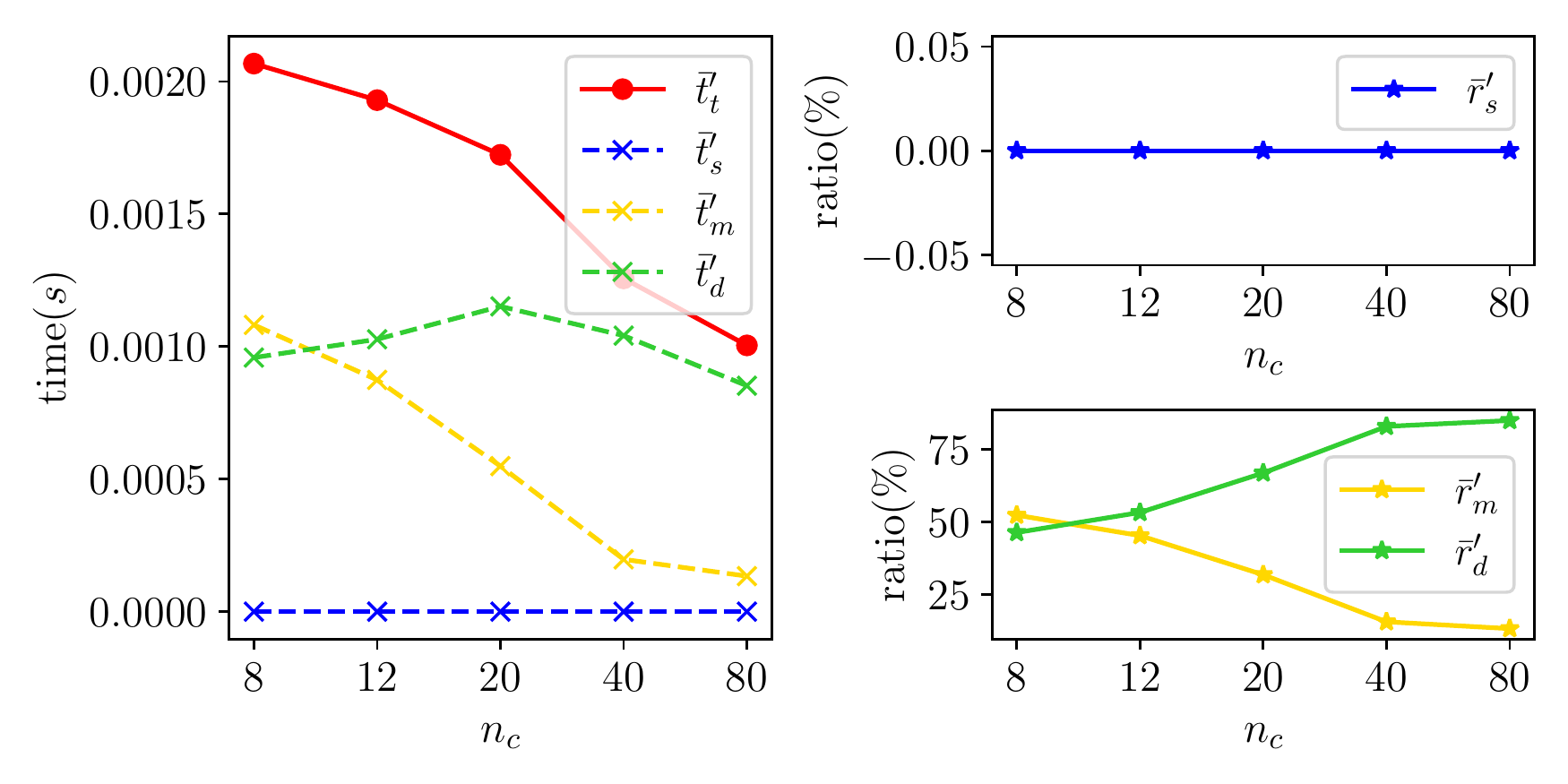}
    \caption{\centering Timing statistics of the test with pre-assembly and with data-driven model. The mesh resolution is fixed. Left: average cost per step  vs. number of partitioning. Right: average ratio per step vs. number of partitioning.}
    \label{fig:w-pre-mesh-dnn}
\end{figure}

\begin{figure}[!ht]
    \centering
    \includegraphics[scale=0.6]{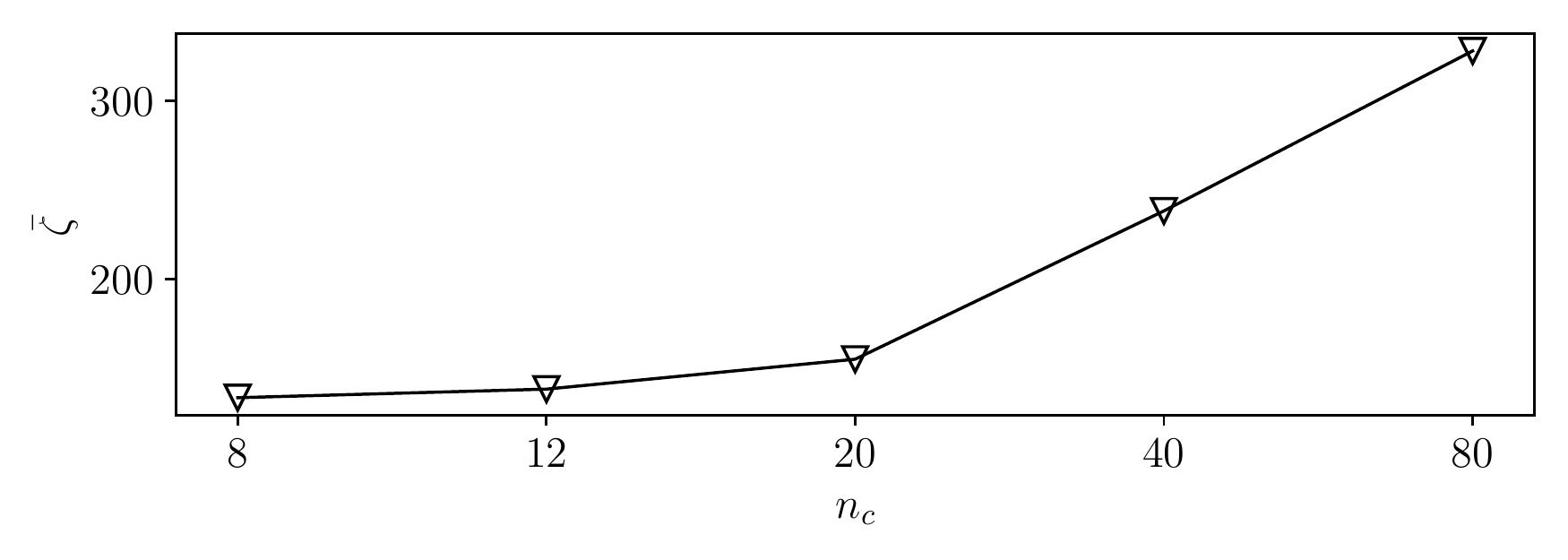}
    \caption{\centering Data driven model speed-up with pre-assembly for an increasing number of processors.}
    \label{fig:w-pre-mesh-sp}
\end{figure}

A significant reduction in the communication costs is also achieved for this scenario, as shown in Figure~\ref{fig:w-pre-mesh-dnn} and Figure~\ref{fig:w-pre-mesh-sp}. 
It is not surprising that the model cost $\bar{t}_d^{\prime}$ stops to increase as more partitions are considered, since the number of shared nodes saturates for the selected coarse mesh. 
Also, promising speed-up factors $\bar{\zeta}$ are shown in Figure~\ref{fig:w-pre-mesh-sp}.

\subsection{Test without pre-assembly of element-level quantities}
%
For general nonlinear structural simulations, the element-level matrices and vectors need to be re-computed at each iteration in time, with a cost that dominates over the remaining components. Here, for simplicity, we mimic this scenario via Algorithm~\ref{alg-para} but still solving a linear problem.

\vspace{3pt}

\begin{figure}[!ht]
    \centering
    \includegraphics[scale=0.6]{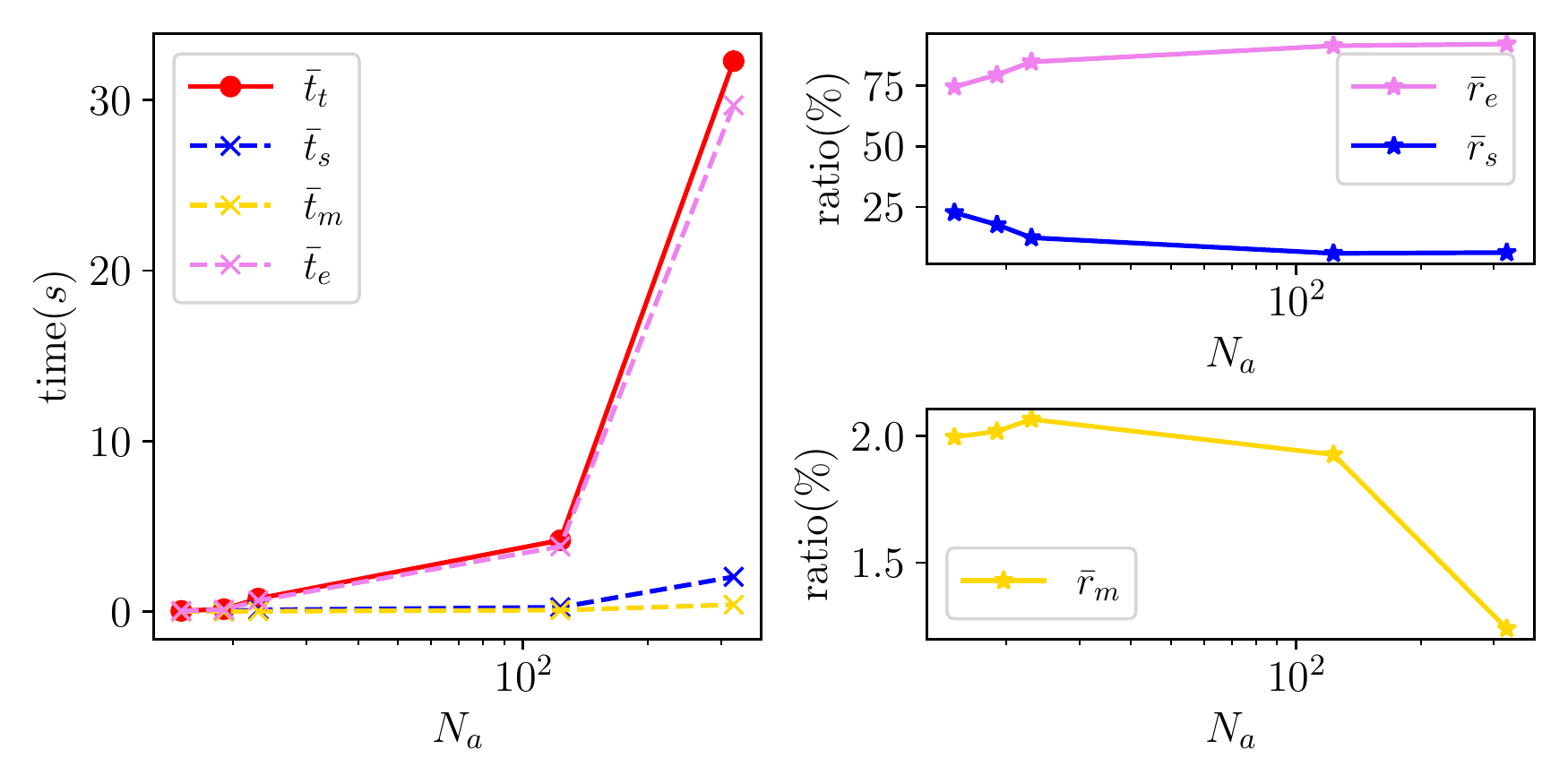}
    \caption{\centering Timing statistics without pre-assembly and without data-driven model. The number of partitioning is fixed at $n_c=8$. Left: average cost per step  vs. number of shared nodes. Right: average ratio per step vs. number of shared nodes.}
    \label{fig:wo-pre-8cpu-syn}
\end{figure}

\begin{figure}[!ht]
    \centering
    \includegraphics[scale=0.6]{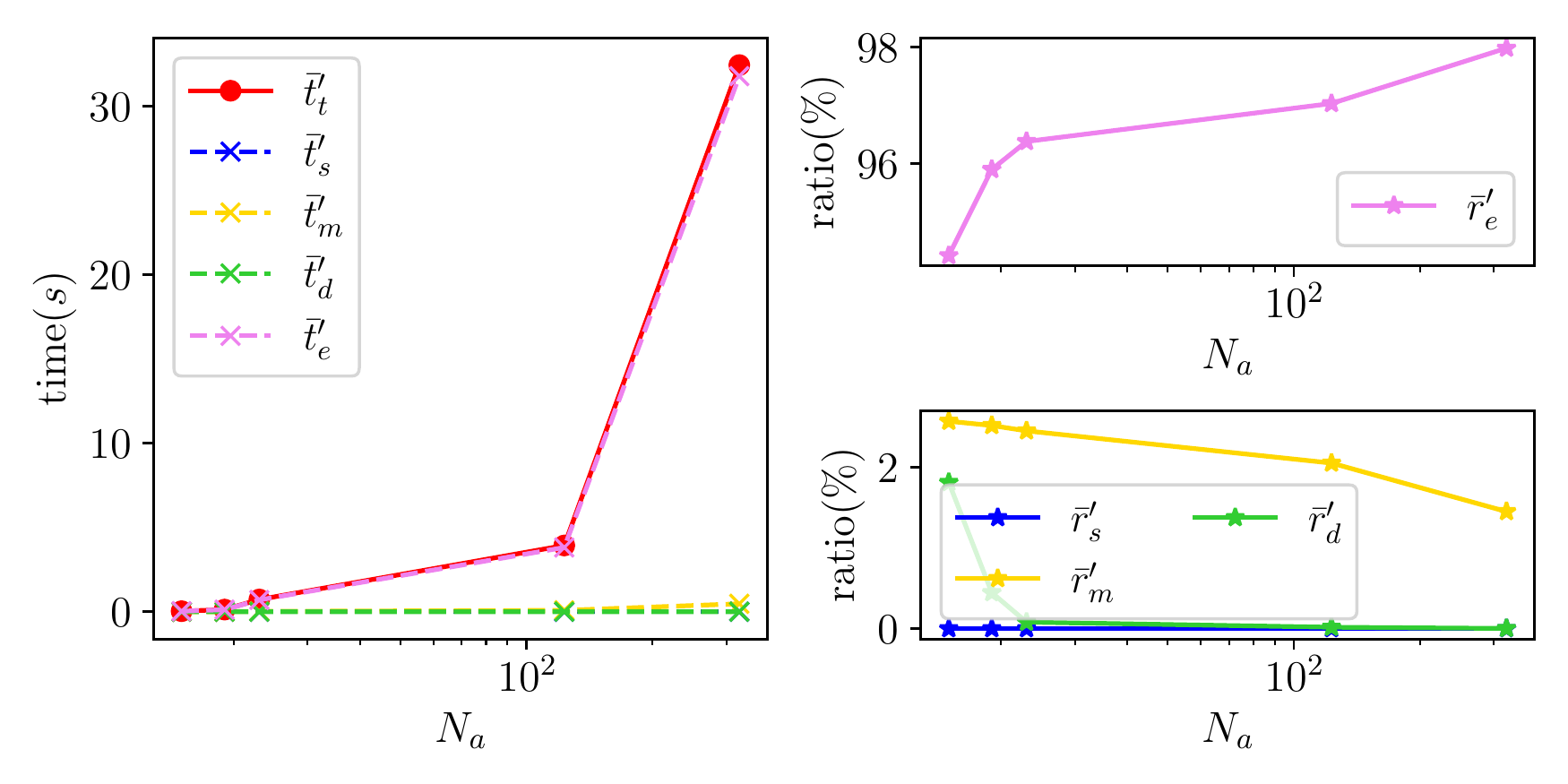}
    \caption{\centering Timing statistics without pre-assembly and with data-driven model. The number of partitioning is fixed at $n_c=8$. Left: average cost per step vs. number of shared nodes. Right: average ratio per step vs. number of shared nodes.}
    \label{fig:wo-pre-8cpu-dnn}
\end{figure}

\begin{figure}[!ht]
    \centering
    \includegraphics[scale=0.6]{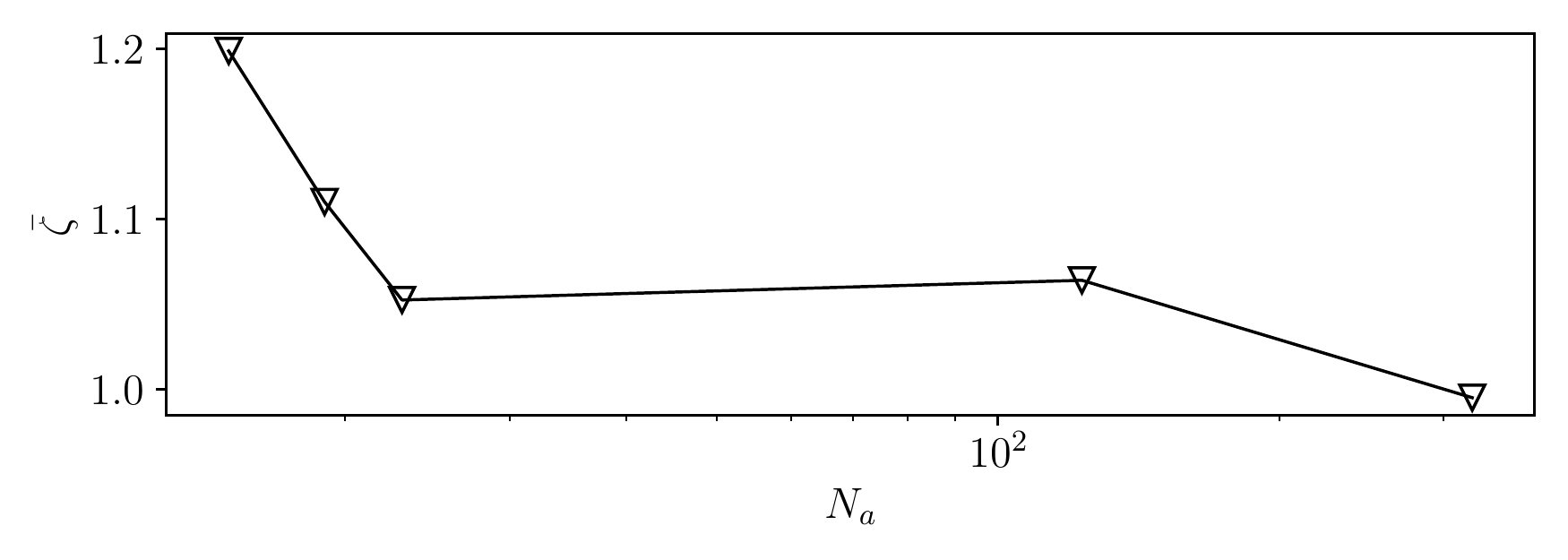}
    \caption{\centering Data driven model speed-up without pre-assembly for $n_c=8$.}
    \label{fig:wo-pre-8cpu-sp}
\end{figure}

\noindent{\bf Fixed number of partitions} - Without the pre-assembly, from Figure~\ref{fig:wo-pre-8cpu-syn} and Figure~\ref{fig:wo-pre-8cpu-dnn}, the majority of cost is occupied by the evaluation of element quantities. 
Again, due to load imbalance and the presence of a barriers in the code preceding inter-processor communication, the synchronization time also increases. 
Finally, for an increasing mesh size, the relative importance of the synchronization cost $\bar{r}'_{s}$ is reduced while $\bar{r}'_{e}$ increases, clearly reducing the performance of our surrogate model (e.g. see Figure~\ref{fig:wo-pre-8cpu-sp}).

\vspace{3pt}

\begin{figure}[!ht]
    \centering
    \includegraphics[scale=0.6]{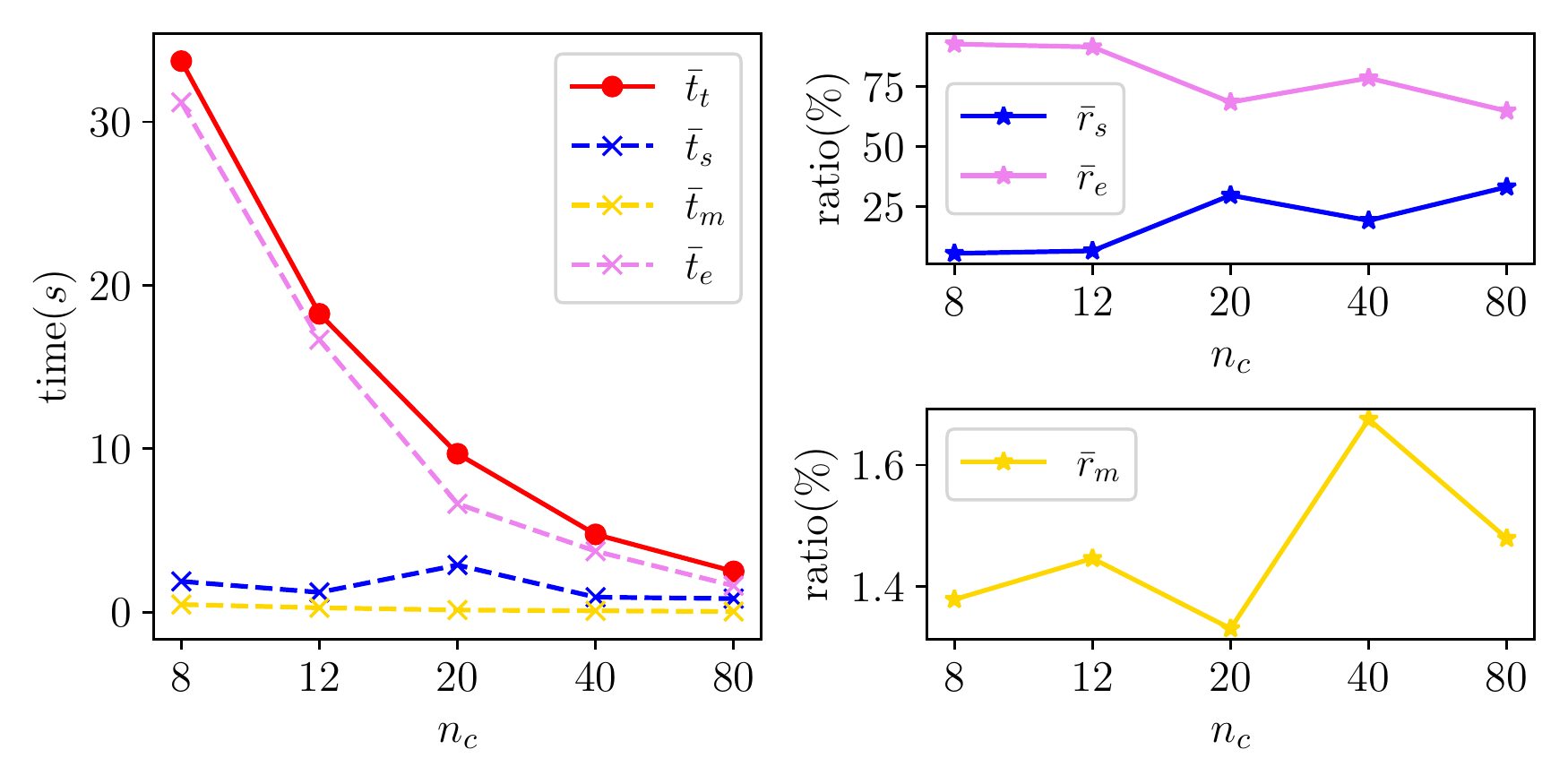}
    \caption{\centering Timing statistics of the test without pre-assembly and without data-driven model. The mesh resolution is fixed. Left: average cost per step  vs. number of partitioning. Right: average ratio per step vs. number of partitioning.}
    \label{fig:wo-pre-mesh-syn}
\end{figure}
\begin{figure}[!ht]
    \centering
    \includegraphics[scale=0.6]{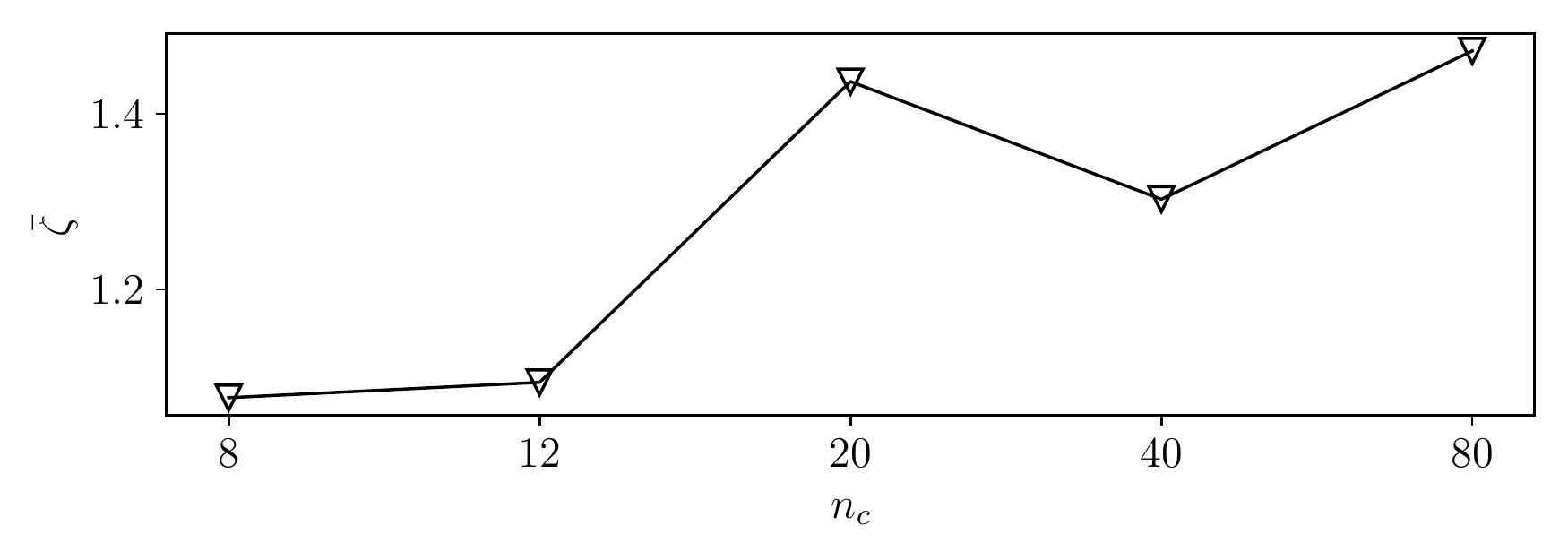}
    \caption{\centering Data driven model speed-up without pre-assembly for an increasing number of processors.}
    \label{fig:wo-pre-mesh-sp}
\end{figure}

\noindent{\bf Fixed mesh size} - For an increasing number of processors, load imbalance is more evident from $\bar{r}_s$ and $\bar{r}_e$ in Figure~\ref{fig:wo-pre-mesh-syn}, where a more efficient evaluation for the element quantities is accompanied by an increasing synchronization cost. 
The speed-up in this case, shown in Figure~\ref{fig:wo-pre-mesh-sp}, is roughly monotonic with the increasing number of partitions.
\begin{figure}[!ht]
    \centering
    \includegraphics[scale=0.6]{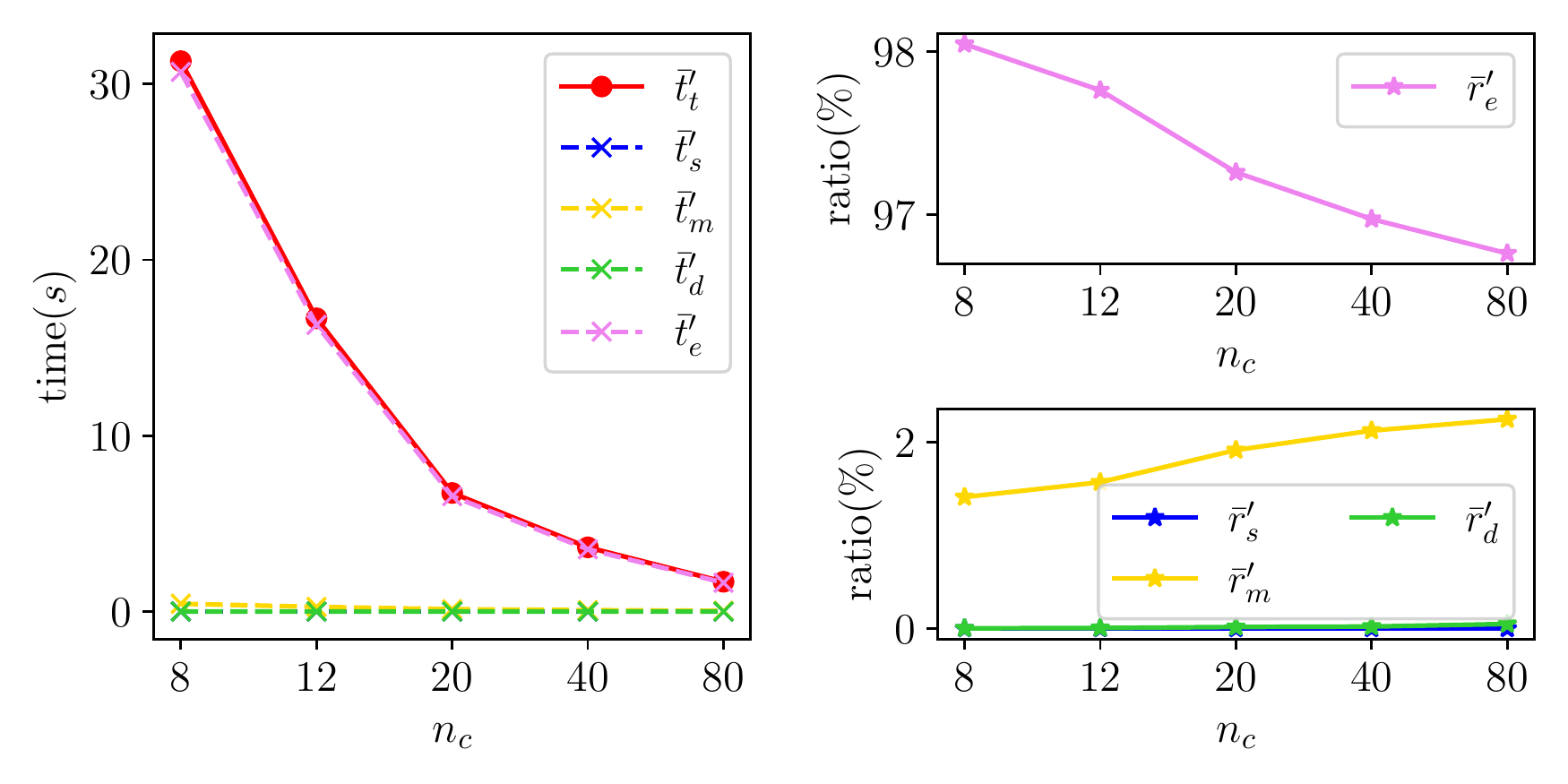}
    \caption{\centering Timing statistics of the test without pre-assembly and with data-driven model. The mesh resolution is fixed. Left: average cost per step  vs. number of partitioning. Right: average ratio per step vs. number of partitioning.}
    \label{fig:wo-pre-mesh-dnn}
\end{figure}
\section{Conclusions and future work}\label{chp:conclusion}

In this paper, we have developed a novel data-driven approach to speed up the structural analysis of soft biological tissue. 
Our approach minimizes the amount inter-processor communication by replacing shared node synchronization tasks with predictions from an optimally trained artificial neural network.
As verified through extensive numerical experiments, the LSTM encoder-decoder network model proposed in this study accurately approximates the displacement values at the shared nodes and maintains stability and accuracy even for a long time horizon. 

The proposed network significantly reduces synchronization times in large scale simulations. This is demonstrated in Section~\ref{cardio} using a realistic coronary artery model used in previous studies for fluid-structure interaction problems. In addition, when modified to include \emph{conditional} predictions and trained accordingly, our model is robust to the choice of initial conditions and external loading. 

Reduction of synchronization times is crucial for \emph{ensemble} distributed finite element solvers where multiple realizations of geometry, boundary conditions or material properties are solved simultaneously. An explicit-in-time multi-GPU implementation of an ensemble solver has been proposed in our previous work~\cite{li2021ensemble}, where GPU-to-CPU synchronization represents a challenging computational bottleneck. Therefore, the proposed approach represents an interesting acceleration, particularly for the periodic response of cardiovascular models.

In addition, the well known poor bending performance of constant strain tetrahedral elements may significantly degrade the reliability of patient-specific cardiovascular structural models. To overcome this problem, we will explore recently proposed stabilized and variational multiscale finite element formulations~\cite{Scovazzi2016}.

\section*{Acknowledgments}

This work was supported by a NSF CAREER award \#1942662 (PI DES), a NSF CDS\&E award \#2104831 (PI DES) and used computational resources provided through the Center for Research Computing at the University of Notre Dame. 
The authors would like to thank Prof. Zhiliang Xu and Prof. Guosheng Fu for their comments and suggestions that contributed to improve the quality of the present manuscript. 


\bibliographystyle{plain}
\bibliography{refs}

\begin{thebibliography}{10}

\bibitem{aslam2020performance}
M.~Aslam, O.~Riaz, S.~Mumtaz, and A.D. Asif.
\newblock Performance comparison of {GPU}-based jacobi solvers using {CUDA}
  provided synchronization methods.
\newblock {\em IEEE Access}, 8:31792--31812, 2020.

\bibitem{bakarji2022discovering}
J.~Bakarji, K.~Champion, J.N. Kutz, and S.L. Brunton.
\newblock Discovering governing equations from partial measurements with deep
  delay autoencoders.
\newblock {\em arXiv preprint arXiv:2201.05136}, 2022.

\bibitem{bartezzaghi2015explicit}
A~Bartezzaghi, M.~Cremonesi, N.~Parolini, and U.~Perego.
\newblock An explicit dynamics {GPU} structural solver for thin shell finite
  elements.
\newblock {\em Computers \& Structures}, 154:29--40, 2015.

\bibitem{belytschko2014nonlinear}
T.~Belytschko, W.K. Liu, B.~Moran, and K.~Elkhodary.
\newblock {\em Nonlinear finite elements for continua and structures}.
\newblock John wiley \& sons, Hoboken, NJ, 2014.

\bibitem{PhysRevE.105.044205}
U.~Bhat and S.B. Munch.
\newblock Recurrent neural networks for partially observed dynamical systems.
\newblock {\em Phys. Rev. E}, 105:044205, Apr 2022.

\bibitem{sindy}
S.L. Brunton, J.L. Proctor, and J.N. Kutz.
\newblock Discovering governing equations from data by sparse identification of
  nonlinear dynamical systems.
\newblock {\em Proceedings of the National Academy of Sciences},
  113(15):3932--3937, 2016.

\bibitem{RNN-anti}
B.~Chang, M.M. Chen, E.~Haber, and E.H. Chi.
\newblock Antisymmetric{RNN}: A dynamical system view on recurrent neural
  networks, 2019.

\bibitem{mgmetisChen}
Q.~Chen.
\newblock {MGMETIS}\text{---}mesh \& graph {METIS} partitioning.
\newblock {\em GitHub repository:
  \normalfont{\url{https://github.com/chiao45/mgmetis}}}, 2020.

\bibitem{CHEN2022110782}
Z.~Chen, V.~Churchill, K.L. Wu, and D.B. Xiu.
\newblock Deep neural network modeling of unknown partial differential
  equations in nodal space.
\newblock {\em Journal of Computational Physics}, 449:110782, 2022.

\bibitem{churchill2022deep}
V.~Churchill and D.B. Xiu.
\newblock Deep learning of chaotic systems from partially-observed data.
\newblock {\em arXiv preprint arXiv:2205.08384}, 2022.

\bibitem{clough1993dynamics}
R.W. Clough and J.~Penzien.
\newblock {\em Dynamics of Structures}.
\newblock Civil engineering series. McGraw-Hill, 1993.

\bibitem{fu2020learning}
X.H. Fu, L.B. Chang, and D.B. Xiu.
\newblock Learning reduced systems via deep neural networks with memory.
\newblock {\em Journal of Machine Learning for Modeling and Computing}, 1(2),
  2020.

\bibitem{FUNAHASHI1993801}
K.~Funahashi and Y.~Nakamura.
\newblock Approximation of dynamical systems by continuous time recurrent
  neural networks.
\newblock {\em Neural Networks}, 6(6):801--806, 1993.

\bibitem{pmlr-v80-garnelo18a}
M.~Garnelo, D.~Rosenbaum, C.~Maddison, T.~Ramalho, D.~Saxton, M.~Shanahan,
  Y.~W. Teh, D.~Rezende, and S.M.A. Eslami.
\newblock Conditional neural processes.
\newblock In {\em Proceedings of the 35th International Conference on Machine
  Learning}, volume~80 of {\em Proceedings of Machine Learning Research}, pages
  1704--1713. PMLR, 10--15 Jul 2018.

\bibitem{goodfellow2016deep}
Ian Goodfellow, Yoshua Bengio, and Aaron Courville.
\newblock {\em Deep learning}.
\newblock MIT Press, Cambridge, MA, 2016.

\bibitem{he2016deep}
K.M. He, X.Y. Zhang, S.Q. Ren, and J.~Sun.
\newblock Deep residual learning for image recognition.
\newblock In {\em Proceedings of the IEEE conference on computer vision and
  pattern recognition}, pages 770--778, 2016.

\bibitem{lstm}
S.~Hochreiter and J.~Schmidhuber.
\newblock Long short-term memory.
\newblock {\em Neural Comput.}, 9(8):1735–1780, nov 1997.

\bibitem{Hu2022NeuralPDEAR}
Y.H. Hu, T.~Zhao, S.X. X{\'u}, Z.L. Xu, and L.Z. Lin.
\newblock Neural-{PDE}: a {RNN} based neural network for solving time dependent
  {PDE}s.
\newblock {\em Communications in Information and Systems}, 2022.

\bibitem{hughes2012finite}
T.J.R. Hughes.
\newblock {\em The finite element method: linear static and dynamic finite
  element analysis}.
\newblock Dover Publications, INC., Mineola, New York, 2012.

\bibitem{huthwaite2014accelerated}
P.~Huthwaite.
\newblock Accelerated finite element elastodynamic simulations using the {GPU}.
\newblock {\em Journal of Computational Physics}, 257:687--707, 2014.

\bibitem{joldes2010real}
G.R. Joldes, A.~Wittek, and K.~Miller.
\newblock Real-time nonlinear finite element computations on {GPU}--application
  to neurosurgical simulation.
\newblock {\em Computer methods in applied mechanics and engineering},
  199(49-52):3305--3314, 2010.

\bibitem{sindy2}
K.~Kaheman, J.N. Kutz, and S.L. Brunton.
\newblock {SINDy-PI}: a robust algorithm for parallel implicit sparse
  identification of nonlinear dynamics.
\newblock {\em Proceedings of the Royal Society A: Mathematical, Physical and
  Engineering Sciences}, 476(2242):20200279, 2020.

\bibitem{Kaiser_2021}
E.~Kaiser, J.N. Kutz, and S.L Brunton.
\newblock Data-driven discovery of {K}oopman {E}igenfunctions for control.
\newblock {\em Machine Learning: Science and Technology}, 2(3):035023, june
  2021.

\bibitem{METIS}
George Karypis and Vipin Kumar.
\newblock {MeTis: Unstructured Graph Partitioning and Sparse Matrix Ordering
  System, Version 4.0}.
\newblock \url{http://www.cs.umn.edu/~metis}, 2009.

\bibitem{kim2010patient}
H.J. Kim, I.E. Vignon-Clementel, J.S. Coogan, C.A. Figueroa, K.E. Jansen, and
  C.A. Taylor.
\newblock Patient-specific modeling of blood flow and pressure in human
  coronary arteries.
\newblock {\em Annals of biomedical engineering}, 38(10):3195--3209, 2010.

\bibitem{kingma2014adam}
D.P. Kingma and J.~Ba.
\newblock Adam: A method for stochastic optimization.
\newblock {\em arXiv preprint arXiv:1412.6980}, 2014.

\bibitem{komatitsch2010high}
D.~Komatitsch, G.~Erlebacher, D.~G{\"o}ddeke, and D.~Mich{\'e}a.
\newblock High-order finite-element seismic wave propagation modeling with
  {MPI} on a large {GPU} cluster.
\newblock {\em Journal of computational physics}, 229(20):7692--7714, 2010.

\bibitem{2022-conditional}
A.~Kovacs, L.~Exl, A.~Kornell, J.~Fischbacher, M.~Hovorka, M.~Gusenbauer,
  L.~Breth, H.~Oezelt, M.~Yano, N.~Sakuma, A.~Kinoshita, T.~Shoji, A.~Kato, and
  T.~Schrefl.
\newblock Conditional physics informed neural networks.
\newblock {\em Communications in Nonlinear Science and Numerical Simulation},
  104:106041, Jan 2022.

\bibitem{10.1145/3322813}
M.~Kronbichler and K.~Ljungkvist.
\newblock Multigrid for matrix-free high-order finite element computations on
  graphics processors.
\newblock {\em ACM Trans. Parallel Comput.}, 6(1), may 2019.

\bibitem{lstm-ed}
L.~Kulowski.
\newblock {LSTM}\_encoder\_decoder.
\newblock {\em GitHub repository:
  {\normalfont{\url{https://github.com/lkulowski/LSTM_encoder_decoder}}}},
  2020.

\bibitem{kutz2016dynamic}
J.N. Kutz, S.L. Brunton, B.W. Brunton, and J.L. Proctor.
\newblock {\em Dynamic mode decomposition: data-driven modeling of complex
  systems}.
\newblock SIAM, 2016.

\bibitem{li2021ensemble}
X.~Li and D.E. Schiavazzi.
\newblock An ensemble solver for segregated cardiovascular {FSI}.
\newblock {\em Computational Mechanics}, 68(6):1421--1436, 2021.

\bibitem{lu2021learning}
Lu~Lu, Pengzhan Jin, Guofei Pang, Zhongqiang Zhang, and George~Em Karniadakis.
\newblock Learning nonlinear operators via deeponet based on the universal
  approximation theorem of operators.
\newblock {\em Nature Machine Intelligence}, 3(3):218--229, 2021.

\bibitem{lstm1}
P.~Malhotra, A.~Ramakrishnan, G.~Anand, L.~Vig, P.~Agarwal, and G.~Shroff.
\newblock {LSTM}-based encoder-decoder for multi-sensor anomaly detection,
  2016.

\bibitem{mccaslin2012closed}
S.E. McCaslin, P.S. Shiakolas, B.H. Dennis, and K.L. Lawrence.
\newblock Closed-form stiffness matrices for higher order tetrahedral finite
  elements.
\newblock {\em Advances in Engineering Software}, 44(1):75--79, 2012.

\bibitem{olovsson2005selective}
L.~Olovsson, K.~Simonsson, and M.~Unosson.
\newblock Selective mass scaling for explicit finite element analyses.
\newblock {\em International Journal for Numerical Methods in Engineering},
  63(10):1436--1445, 2005.

\bibitem{park2018sequence}
S.H. Park, B.D. Kim, C.M. Kang, C.C. Chung, and J.W. Choi.
\newblock Sequence-to-sequence prediction of vehicle trajectory via {LSTM}
  encoder-decoder architecture.
\newblock In {\em 2018 IEEE Intelligent Vehicles Symposium (IV)}, pages
  1672--1678. IEEE, 2018.

\bibitem{lauren_mf}
Lauren Partin, Gianluca Geraci, Ahmad Rushdi, Michael~S. Eldred, and Daniele~E.
  Schiavazzi.
\newblock Multifidelity data fusion in convolutional encoder/decoder networks,
  2022.

\bibitem{paszke2019pytorch}
A.~Paszke, S.~Gross, F.~Massa, A.~Lerer, J.~Bradbury, G.~Chanan, T.~Killeen,
  Z.M. Lin, N.~Gimelshein, L.~Antiga, et~al.
\newblock Pytorch: An imperative style, high-performance deep learning library.
\newblock {\em Advances in neural information processing systems},
  32:8026--8037, 2019.

\bibitem{QIN2019620}
T.~Qin, K.L. Wu, and D.B. Xiu.
\newblock Data driven governing equations approximation using deep neural
  networks.
\newblock {\em Journal of Computational Physics}, 395:620--635, 2019.

\bibitem{raissi2019physics}
Maziar Raissi, Paris Perdikaris, and George~E Karniadakis.
\newblock Physics-informed neural networks: A deep learning framework for
  solving forward and inverse problems involving nonlinear partial differential
  equations.
\newblock {\em Journal of Computational physics}, 378:686--707, 2019.

\bibitem{reduced}
D.~Schillinger, J.A. Evans, F.~Frischmann, R.R. Hiemstra, M.C. Hsu, and T.J.R.
  Hughes.
\newblock A collocated {C0} finite element method: {R}educed quadrature
  perspective, cost comparison with standard finite elements, and explicit
  structural dynamics.
\newblock {\em International Journal for Numerical Methods in Engineering},
  102(3-4):576--631, 2015.

\bibitem{Scovazzi2016}
G.~Scovazzi, B.~Carnes, X.~Zeng, and S.~Rossi.
\newblock A simple, stable, and accurate linear tetrahedral finite element for
  transient, nearly, and fully incompressible solid dynamics: a dynamic
  variational multiscale approach.
\newblock {\em International Journal for Numerical Methods in Engineering},
  106(10):799--839, 2016.

\bibitem{seo2020effects}
J.~Seo, D.E. Schiavazzi, A.M. Kahn, and A.L. Marsden.
\newblock The effects of clinically-derived parametric data uncertainty in
  patient-specific coronary simulations with deformable walls.
\newblock {\em International journal for numerical methods in biomedical
  engineering}, 36(8):e3351, 2020.

\bibitem{seo2019performance}
J.~Seo, D.E. Schiavazzi, and A.L. Marsden.
\newblock Performance of preconditioned iterative linear solvers for
  cardiovascular simulations in rigid and deformable vessels.
\newblock {\em Computational mechanics}, 64(3):717--739, 2019.

\bibitem{PhysRevResearch.3.023255}
D.E. Shea, S.L. Brunton, and J.N. Kutz.
\newblock {SINDy-BVP}: Sparse identification of nonlinear dynamics for boundary
  value problems.
\newblock {\em Phys. Rev. Research}, 3:023255, Jun 2021.

\bibitem{sherstinsky2020fundamentals}
A.~Sherstinsky.
\newblock Fundamentals of recurrent neural network ({RNN}) and long short-term
  memory ({LSTM}) network.
\newblock {\em Physica D: Nonlinear Phenomena}, 404:132306, 2020.

\bibitem{shiakolas1992closed}
P.S. Shiakolas, R.V. Nambiar, K.L. Lawrence, and W.A. Rogers.
\newblock Closed-form stiffness matrices for the linear strain and quadratic
  strain tetrahedron finite elements.
\newblock {\em Computers \& structures}, 45(2):237--242, 1992.

\bibitem{SchillingerMassScaling2022}
S.K.F. Stoter, T.H. Nguyen, R.R. Hiemstra, and D.~Schillinger.
\newblock Variationally consistent mass scaling for explicit time-integration
  schemes of lower- and higher-order finite element methods, 2022.

\bibitem{strbac2017gpgpu}
V.~Strbac, D.M. Pierce, J.~Vander~Sloten, and N.~Famaey.
\newblock {GPGPU}-based explicit finite element computations for applications
  in biomechanics: the performance of material models, element technologies,
  and hardware generations.
\newblock {\em Computer Methods in Biomechanics and Biomedical Engineering},
  20(16):1643--1657, 2017.

\bibitem{4388142}
Z.A. Taylor, M.~Cheng, and S.~Ourselin.
\newblock High-speed nonlinear finite element analysis for surgical simulation
  using graphics processing units.
\newblock {\em IEEE Transactions on Medical Imaging}, 27(5):650--663, 2008.

\bibitem{chaotic17}
P.R. Vlachas, W.~Byeon, Z.Y. Wan, T.P. Sapsis, and P.~Koumoutsakos.
\newblock Data-driven forecasting of high-dimensional chaotic systems with long
  short-term memory networks.
\newblock {\em Proceedings of the Royal Society A: Mathematical, Physical and
  Engineering Sciences}, 474(2213):20170844, 2018.

\bibitem{WU2020109307}
K.L. Wu and D.B. Xiu.
\newblock Data-driven deep learning of partial differential equations in modal
  space.
\newblock {\em Journal of Computational Physics}, 408:109307, 2020.

\bibitem{reduced2}
O.C. Zienkiewicz, R.L. Taylor, and J.M. Too.
\newblock Reduced integration technique in general analysis of plates and
  shells.
\newblock {\em International Journal for Numerical Methods in Engineering},
  3(2):275--290, 1971.

\end{thebibliography}

\end{document}